\documentclass[preprint,12pt]{elsarticle}

\usepackage{amssymb}

\usepackage{amsmath,amssymb,amsfonts}
\usepackage{array}
\newcolumntype{P}[1]{>{\centering\arraybackslash}p{#1}}
\usepackage{tabularx}
\usepackage{booktabs}
\usepackage{supertabular}
\usepackage{epstopdf}
\usepackage{longtable}
\usepackage{fancyhdr}
\usepackage{mathtools}
\usepackage{float}
\usepackage{subfigure}
\usepackage{graphics}
\usepackage{graphicx}
\usepackage{subfigure}
\usepackage{epstopdf}
\usepackage{algorithm}
\usepackage{algpseudocode}
\usepackage{multirow}
\usepackage{multicol}
\usepackage{adjustbox}
\usepackage{hhline}
\usepackage{color}
\usepackage[nohyperlinks]{acronym}
\usepackage{listings}
\usepackage[table,xcdraw]{xcolor}
\usepackage{textcomp}
\usepackage{xcolor}
\usepackage{comment}
\usepackage{makecell}
\usepackage{longtable}
\usepackage{caption}
\usepackage{hyperref}
\usepackage{enumitem}
\usepackage{url}
\usepackage{ragged2e}

\usepackage{enumitem}

\newlist{tabitemize}{itemize}{1}
\setlist[tabitemize]{label=\tiny\textbullet, 
                     leftmargin=*,
                     nosep, 
                     before=\begin{minipage}[t]{\hsize}\raggedright, 
                     after=\end{minipage}}
\definecolor{codegreen}{rgb}{0,0.6,0}
\definecolor{codegray}{rgb}{0.5,0.5,0.5}

\lstdefinestyle{mystyle}{
    backgroundcolor=\color{white},   
    keywordstyle=\color{magenta},
    numberstyle=\tiny\color{codegray},
    stringstyle=\color{codegreen},
    basicstyle=\ttfamily\footnotesize,
    keywordstyle=\bfseries,
    morekeywords={allocate, set, prioritize, for, on, over},
    breakatwhitespace=false,         
    breaklines=true,                 
    captionpos=b,                    
    keepspaces=true,                 
    numbers=left,                    
    numbersep=5pt,                  
    showspaces=false,                
    showstringspaces=false,
    showtabs=false,                  
    tabsize=2,
    morestring=*[d]{"},
}

\journal{Computer Networks}

\begin{document}
\begin{frontmatter}
\title{Zero-Touch Networks: Towards Next-Generation Network Automation}

\author{Mirna El Rajab}
\ead{melrajab@uwo.ca}

\author{Li Yang}
\ead{lyang339@uwo.ca}

\author{Abdallah Shami}
\ead{abdallah.shami@uwo.ca}

\address{Department of Electrical and Computer Engineering, Western University, 1151 Richmond St., London, Ontario, Canada N6A 3K7}

\date{}

\begin{abstract}
The Zero-touch network and Service Management (ZSM) framework represents an emerging paradigm in the management of the fifth-generation (5G) and Beyond (5G+) networks, offering automated self-management and self-healing capabilities to address the escalating complexity and the growing data volume of modern networks. ZSM frameworks leverage advanced technologies such as Machine Learning (ML) to enable intelligent decision-making and reduce human intervention. This paper presents a comprehensive survey of Zero-Touch Networks (ZTNs) within the ZSM framework, covering network optimization, traffic monitoring, energy efficiency, and security aspects of next-generational networks. The paper explores the challenges associated with ZSM, particularly those related to ML, which necessitate the need to explore diverse network automation solutions. In this context, the study investigates the application of Automated ML (AutoML) in ZTNs, to reduce network management costs and enhance performance. AutoML automates the selection and tuning process of a ML model for a given task. Specifically, the focus is on AutoML's ability to predict application throughput and autonomously adapt to data drift. Experimental results demonstrate the superiority of the proposed AutoML pipeline over traditional ML in terms of prediction accuracy. Integrating AutoML and ZSM concepts significantly reduces network configuration and management efforts, allowing operators to allocate more time and resources to other important tasks. The paper also provides a high-level 5G system architecture incorporating AutoML and ZSM concepts. This research highlights the potential of ZTNs and AutoML to revolutionize the management of 5G+ networks, enabling automated decision-making and empowering network operators to achieve higher efficiency, improved performance, and enhanced user experience.
\end{abstract}


\end{frontmatter}

\section{Introduction}
In today's digital world, we rely on telecommunication networks for more than just phone calls. From streaming movies to controlling smart home devices, these networks have revolutionized the way we live, work, and communicate. As we move towards a world where the Internet of Things (IoT) is becoming increasingly widespread, the need for Next-Generation Networks (NGNs) has only grown more indispensable.

NGNs, such as the fifth-generation (5G) and the upcoming sixth-generation (6G) networks, mark a landmark in telecommunications history; these networks represent not only an upgrade from their predecessors but also a paradigm shift in terms of speed, latency, capacity, and reliability - unlocking new possibilities for emerging applications and service areas. According to the International Telecommunication Union IMT-2020, three core service areas for 5G networks include enhanced Mobile Broadband (eMBB), ultra-Reliable Low-Latency Communication (uRLLC), and massive Machine-Type Communication (mMTC) \cite{rancy2016imt}. Each service area addresses specific use cases such as multimedia content access (eMBB), mission-critical applications (uRLLC), or smart cities (mMTC).

NGNs have the potential to unlock the full potential and meet the challenging requirements of future use cases, but to fully realize this potential, they must be designed as highly-flexible and programmable infrastructures that are context-aware and service-aware. Advancements such as Software Defined Networking (SDN), Network Function Virtualization (NFV), Multi-access Edge Computing (MEC), and network slicing play a pivotal role in the network architecture \cite{SDNNFV2}. These technologies will open up new business models, such as multi-domain, multi-service, and multi-tenancy models, to support new markets.

The growth of NGNs has brought with it new challenges, particularly in terms of network management. As networks become more intricate, traditional manual methods for configuring, deploying, and maintaining them become cumbersome, time-consuming, and error prone \cite{aizsm5gp}. To tackle this issue, various efforts have been made to introduce intelligence and reasoning into mobile networks for automation and optimization purposes. These efforts include active networks \cite{activenetworks}, self-organizing networks \cite{SON}, autonomic network management \cite{autonomicmanagement}, and Zero-Touch Networks (ZTNs) \cite{zsmsurvey}.

The ZTN approach has emerged as a fully automated management solution, enabling the network to analyze its current state, interpret it, and provide suggestions for possible reconfigurations - while leaving validation and acceptance up to a human operator \cite{zsm5g6g}. Implementing ZTN concepts and technologies will be essential for operators to achieve greater levels of automation, improve network performance, and reduce time-to-market for new features. ZTN-based solutions are available for a diverse set of problems, from managing resources to ensuring network security and privacy. European Telecommunications Standards Institute (ETSI) has gained interest in shifting towards ZTN-based solutions. In 2017, ETSI created a Zero-touch network and Service Management (ZSM) Industry Specification Group (ISG), to define the requirements and architecture for a network automation framework based on ZTN concepts \cite{etsi002}. ZSM will ensure that NGNs remain responsive to evolving user needs and demands via proactive network management techniques. These techniques leverage the power of Artificial Intelligence (AI) and Machine Learning (ML) to automate and optimize network operations, enabling efficient resource allocation, dynamic service provisioning, and predictive maintenance. AI and ML are technologies that enable systems to simulate human intelligence, learn from data, and make intelligent decisions or predictions.

ZSM still faces significant ML challenges, such as the need for effective feature engineering, algorithm selection, and hyperparameter tuning. Thus, there is a need to explore other network automation solutions that can complement ZSM in order to achieve higher levels of automation and efficiency. One such solution is Automated ML (AutoML), which helps address these challenges by automating the ML pipeline and improving the efficiency and effectiveness of the ZSM solution. AutoML handles crucial tasks such as data preprocessing, feature engineering, model selection, hyperparameter tuning, model evaluation, and even model updating. By automating these processes, AutoML significantly reduces the manual effort needed to develop high-performing ML models. 

Accordingly, this survey aims to provide a comprehensive overview of ZSM in 5G and Beyond (5G+) networks, with a focus on network optimization, energy efficiency, network security, and traffic control. By highlighting the ML challenges in ZSM and exploring the potential of AutoML, this survey aims to contribute to the development of more effective network automation solutions. In particular, this survey offers the following notable contributions:
\begin{enumerate}
    \item Review of current standards, architectures, and projects: The paper examines the existing ZSM standard, its reference architecture, and other related projects. This analysis serves as a valuable reference and enhances the understanding of the current landscape in ZSM.
    \item In-depth analysis of ZSM in 5G+ networks: The paper comprehensively examines various aspects of ZSM in NGNs. It delves into network optimization, energy efficiency, network security, and traffic control, providing a comprehensive understanding of zero-touch applications.
    \item Identification of ML challenges in ZSM: The paper highlights the challenges associated with applying ML techniques in the context of ZSM.
    \item Exploration of network automation solutions: The paper explores the potential of AutoML and DTs as viable solutions for addressing the ML-related challenges in ZSM.
    \item Detailed breakdown of the AutoML pipeline: The paper provides a comprehensive understanding of the steps involved in the AutoML pipeline and reviews existing techniques for each step. This breakdown aids in the implementation and application of AutoML in NGNs.
    \item Application of AutoML in ZSM: Through a detailed case study, the paper demonstrates the real-world application of an online AutoML pipeline for network traffic tasks within the ZSM framework.
    \item Discussion of research challenges and future directions: The paper discusses the research challenges in ZSM and AutoML, suggesting future directions for further exploration and advancement.
\end{enumerate}

In comparison to earlier review papers on the subjects of ZSM \cite{survey5GP, zsmsurvey} and network automation solutions \cite{zsm5g6g, MLCH1}, this survey stands out with the following differences and improvements:
\begin{enumerate}
    \item Comprehensive survey of zero-touch applications in NGNs: It is the first paper to comprehensively explore Zero-Touch Network Operation (ZNO) applications, spanning network optimization, traffic control, energy efficiency, and security in 5G+ networks.
    \item Addressing ML challenges with innovative solutions: Unlike previous papers, this paper tackles ML challenges in ZSM and proposes AutoML and Digital Twins (DTs) as potential automation solutions.
    \item Practical case study of online AutoML: This paper presents the first case study applying an online AutoML pipeline to a real network traffic task within the ZSM context. Additionally, it outlines a high-level architecture integrating AutoML and ZSM concepts into a 5G system.
\end{enumerate}

\begin{figure*}[htbp]
	\centering
	\includegraphics[height=0.95\textheight,keepaspectratio]{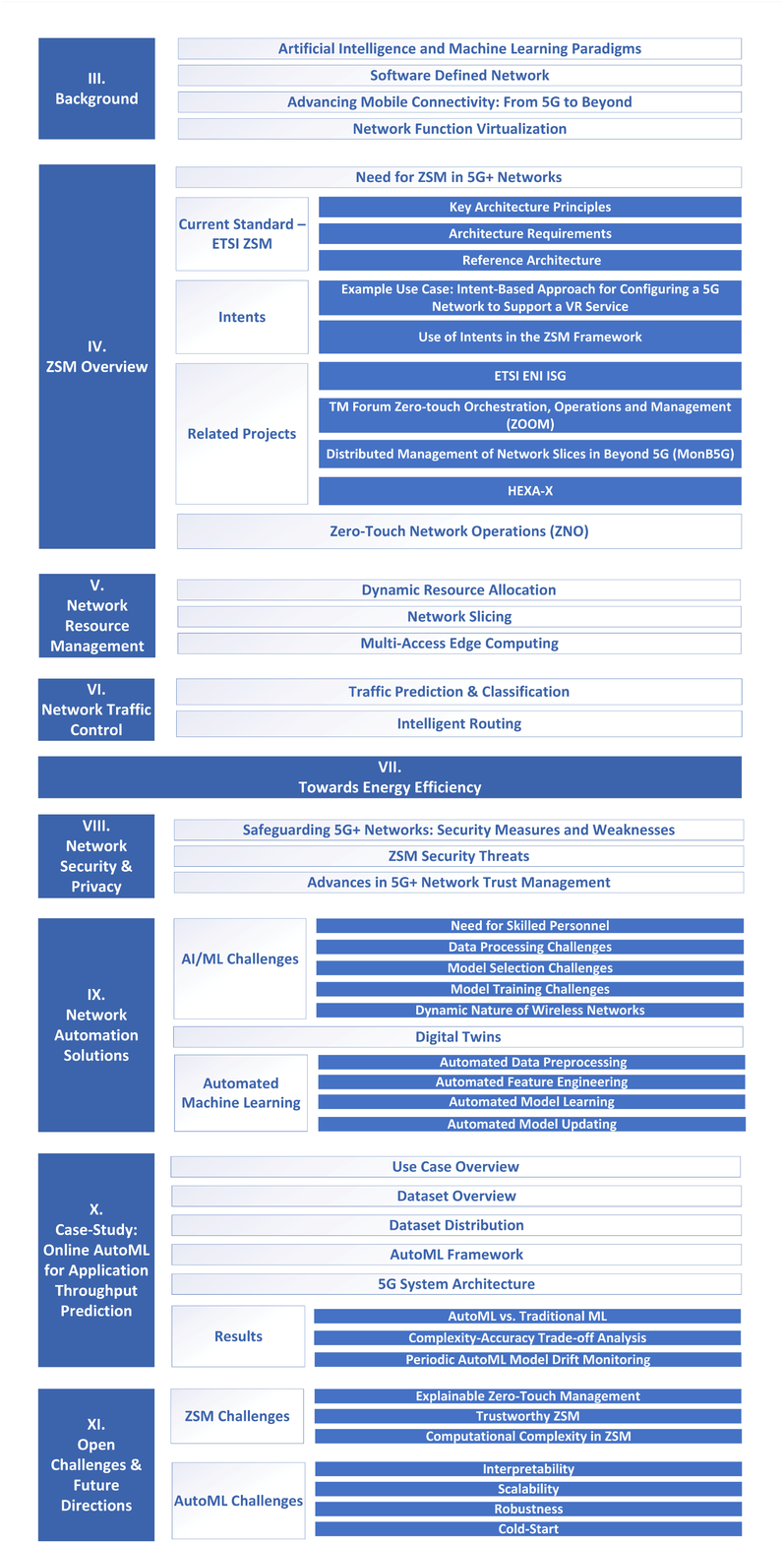}
	\caption[]{Survey Outline}
	\label{fig:outline}
\end{figure*}

The remainder of this paper is organized into nine sections, as illustrated in Figure \ref{fig:outline}, providing a comprehensive analysis of ZSM in NGNs. Section \ref{sec:acronyms} lists all the acronyms used in this paper. Section \ref{sec:background} lays the foundation for ZSM by exploring the pillars of the framework, including the ML paradigm and 5G+ networks along with their enablers, SDN and NFV. Section \ref{sec:overview} takes a deep dive into ZSM, examining its role in managing the complexities of 5G+ networks. We explore the current standard's requirements, reference architecture, and use of intents, as well as other related projects. In Section \ref{sec:resourcemanagement}, we analyze network resource management and optimization, looking at dynamic resource allocation, network slicing, and MEC. Section \ref{sec:trafficcontrol} focuses on network traffic control, from traffic prediction and classification to intelligent routing. Additionally, in Sections \ref{sec:energyefficiency} and \ref{sec:networksecurity}, we analyze ZSM's potential in terms of energy efficiency and network security, respectively, addressing 5G+ security measures and weaknesses, ZSM security threats, and recent advances in 5G+ network trust management in the latter section. In Section \ref{sec:automationsolutions}, we explore different network automation solutions to tackle certain ML challenges. In particular, automation solutions, such as DTs and AutoML, have significant importance in a world where communication networks are integral to daily life. Section \ref{sec:casestudy} further studies AutoML through a use-case to predict application throughput and see how it fits in a ZSM framework. In Section \ref{sec:challenges}, we discuss the research challenges in this field and future lines of work. Finally, Section \ref{sec:conc} concludes the survey.

\section{List of Acronyms}
\label{sec:acronyms}
\subsection*{Artificial Intelligence Acronyms}
\begin{acronym}[Seq2SeqQ]
\acro{AI}{Artificial Intelligence}
\acro{AutoML}{Automated Machine Learning}
\acro{ANN}{Artificial Neural Network}
\acro{CNN}{Convolutional Neural Network}
\acro{DDPG}{Deep Deterministic Policy Gradient}
\acro{DL}{Deep learning}
\acro{DRL}{Deep Reinforcement Learning}
\acro{DQN}{Deep Q-Network}
\acro{GRU}{Gated Recurrent Unit}
\acro{LSTM}{Long Short-Term Memory}
\acro{MAE}{Mean Absolute Error}
\acro{MAPE}{Mean Absolute Percentage Error}
\acro{ML}{Machine Learning}
\acro{MLP}{Multi-Layer Perceptron}
\acro{NAS}{Neural Architecture Search}
\acro{PCA}{Principle Component Analysis}
\acro{Seq2Seq}{Sequence-to-Sequence}
\acro{SL}{Supervised Learning}
\acro{SVM}{Support Vector Machine}
\acro{RL}{Reinforcement Learning}  
\acro{RNN}{Recurrent Neural Network}
\acro{XAI}{Explainable Artificial Intelligence}
\end{acronym}

\subsection*{Network Management Acronyms}
\begin{acronym}[AREL3PP]
\acro{AREL3P}{Adapted REinforcement Learning VNF Performance Prediction module for Autonomous VNF Placement}
\acro{CLARA}{Closed Loop-based zero-touch network mAnagement fRAmework}
\acro{DASA}{Dynamic Auto-Scaling Algorithm}
\acro{DLT}{Distributed Ledger Technology}
\acro{DT}{Digital Twin}
\acro{HARNESS}{High Availability supportive self-Reliant NEtwork Slicing System}
\acro{MANO}{Management and Orchestration}
\acro{MD}{Management Domain}
\acro{MonB5G}{Distributed Management of Network Slices in Beyond 5G}
\acro{NAP}{Novel Autonomous Profiling}
\acro{NSOS}{Network Slicing Orchestration System}
\acro{PFR}{Proactive Failure Recovery}
\acro{RIRM}{Reliable Intelligent Routing Mechanism}
\acro{TM}{TeleManagement}
\acro{ZNO}{Zero-Touch Network Operation}
\acro{ZOOM}{Zero-touch Orchestration, Operations and Management}
\acro{ZSM}{Zero-Touch Network and Service Management}
\acro{ZTN}{Zero-Touch Network}
\acro{ZT-PFR}{Zero-Touch PFR}
\end{acronym}

\subsection*{Performance Metrics Acronyms}
\begin{acronym}[CAPEXX]
\acro{CAPEX}{Capital Expenditure}
\acro{CPU}{Central Processing Unit}
\acro{E2E}{End-to-End}
\acro{KPI}{Key Performance Indicator}
\acro{OPEX}{Operational Expenditure}
\acro{QoE}{Quality of Experience}
\acro{QoS}{Quality of Service}
\acro{SLA}{Service Level Agreement}    
\end{acronym}

\subsection*{Next-Generation Networks Acronyms}
\begin{acronym}[mmWaveE]
\acro{5G+}{5G and Beyond}
\acro{AES}{Application Edge Slice}
\acro{AMF}{Application Management Function}
\acro{B5G}{Beyond 5G}
\acro{CN}{Core Network}
\acro{eMBB}{enhanced Mobile Broadband}
\acro{MEC}{Multi-access Edge Computing}
\acro{mMTC}{massive Machine-Type Communication}
\acro{gNB}{gNodeB}
\acro{NFV}{Network Function Virtualization}
\acro{NGN}{Next-Generation Network}
\acro{NSSF}{Network Slice Selection Function}
\acro{NSSMF}{Network Slice Subnet Management Function}  
\acro{NWDAF}{Network Data Analytics Function}
\acro{RAN}{Radio Access Network}
\acro{SDN}{Software Defined Network}
\acro{UE}{User Equipment}
\acro{UPF}{User Plane Function}
\acro{uRLLC}{ultra-Reliable Low-Latency Communication}
\acro{VNF}{Virtual Network Function}
\acro{mmWave}{millimeter wave}
\end{acronym}

\subsection*{Network Security Acronyms}
\begin{acronym}[DDoSS]
\acro{DDoS}{Distributed Denial of Service}
\acro{DOS}{Denial of Service}
\acro{IDS}{Intrusion Detection System}
\acro{MitM}{Man in the Middle}
\acro{MUD}{Manufacturer Usage Description}
\acro{TEE}{Trusted Execution Environment}
\acro{TRM}{Trust and Reputation Manager}
\end{acronym}

\subsection*{Organizations and Programs}
\begin{acronym}[H20200]
\acro{ENI}{Experiential Network Intelligence}
\acro{ETSI}{European Telecommunications Standards Institute}
\acro{H2020}{Horizon 2020}
\acro{ISG}{Industry Specification Group}
\end{acronym}

\subsection*{General Telecommunication Acronyms}
\begin{acronym}[FANETT]
\acro{API}{Application Programming Interface}
\acro{CSP}{Communication Service Provider}
\acro{IoT}{Internet of Things}
\acro{RSRP}{Reference Signal Received Power}
\acro{RSRQ}{Reference Signal Received Quality}
\acro{RSSI}{Received Signal Strength Indicator}
\acro{FANET}{Flying Ad Hoc Network}
\acro{V2X}{Vehicle-to-Everything}
\acro{VPN}{Virtual Private Network}
\acro{VM}{Virtual Machine}
\acro{VR}{Virtual Reality}
\acro{USR}{User Service Request}
\acro{WLAN}{Wireless Local Area Network}
\end{acronym}

\section{Background} \label{sec:background}
As the telecommunications industry moves towards the deployment of 5G+ networks and the implementation of ZSM, advanced technologies such as AI/ML, SDN, and NFV are becoming essential components of network infrastructure. Such advancements support intelligent, flexible, and automated network management which in turn allow for efficient and scalable operations for network operators and service providers. AI/ML optimizes network performance, predicts and prevents network failures, and automates network management tasks. SDN provides a simpler approach to network management by dividing the control and data planes and enabling the dynamic allocation of network resources. NFV enables the deployment of network functions as the software on general-purpose hardware and eliminates the need for proprietary hardware by decoupling network functions.

\subsection{Artificial Intelligence and Machine Learning Paradigms}
AI refers to "the simulation of human intelligence in machines that are programmed to think like humans and mimic their actions" \cite{AIin5G}. In other words, AI refers to the ability of systems to imitate human cognitive functions such as learning. ML is an application of AI that enables machines to learn from large volumes of data and make predictions without directly being instructed. ML is considered a subset of AI. ML can be further divided into three main categories: supervised, unsupervised, and reinforcement learning \cite{MLterms}. Table \ref{MLtechniques} provides an overview of common techniques used in each ML category, outlining their strengths and weaknesses to provide a comprehensive understanding of each approach \cite{autoML,compareML, compareML2, compareML3, compareCNN, compareQlearning}. 

\begin{center}
\renewcommand{\arraystretch}{1.5}
\scriptsize
\begin{longtable}{|P{2cm} P{3.5cm} P{3cm} P{3cm}|}
\caption{An Overview of Traditional ML Algorithms \cite{autoML,compareML,compareML2,compareML3,compareCNN, compareQlearning}} \label{MLtechniques} 
\\

\hline \multicolumn{1}{|c}{\textbf{ML Algorithm}} & \multicolumn{1}{c}{\textbf{Description}} & \multicolumn{1}{c}{\textbf{Advantages}} & \multicolumn{1}{c|}{\textbf{Limitations}} \\ \hline 
\endfirsthead

\multicolumn{4}{c}%
{{\bfseries \tablename\ \thetable{} -- continued from previous page}} \\
\hline \multicolumn{1}{|c}{\textbf{ML Algorithm}} &
\multicolumn{1}{c}{\textbf{Description}} &
\multicolumn{1}{c}{\textbf{Advantages}} & \multicolumn{1}{c|}{\textbf{Limitations}} \\ \hline 
\endhead

\hline \multicolumn{4}{|r|}{{Continued on next page}} \\ \hline
\endfoot

\hline
\endlastfoot

\textbf{Linear Regression} & \justifying A statistical model for predicting continuous variables based on one or more independent variables  & \begin{tabitemize}
    \item Easy to understand \item Avoid overfitting by regularization
\end{tabitemize} & \begin{tabitemize}
    \item Oversimplification of real world problems
    \item Limited to linear relationships
\end{tabitemize} \\ \hline

\textbf{Logistic Regression} & \justifying A statistical model for binary classification based on input features  & \begin{tabitemize}
    \item Simple to implement \item Computationally Efficient \item No need for feature scaling
\end{tabitemize} & \begin{tabitemize}  
    \item High reliance on proper presentation of data
    \item Inability to solve non-linear problem
\end{tabitemize} \\ \hline

\textbf{Decision Trees} & \justifying A tree-based model that partitions data into subsets based on feature values and predicts the class/value of a new instance by traversing the tree from the root to a leaf node &  \begin{tabitemize}
    \item Easy to interpret and visualize \item Process categorical and continuous features \item Handle missing values
\end{tabitemize} & \begin{tabitemize}  
    \item Prone to overfitting
    \item Sensitive to data
    \item Potential to produce a locally optimal solution
\end{tabitemize} \\ \hline

\textbf{Random Forest} & \justifying An ensemble method combining multiple decision trees through averaging to improve the accuracy and robustness of classification and regression tasks  & \begin{tabitemize}
    \item Robust performance with high-dimensional datasets
    \item Deal with imbalanced datasets
    \item Extract feature importance 
\end{tabitemize} & \begin{tabitemize}
    \item Slow in predictions
    \item Appear as a black box
\end{tabitemize} \\ \hline

\textbf{Naïve Bayes} & \justifying A probabilistic algorithm that uses Bayes' theorem to predict the class of new data based on the conditional probability of features given a class  & \begin{tabitemize} 
    \item Address multi-class classification problems
    \item Insensitive to irrelevant features
\end{tabitemize} & \begin{tabitemize}
    \item Assume independent features
    \item Handle discrete datasets better than continuous datasets
\end{tabitemize} \\ \hline

\textbf{Support Vector Machine} & \justifying A linear model for classification and regression that finds the best hyperplane to separate data points into different classes in a high-dimensional feature space  & \begin{tabitemize} 
    \item Work with high-dimensional data
    \item Handle non-linear relationships through kernel functions    
\end{tabitemize} & \begin{tabitemize}
    \item Slow training with large datasets
    \item Poor performance with noisy data
\end{tabitemize} \\ \hline

\textbf{K-Nearest Neighbors} & \justifying A non-parametric method for classifying data points based on the majority class among the \textit{K} training instances closest to it in the feature space & \begin{tabitemize}
    \item Simple
    \item Handle multi-class problems
    \item Make no assumption about the data
\end{tabitemize} & \begin{tabitemize}
    \item Slow for large datasets
    \item Suffer the curse of dimensionality
    \item Sensitive to outliers
\end{tabitemize} \\ \hline

\textbf{K-Means} & \justifying A clustering algorithm that partitions \textit{n} observations into \textit{k} clusters based on the similarity of features & \begin{tabitemize} 
    \item Simple and easy to implement
    \item Computationally efficient, making it suitable for large datasets
\end{tabitemize} & \begin{tabitemize}
    \item Assume spherical, equally-sized clusters
    \item Sensitive to the initial position of centroids
\end{tabitemize} \\ \hline

\textbf{Principle Component Analysis} & \justifying An orthogonal linear transformation technique that transforms the data into a lower-dimensional space & \begin{tabitemize} 
    \item Effective feature extraction
    \item Reduce dimensionality for better analysis
\end{tabitemize} & \begin{tabitemize}
    \item May result in loss of information
    \item Assume a linear relationship between features
\end{tabitemize} \\ \hline

\textbf{Feed-Forward Neural Networks} & \justifying A network of interconnected processing nodes that learns to recognize patterns in data by adjusting the inter-node weights during training & \begin{tabitemize}
    \item Can handle complex non-linear relationships
    \item Work well with large datasets
\end{tabitemize} & \begin{tabitemize}
    \item Need abundant training data
    \item Prone to overfitting
    \item Difficult to interpret
\end{tabitemize} \\ \hline

\textbf{Convolutional Neural Networks} & \justifying An ANN that uses convolutional layers to automatically learn spatial hierarchies and extract features from image and video data  & \begin{tabitemize}
    \item Effective in image and video analysis
    \item Reduce the number of parameters through weight sharing
    \item Learn hierarchical representations
\end{tabitemize} & \begin{tabitemize}
    \item Complex computations for convolutional and pooling operations
    \item Require large amounts of data
\end{tabitemize} \\ \hline

\textbf{Recurrent Neural Network} &\justifying  An ANN that can handle sequential data by using loops to maintain a hidden state that incorporates past information  & \begin{tabitemize}
    \item Well-suited for sequence-related tasks and time-series data
    \item Can handle variable-length inputs and outputs
\end{tabitemize} & \begin{tabitemize}
    \item Incapable of capturing long-term dependencies
    \item Sensitive to the exploding gradient and vanishing problems
\end{tabitemize} \\ \hline

\textbf{Long Short-Term Memory RNN} & \justifying A RNN that can handle long-term dependencies by using a memory cell to selectively forget and remember information  & \begin{tabitemize}
    \item Can handle long-term dependencies
    \item Address the exploding and vanishing gradient problems
\end{tabitemize} & \begin{tabitemize}
    \item Computationally expensive in terms of memory bandwidth
    \item Not well-suited for parallelization
\end{tabitemize} \\ \hline

\textbf{Autoencoders} & \justifying Unsupervised NNs that learn data representations by compressing high-dimensional data into a lower-dimensional latent space and then reconstructing it with a decoder  & \begin{tabitemize}
    \item Can handle unlabeled data
    \item Low complexity
    \item Preserve useful patterns
\end{tabitemize} & \begin{tabitemize}
    \item High reconstruction error for complex and noisy data
    \item Prone to layer-by-layer errors
    \item Limited interpretability
\end{tabitemize} \\ \hline

\textbf{Q-Learning} & \justifying A value-based approach based on the Q-Table, which calculates the maximum expected future reward, for each action at each state, to later learn an optimal action-value function.
& \begin{tabitemize}
    \item Suitable when a training set is not available
    \item Model-free algorithm
\end{tabitemize} & \begin{tabitemize}
    \item Only suitable for problems with small state spaces.
    \item Slow convergence due to initial zero Q-values
\end{tabitemize}

\end{longtable}
\end{center}

Supervised Learning (SL) involves training a model on labeled data to make predictions based on input-output relationships \cite{MLterms}. Common SL algorithms include linear and logistic regression, K-nearest neighbors, Support Vector Machine (SVM), naïve Bayes, decision trees, and random forests. Deep learning (DL), a subset of SL, involves the use of Artificial Neural Networks (ANNs) to model intricate input-output mappings. DL models, such as Convolutional Neural Networks (CNNs), are particularly effective for image processing, whereas Recurrent Neural Networks (RNNs), such as Long Short-Term Memory (LSTM), are commonly used for sequence modeling in natural language processing tasks.

Unsupervised learning involves training a model on unlabeled data where the algorithm learns to identify patterns and structure within the data \cite{MLterms}. Common techniques used in unsupervised learning include K-means clustering and Principle Component Analysis (PCA) for dimensionality reduction. DL also has applications in unsupervised learning, such as leveraging autoencoders to learn the underlying representation of data.

Reinforcement Learning (RL) is a feedback-based, environment-driven approach in which an agent learns to behave in an environment through trial and error. The ultimate objective is to improve performance by maximizing a reward signal \cite{MLterms}. Common techniques used in RL include Q-learning and policy gradient methods. The latter optimizes policy parameters directly using gradient-based optimization, utilizing the policy gradient theorem to compute gradients of the expected cumulative reward. While useful for tasks involving continuous action spaces like robotics control, these methods may have slow convergence rates and suffer from high variance. RL has evolved towards Deep Reinforcement Learning (DRL), where deep neural networks are utilized to model the value function (value-based), the agent's policy (policy-based), or both (actor-critic). DRL is most beneficial in problems with high-dimensional state space.

AI/ML technologies are seen as foundational pillars for network automation in terms of development, configuration, and management \cite{AIin5G}. They will play a crucial role in achieving a new level of automation and intelligence toward network and service management. Additionally, they will enhance network performance, reliability, and adaptability through a series of real-time and robust decisions based on predictions of network and user behavior, such as user traffic.

\subsection{Software Defined Network}
Softwarization refers to the concept of running a specific functionality in software instead of hardware, thus breaking its relationship with the underlying hardware \cite{SDNNFV}. The benefits of softwarization lie in terms of decreased deployment time in addition to reduced Capital Expenditure (CAPEX) and Operational Expenditure (OPEX) when new functions are introduced. Softwarization also ensures high degrees of flexibility and reconfigurability.

One primary example of softwarization is SDN. SDN consists of SDN applications, an SDN controller, and an SDN datapath \cite{SDNNFV}. The SDN application is the software that runs network functionalities, such as routing. It receives an abstract view of the network through an SDN controller, which is a logically centralized unit. The controller also translates the inputs received from the applications down to the switches (\emph{i.e.}, physical network). Another function of the controller is instructing and configuring a set of network devices to perform certain actions, such as packet forwarding. This set of network devices is known as the SDN datapath.

\subsection{Network Function Virtualization}
Virtualization improves the software/hardware decoupling of the softwarization paradigm by creating virtual instances of dedicated hardware platforms, operating systems, storage devices, and computer networking resources \cite{SDNNFV}. In other words, software can run on commercial off-the-shelf equipment by utilizing a Virtual Machine (VM) or a Docker container. This is similar to softwarization, as virtualization provides flexibility (VMs can be migrated across different platforms) and reduces CAPEX and OPEX (introducing new services involves creating new VMs without any effort from the hardware side) \cite{NFV3}.

From a network perspective, virtualization is associated with the concept of NFVs, where network node functions (such as firewalls and switches) are virtualized and decoupled from the underlying hardware \cite{SDNNFV2}. These softwarized network functions are known as Virtual Network Functions (VNFs) and can consist of one or more VMs/containers. 

\subsection{Advancing Mobile Connectivity: From 5G to Beyond}
The fifth generation of cellular technology has prevailed research in the broad information and communication technology field. It was designed to increase speed, reduce latency, and improve the flexibility and adaptability of wireless services. Additionally, 5G supports and enhances a wide range of applications, including autonomous vehicles, online gaming, and voice over IP. 5G is expected to meet a diverse set of Key Performance Indicators (KPIs) for eMBB, mMTC, and uRLLC use cases \cite{5Gusecase, 5Gusecase2}. The 5G eMBB service is characterized by its peak data rate, which ranges from 10 to 20 Gbps and is suitable for applications such as 4K media. Another 5G use case is the mMTC service, which supports a high device density of up to 1 million per square kilometer for applications such as a smart city. Regarding uRLLC applications, such as mission-critical applications and self-driving cars, 5G uRLLC service is expected to offer 1 ms air interface latency and achieve six nines network availability.

Softwarization and virtualization are two complementary key enablers for providing flexibility in 5G networks \cite{SDNNFV}. Cloud-based environments run and move on-demand VNFs, while SDN dynamically changes the network topology according to the load and service requirements.

As we move towards the future,  we are witnessing the emergence of Beyond 5G (B5G), also known as 6G. This revolutionary technology is set to take mobile communications to unprecedented heights by building upon the capabilities of 5G \cite{e2e6g}. B5G promises to offer even faster speeds, lower latency, and greater capacity than 5G \cite{6gkpi}. Specifically, B5G is envisioned to support data transfer rates of up to 1 Tbps, a monumental leap forward from 5G's maximum data transfer rate of 20 Gbps. It also aims to drastically reduce latency to a sub-millisecond range of 10 $\mu$s to 100 $\mu$s, which is a minuscule fraction of 5G's latency of less than 1 ms. Additionally, B5G is expected to support a staggering number of connected devices, with an expected capacity of up to 10 million/km\textsuperscript{2}, a tenfold increase from 5G's capacity of up to 1 million/km\textsuperscript{2}.

B5G intends to support a plethora of new and emerging use cases, such as terahertz communication providing ultra-high data rates and low latency \cite{5gto6g,6g}. In addition to these performance enhancements, B5G will bring new capabilities such as advanced network slicing, edge computing, and network intelligence. These capabilities will enable more efficient and flexible network operations, as well as new business models and revenue streams. B5G will also focus on energy efficiency, sustainability, and security, ensuring that it is not only technologically advanced but also environmentally conscious and secure \cite{5gto6g}. This represents a leap forward, not just in terms of technology but also in terms of societal impact.

\section{Zero-Touch Network and Service Management Overview} \label{sec:overview}

\subsection{Need for ZSM in 5G+ Networks} \label{sec:zsmin5g}
Society's constant desire for seamless connectivity, high capacity, and new services has paved the way for 5G+ networks. The 5G+ concept represents a generational leap in NGNs and aims to revolutionize the telecommunications industry with new spectrum frequencies, a new Core Network (CN), a new Radio Access Network (RAN), and its adopted new radio. Softwarization and virtualization, enabled by technologies such as SDN and NFV, which decouple network functions from the underlying hardware, are considered the foundations of 5G+ networks. While this approach guarantees a high degree of flexibility and reconfigurability while reducing CAPEX and OPEX, it also results in a complex 5G/6G architecture that network operators find challenging to manage and operate due to the static nature of traditional Management and Orchestration (MANO) techniques \cite{aizsm5gp}. Therefore, there is a need to realize the vision of zero-touch networks and service management to enable automated orchestration and management of network resources and to ensure End-to-End (E2E) Quality of Experience (QoE) guarantees for end-users. The goal is to govern the services driven through an autonomous network by high-level policies and rules (aka intents), which is capable of offering Self-X life cycle operations (self-serving, self-fulfilling, and self-assuring) with minimal, if any, human intervention \cite{etsiurl}. To achieve this, ETSI established the ZSM ISG in 2017. The group's objective is to create a framework to enable fully-autonomous network operation and service management for 5G+ networks capable of self-\{configuration, monitoring, healing, optimization\} \cite{zsmsurvey}.
\subsection{Current Standard - ETSI ZSM}
The ultimate goal of ETSI ZSM is to create a framework that enables full E2E automation of network and service management in a multi-domain environment. The ZSM framework comprises operational processes that are automatically executed without human intervention, such as \cite{zsm5g6g}:
\begin{itemize}
    \item Design and planning to create new services that meet users' needs.
    \item Delivery to enable the on-demand delivery of services while satisfying requirements.
    \item Deployment to enhance network and resource utilization.
    \item Provisioning to reduce manual configuration errors.
    \item Monitoring and optimizing to avoid service degradation and ensure a fast recovery.
\end{itemize}
The ZSM ISG has already released a reference architecture that moves away from rigid management systems and towards more flexible services \cite{etsi002}. The key principles, requirements and components are elaborated next in Sections \ref{sec:principles}, \ref{sec:reqs} and \ref{sec:architecture}, respectively.
\subsubsection{Key Architecture Principles} \label{sec:principles}
The ZSM reference architecture defines a set of building blocks, as shown in Figure \ref{fig:etsizsm}, that can be integrated to build more complex management services and functions following a set of composition and interoperation patterns. This approach makes the architecture modular, scalable, and extensible. It is also resilient to failure as management services are devised to cope with the degradation of other services and/or the infrastructure. These services can also be combined to create new management services, which is referred to as service composability. In terms of management functions, stateless functions that separate processing from data storage are also supported. 
\par
Separation of concern in management is another key principle behind ZSM. This principle differentiates two management concerns: Management Domain (MD) and E2E cross-domain service management (\emph{i.e.}, across MDs). Within the former, services are managed based on their respective resources. In the latter, E2E services that span multiple MDs are managed, and coordination between MDs is orchestrated. This principle ensures non-monolithic systems and reduces the complexity of the E2E service. To automate service assurance, closed-loop management automation is used to achieve and maintain a set of objectives without any external disruption. The architecture is also coupled with intent-based interfaces that express consumer requests in an interpretable form and offer high-level abstraction. Overall, the architecture is of minimal complexity and meets all the functional and non-functional requirements that are discussed next \cite{etsi002,survey5GP}.

\subsubsection{Architecture Requirements} \label{sec:reqs}
The ZSM reference architecture follows a set of requirements that are divided into non-functional, functional, and security requirements. Functional requirements define what the system must do, such as automating the deployment and management of network functions and services. Non-functional requirements specify how well the system must perform, such as scalability and reliability. Security requirements dictate how to protect the system and its data from cyber threats by implementing measures such as encryption and access control. These requirements are extracted from ETSI GS ZSM 002 \cite{etsi002} and summarized in Table \ref{table:zsm_requirements}.

\begin{table}[htbp]
\centering
\caption{ETSI ZSM Framework Requirements}
\label{table:zsm_requirements}
\bgroup
\def\arraystretch{2}
\scriptsize
\begin{tabular}{|P{2cm}|  P{3cm}  P{7.3cm}|}
\hline
\textbf{Requirement Type} & \textbf{Category} & \textbf{Examples of Requirements} \\
\hline
\multirow{3}{*}{\begin{tabular}[]{p{0.85\linewidth}@{}@{}}\end{tabular}}  & General & Availability, energy efficiency, independence from vendors, monitoring requirements \\
\cline{2-3} \textbf{Non-Functional}
& Cross-Domain Data Services & High data availability, QoS support, task completion \\
\cline{2-3}
& Cross-Domain Service Integration & On-demand service addition/removal, service versions co-existence, seamless integration of new/legacy functions \\
\hline
\multirow{5}{*}{\begin{tabular}[]{p{0.85\linewidth}@{}@{}}\end{tabular}}& General & Resource/service management, closed-loop management, E2E services support, automation of processes \\
\cline{2-3}
& Data Collection & Real-time data collection, data processing and governance, metadata attachment, data sharing, and access \\
\cline{2-3} \textbf{Functional}
& Cross-Domain Data Services & Data storage/processing/sharing management, automation of data management processes (recovery, redundancy, overload, service failover, processing, and services with distinct data types) \\
\cline{2-3}
& Cross-Domain Service Integration and Access & Service discovery and registration, synchronous/asynchronous communication, service invocation \\
\cline{2-3}
& Lawful Intercept & Uninterrupted lawful interception \\
\hline
\multirow{3}{*}{\begin{tabular}[]{p{0.85\linewidth}@{}@{}}\end{tabular}} & Security and Privacy & Prioritizing privacy of personal data, ensuring the security of data and resources, applying security policies based on compliance status \\
\cline{2-3} \textbf{Security}
& Availability & Availability of data and services, Authorized access to services by authenticated users  \\
\cline{2-3}
& Attack Prevention & Automation of attack detection and prevention, supervision of ML privacy decisions \\
\hline
\end{tabular}
\egroup
\end{table}

\begin{enumerate}[label=(\roman*)]
    \item Non-functional requirements refer to the qualities or characteristics that the ETSI ZSM framework must possess in order to operate effectively and efficiently.
    \begin{itemize}
        \item General Requirements: \begin{enumerate}
            \item Realize a certain degree of availability.
            \item Become energy efficient.
            \item Achieve independence from vendors, operators, and service providers.
            \item Follow specific monitoring requirements. \end{enumerate}
        \item Requirements for Cross-Domain Data Services: \begin{enumerate}
            \item Realize high data availability.
            \item Support Quality of Service (QoS) specifications for data services within and outside the framework.
            \item Complete tasks within a preset timeframe. \end{enumerate}
        \item Requirements for Cross-Domain Service Integration: \begin{enumerate}
            \item Support the on-demand addition and removal of services.
            \item Support the co-existence of different service versions simultaneously.
            \item Avoid any changes to the management functions when integrating services.
            \item Allow integration of new and legacy functions.
        \end{enumerate}
    \end{itemize}
    \item Functional requirements refer to the features and capabilities that the ETSI ZSM framework must have in order to perform its intended functions.
        \begin{itemize}
        \item General Requirements: \begin{enumerate}
            \item Manage resources and services provided by the MDs.
            \item Support cross-domain management of E2E services.
            \item Support closed-loop management.
            \item Support technology domains needed for an E2E service.
            \item Support access control to services within the MD.
            \item Support open interfaces.
            \item Support hiding the management complexity of MDs and E2E services.
            \item Automate constrained decision-making processes.
            \item Promote automation of operational life-cycle management functions.
        \end{enumerate}
        \item Requirements for Data Collection: \begin{enumerate}
            \item Allow the collection and storage of real-time data.
            \item Enable the preprocessing and filtering of collected data.
            \item Support attaching metadata to collected data.
            \item Allow common access to the collected data across the MDs.
            \item Support the aggregation of collected data cross-domain.
            \item Enforce data governance by supporting various degrees of data sharing/collection velocity and volume.
            \item Manage the data distribution to maintain consistency.
            \item Provide data to the consumer based on their requirements.
        \end{enumerate}
        \item Requirements for Cross-Domain Data Services:\begin{enumerate}
            \item Allow separation of data storage and processing.
            \item Logically centralize the storage/processing of data. 
            \item Enable data sharing within the framework.
            \item Automate management of redundant data.
            \item Automate overload handling of data services.
            \item Automate data service failover.
            \item Automate data recovery.
            \item Automate policy-based data processing.
            \item Automate the processing of data services with distinct data types.
        \end{enumerate}
        \item Requirements for Cross-Domain Service Integration and Access: \begin{enumerate}
            \item Enable the discovery and registration of management services.
            \item Provide information on accessing the discovered service.
            \item Invoke services indirectly or directly (by the consumer).
            \item Support both synchronous and asynchronous communication between the consumer and the service provider.
        \end{enumerate}
        \item Requirements for Lawful Intercept: ZSM architecture must ensure that lawful interception is not interrupted regardless of any management service performed by the framework.
    \end{itemize}
    \item Security requirements refer to the measures that must be taken to ensure the security and privacy of the network and its data.
    \begin{enumerate}
        \item Ensure the security of data, whether at rest, in transit, or in use.
        \item Ensure the security of resources in addition to management services and functions.
        \item Provide special attention to the privacy of personal data by utilizing mechanisms such as privacy-by-design or privacy-by-default.
        \item Ensure the availability of data, resources, functions, and services.
        \item Apply relevant security policies based on the compliance status of services regarding security requirements.
        \item Allow authorized access to services by authenticated users.
        \item Automatically detect, identify, prevent, and migrate attacks.
        \item Supervise decisions of ML/AI regarding privacy and security to prevent attacks from spreading.
    \end{enumerate}
\end{enumerate}

\subsubsection{Reference Architecture} \label{sec:architecture}
\begin{figure*}[h!]
	\centering
	\includegraphics{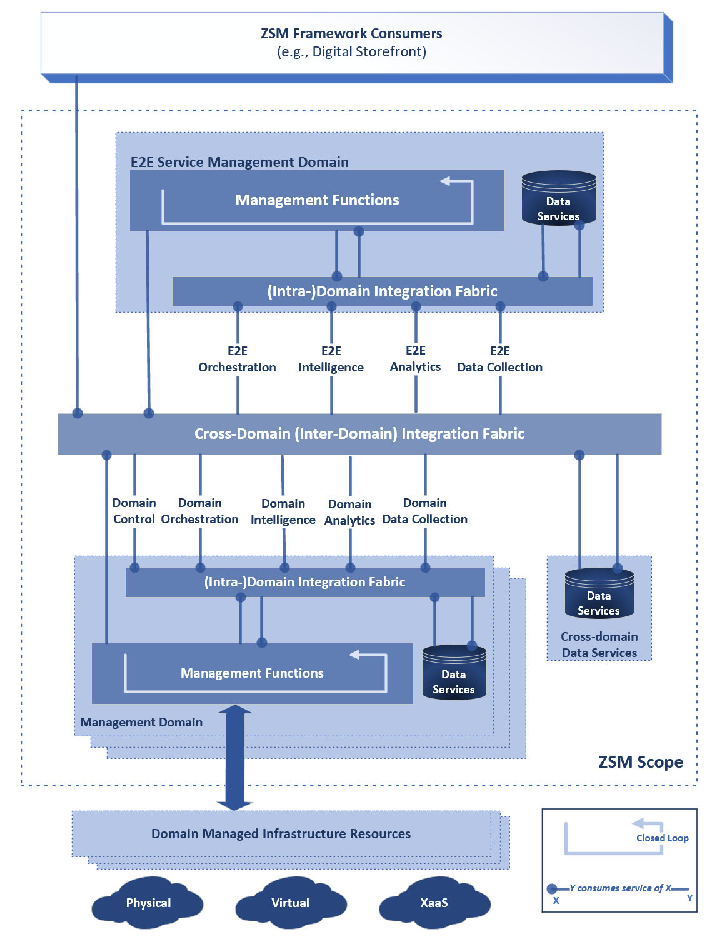}
	\caption[]{ETSI ZSM Framework Reference Architecture \cite{etsi002}}
	\label{fig:etsizsm}
\end{figure*}
The ZSM framework reference architecture, as shown in Figure \ref{fig:etsizsm}, integrates MDs, E2E service MD, cross-domain data services, and intra- and inter- integration fabrics. Self-contained and loosely-coupled services are found within the MD. Each MD handles the automation of orchestration, control, and guarantee of resources and services within its domain. These managed resources might be physical (\emph{e.g.}, physical network functions), virtual (\emph{e.g.}, VNFs) and/or cloud resources (\emph{e.g.}, "X-as-a-service" resources) \cite{aizsm5gp}. E2E services that span multiple MDs coordinate across domains through orchestration. The E2E service MD oversees the management of such services. Each MD, including E2E service MD, is composed of logically grouped management functions (data collection services, intelligence services, analytics services, control services and orchestration services), intra-domain integration fabric in addition to data services that enable data sharing and authorized access management across services within the MD. The management functions produce and use management services; thus, they can be both producers and consumers of the service. Each MD provides a set of management services through service interfaces. Services that can only be consumed locally within the domain are provided via the intra-domain integration fabric, while cross-domain services are enabled through the inter-domain integration fabric.  The inter-domain integration fabric also handles the communication between management functions and ZSM framework consumers. Another building block is the cross-domain data service that oversees data persistence among MDs while also permitting processing tasks to run on the stored data as a means to reach E2E global optimization. The data in the cross-domain data services can be exploited by the intelligence services within E2E service MDs and MDs to support cross-domain and domain-level AI-based closed-loop automation.
\subsection{Intents}
Intents are a crucial component of network automation in zero-touch networks. They provide a high-level, abstract representation of the desired state of the network, making it easier for network administrators to manage and configure large and complex networks \cite{intentsurvey}.

The main goal of intents is to make network configuration and management more efficient, accurate, and scalable. They achieve this by allowing network administrators to describe the network's desired state using a domain-specific language that is then translated into the underlying network configurations \cite{gomes2021intent}. This falls under the intent-driven management paradigm. This paradigm eliminates the need for manual intervention, reducing the risk of human error and freeing up time for more critical tasks.

The benefits behind intents in zero-touch networks are significant \cite{intentbenefits, intentesg}. Some of the key advantages include:
\begin{itemize}
    \item Improved Accuracy: Intents allow network administrators to define the desired state of the network more precisely and consistently, reducing the risk of human error.
    \item Increased Efficiency: Intents automate network configuration and management tasks, reducing the time and effort required to maintain the network.
    \item Enhanced Scalability: Intents can be used to manage large and complex networks, enabling network administrators to quickly and easily make changes to the network as it evolves.
    \item Improved Collaboration: Intents allow multiple stakeholders to work together to define the desired state of the network, improving communication and collaboration among teams.
    \item Better Compliance: Intents provide a clear and consistent representation of the network's desired state, making it easier to ensure that the network is compliant with regulatory requirements.
\end{itemize}

\subsubsection{Example Use Case: Intent-Based Approach for Configuring a 5G Network to Support a Virtual Reality Service}
For instance, suppose a network operator is tasked with configuring a 5G network to support a new Virtual Reality (VR) service. The VR service requires low latency and high bandwidth to provide an immersive experience for users. The network should also be able to dynamically allocate network resources to meet the changing demands of the VR service. In a traditional approach, the network operator would have to manually configure the necessary network settings and policies to meet the VR service's requirements. However, in a zero-touch network, the network operator can use an intent-based approach to simplify the process, as shown in Listing \ref{lst:intentex}.
\begin{minipage}{\linewidth}
\vspace{15pt}
\begin{lstlisting}[frame=tb, caption={Intent Example to Support VR in 5G Network}, label={lst:intentex}]
intent "VR Service" {
    allocate bandwidth "1 Gbps" for "VR Service"
    set latency "5 ms" for "VR Service"
    policy "Resource Management" {
        dynamically allocate bandwidth for "VR Service" based on demand
        prioritize "VR Service" over other network traffic
    }
}
\end{lstlisting}
\end{minipage}
The intent in this example describes the desired state of the network as follows:
\begin{enumerate}
    \item Allocate 1 Gbps of bandwidth for the VR service.
    \item Set a maximum latency of 5 ms for the VR service.
    \item The "Resource Management" policy dynamically allocates bandwidth for the VR service based on demand and prioritizes the VR service over other network traffic.
\end{enumerate}
Once the intent is specified, it can be translated into the necessary network configurations and policies. These can then be automatically implemented and enforced by the zero-touch network management system, resulting in a network optimized to support the VR service with the aforementioned requirements.

\subsubsection{Use of Intents in the ZSM Framework}
Intent should serve as the sole method of communicating requirements between the zero-touch system and human operators, as well as between the different subsystems and layers of the management system. In the ZSM framework, this means that the service specification provided by ZSM framework consumers must be conveyed through an intent object. The E2E service domain is responsible for translating it into sub-intents that specify specific requirements for each MD. Communication based on intent objects is a universal mechanism that can be applied to any MD within the ZSM framework. With intents, domain-specific semantics can be encapsulated in shared information models, and endpoints based on intents can leverage a generic knowledge management service for the life-cycle management of intent objects. In line with this, Gomes \emph{et al.} introduced a cutting-edge framework for the management of autonomous networks within the ZSM framework \cite{gomes2021intent}. This framework leverages the concept of intent-based models, which are translated into a set of rules and constraints that drive the configuration and operation of the network, resulting in a closed-loop control mechanism. One of the key features of the framework is its ability to continuously monitor the network's state and adjust its configuration and operation accordingly, ensuring that it remains aligned with the specified intent. The framework employs feedback mechanisms to collect data from the network and update its configuration and operation in real-time, allowing for rapid adaptation to changing network conditions. In addition to its closed-loop control capabilities, the framework also provides abstraction and simplification, presenting the network operator with a simplified view of the underlying infrastructure. This abstraction reduces the complexity of network management and enhances the efficiency of decision-making. Results show that the framework can significantly improve network performance and stability, compared to traditional manual approaches to network management. Additionally, results show that the framework can significantly reduce operational complexity, streamlining network management and enabling the deployment of more advanced autonomous networks. Future work includes investigating the interoperability of the proposed framework based on the intent meta-models.

Another Closed Loop-based zero-touch network mAnagement fRAmework, CLARA, was developed by Sousa \emph{et al.} \cite{clara}. CLARA's two main components are the closed-loop data plane and closed-loop control plane. The closed-loop data plane is the component of the CLARA framework that realizes and implements the intents defined in the intent definition language. The data plane is designed to be programmable and flexible, allowing it to adapt to changing network conditions and accommodate new network services. The closed-loop control plane, on the other hand, is the component of the CLARA framework that monitors and enforces the intents defined in the intent definition language. It continuously receives feedback from the network and adjusts its behavior to maintain the desired state of the network. The control plane is responsible for detecting deviations from the desired state and triggering corrective actions to restore the network to its desired state. Similar to the results of the previous framework \cite{gomes2021intent}, Sousa \emph{et al.} demonstrate CLARA's superiority over traditional network management approaches in terms of automation, reliability, and scalability. Upcoming initiatives include integrating ML algorithms into the framework to improve its accuracy and efficiency. This could include algorithms for network optimization, fault detection and diagnosis, and proactive network management.

\subsection{Related Projects}
Several organizations and projects are closely related to the ZSM framework, such as the ETSI Experiential Network Intelligence (ENI) ISG and the TeleManagement (TM) Forum. These organizations focus on exploring use cases, architectural components, and interfaces related to network automation. Other projects, such as those funded by the EU's Horizon 2020 (H2020) program, are also working towards network automation. \par

\subsubsection{ETSI ENI ISG}
This group focuses on AI/ML techniques, context-aware policies, and closed-loop mechanisms to design a cognitive network management architecture that provides an effective and adaptive service delivery experience. ENI aims to enhance the entire management cycle of 5G networks, including provisioning, operation, and assurance. Coronado \emph{et al.} state that the outputs of ENI in terms of AI/ML algorithms, intent policies, and Service Level Agreement (SLA) management are to promote service intelligence capabilities in cross-domain cases \cite{zsm5g6g}. Some ENI use cases, such as intelligent network slice management, network fault identification/prediction, and assurance of service requirements, are relevant to ZSM.

\subsubsection{TM Forum Zero-touch Orchestration, Operations and Management (ZOOM)}
TM Forum's ZOOM project aims to define a new management architecture of virtual networks and services through automated configuration, provisioning, and assurance. The guiding principles of ZOOM include near real-time request execution with no human intervention, open standard Application Programming Interfaces (APIs), closed-loop control, and E2E management \cite{aizsm5gp,zsm5g6g}. These principles are also shared by ZSM networks.

\subsubsection{Distributed Management of Network Slices in Beyond 5G (MonB5G)}
This EU-funded H2020 project began on November 1\textsuperscript{st}, 2019, and it is expected to run for three years and a half to end on April 30\textsuperscript{th}, 2023 \cite{monb5g}. MonB5G works towards providing zero-touch network slice orchestration and management at massive scales for 5G+ networks. It proposes a hierarchical, fault-tolerant, and automated data-driven network management system that focuses on security and energy efficiency. The goal is to split the centralized management system into distributed sub-systems where the intelligence and decision-making processes will be split across various components.

\subsubsection{Hexa-X}
This is another EU-funded H2020 project representing a flagship for the 6G vision \cite{hexax}. The objective is to interconnect three worlds, namely the human, physical, and digital worlds, via technology enablers. Over a duration of 36 months, this project will focus on creating 6G use cases, developing essential 6G technologies, and defining a new architecture for an intelligent fabric that weaves together the key technology enablers. In the ZSM domain, the Hexa-X project defines AI/ML-driven orchestration as an essential component for 5G+ networks, which will, in turn, support data-driven and zero-touch approaches.

\subsection{Zero-Touch Network Operations}
ZNO is a key concept in the evolution of 5G+ networks, which aims to automate network operations and reduce the need for manual intervention. ZSM and AI/ML have a wide range of practical applications in ZNO.
\begin{enumerate}
    \item Network resource management and optimization: Dynamic resource allocation, for example, can be automated using ML algorithms to optimize resource utilization based on real-time network demands, while network slicing enables the creation of virtualized network slices that can be automatically customized for specific use cases. MEC can also be leveraged to automate network functions at the edge, reducing latency and improving the overall user experience.
    \item Network traffic control: ML can be used to predict and classify network traffic, enabling automated network management systems to optimize network performance and improve the overall user experience. Intelligent routing can also be used to automate the routing of network traffic based on real-time network conditions, reducing congestion and improving network efficiency.
    \item Energy efficiency: By automating network operations, energy consumption can be optimized based on real-time network demands, which is critical in achieving the sustainability goals of 5G+ networks.
    \item Network security and privacy: ML algorithms can be leveraged to automate network security functions, including threat detection and response, improving the overall security posture of 5G+ networks.
\end{enumerate}
These applications are further discussed in Sections \ref{sec:resourcemanagement}-\ref{sec:networksecurity}. Table \ref{table:refs} presents a comprehensive list of the corresponding schemes and their references.

\begin{table}[htbp]
\centering
\caption{ZNO Applications \& Corresponding Schemes}
\label{table:refs}
\def\arraystretch{2}
\scriptsize
\begin{tabular}{|P{0.2\linewidth} |P{0.3\linewidth}|  P{0.38\linewidth}|}
\hline
\multicolumn{2}{|c|}{\textbf{ZNO Application}} & \textbf{Schemes} \\\hline

\multirow{3}{*}{\begin{tabular}{@{}c@{}}\textbf{Resource}\\[-9pt] \textbf{Management}\end{tabular}}  & Dynamic Resource Allocation & \cite{wlanDRL_RA}, \cite{rlvnfplacement}, \cite{urllcvnfplacement}, \cite{novelnfvorch}, \cite{selfhealing}, \cite{pfr} \\
\cline{2-3} 
& 
Network Slicing & \cite{casale2019autonomic}, \cite{harness}, \cite{manoNS}, \cite{certifiedzsmns}, \cite{nsos}, \cite{csps} \\
\cline{2-3} &
Multi-access Edge Computing & \cite{mecNS}, \cite{AVMEC}, \cite{MECZSM}\\
\hline

\multirow{2}{*}{\textbf{Traffic Control}} & Traffic Prediction \& Classification & \cite{SVMkmeans}, \cite{LSTMTP}, \cite{trafficforecastalawe}, \cite{guptaDL} \\
\cline{2-3}
& 
Intelligent Routing & \cite{ears}, \cite{rirm}, \cite{fanets}, \cite{v2x} \\\hline

\multicolumn{2}{|c|}{\textbf{Energy Efficiency}} & \cite{aseEE}, \cite{sche2ma}, \cite{ZTenergycontrolNS} \\\hline

\multicolumn{2}{|c|}{\textbf{Network Security \& Privacy}} & \cite{trust5Gblockchain}, \cite{MUDtrm}, \cite{insliceddos}, \cite{FLzsm} \\
\hline
\end{tabular}
\end{table}

\section{Network Resource Management} \label{sec:resourcemanagement}
To achieve the full potential of the envisioned pervasive network, current 5G networks need improvements. Specifically, automation is limited as network monitoring via analytics is not fully supported \cite{zorroAIenabled}. The performance requirements specified by 3GPP rely on incorporating technologies such as dynamic resource/spectrum sharing and cognitive zero-touch network orchestration for an optimized network.

Network optimization is the art and science of fine-tuning and configuring a network to achieve the best possible performance, efficiency, and scalability. This includes optimizing the configuration of network devices, such as routers and switches, and adjusting the parameters of different protocols and services that run on the network. The aim is to ensure that the network is running at its best possible performance, which can include factors such as reducing network congestion, increasing network throughput, and improving network availability. It is a continuous process that aims to keep the network running smoothly and efficiently, with the ultimate goal of providing a reliable and high-quality service to the users. Traffic engineering, capacity planning, and network design are among the various techniques and approaches used for network optimization, which can optimize routing, bandwidth allocation, and QoS, among other aspects of the network. Monitoring and troubleshooting tools can also be used to diagnose performance issues in real-time and automate the network optimization process.

In the context of 5G networks, network optimization becomes even more crucial due to their unique requirements. These networks are characterized by high bandwidth, low-latency, and high-concurrency, which demand advanced techniques for network optimization, such as network slicing, edge computing, and advanced resource allocation. 5G networks have to handle a large number of connected devices and services, each with different requirements and characteristics. Resource allocation involves managing the available network resources, such as bandwidth, processing power, and storage, in order to provide an optimal service to each device and service. This can include allocating resources dynamically in response to changing network conditions and user demands, as well as using advanced techniques such as ML and optimization algorithms to improve the efficiency of resource allocation. Another important aspect of network optimization in 5G networks is network slicing, paving the way for efficient and flexible allocation of network resources, as well as the ability to support diverse and dynamic requirements of 5G networks. Edge computing is also a key technology for network optimization in 5G networks. Edge computing involves moving computing and storage resources closer to the network edge, where they can be used to reduce network congestion and improve the responsiveness of the network. This is particularly important for 5G networks, which will support a wide range of low-latency and high-bandwidth services, such as virtual reality and augmented reality applications. Tables \ref{table:res_alloc_schemes}, \ref{table:ns_schemes}, and \ref{table:mec_schemes} provide an overview of different proposed schemes and frameworks addressing dynamic resource allocation, network slicing, and edge computing, respectively.

\subsection{Dynamic Resource Allocation}
5G+ networks require efficient dynamic network optimization to fully utilize resources and achieve higher capacity and better QoS with minimal SLA violations \cite{prosOfResourceAlloc}. Effective resource allocation has always been a critical challenge in wireless communication.

In the context of ZSM, DRL is a promising solution for dynamic resource allocation problems, which are generally formulated as hard optimization problems \cite{DRLimpo}. Iacoboaiea \emph{et al.} studied the deployment of DRL in a large-scale zero-touch Wireless Local Area Network (WLAN) for dynamic radio resource allocation under varying traffic conditions \cite{wlanDRL_RA}. Their actor-critic neural network algorithm aims to maximize the E2E performance, where the WLAN management solution (agent) reconfigures every 10 minutes in a closed-loop fashion (action) based on the telemetry data received (state). Telemetry refers to the automatic measurement and wireless transmission of data from isolated sources. Training on the real WLAN setup would have led to the exploration of bad network configurations, so they have trained the DRL agent on a DT (\emph{i.e.}, a digital replica that behaves identically as its physical WLAN counterpart). This allowed the use of both simulated and existing network data for training, as utilizing the latter data alone was not sufficient due to the large number of samples needed for training.

Another scenario is utilizing RL/DRL in 5G networks based on NFV and SDN technologies. The placement of the VNFs depends on the availability of physical resources, which in turn can affect network performance and service latency. As such, inefficient VNF placement and resource utilization might result in serious performance degradation \cite{urllcvnfplacement}. Following the ENI and ZSM standards, Bunyakitanon \emph{et al.} designed an Adapted REinforcement Learning VNF Performance Prediction module for Autonomous VNF Placement (AREL3P) based on the Q-learning algorithm \cite{rlvnfplacement}. The algorithm predicts the total service time of an E2E application running VNF video transcoding. Results show the resilience of AREL3P to network dynamics in addition to the ability to generalize better than SL algorithms, thus tackling adaptability concerns. However, RL approaches slowly converge to the optimal policy in large-state action sets, rendering it challenging to use in large-scale 5G deployments. This led to DRL, where the intersection of RL and deep learning helps overcome this limitation \cite{drlandrl}. Subsequently, Dalgkitsis \emph{et al.} proposed an intelligent VNF placement solution using a deep deterministic policy gradient RL algorithm \cite{urllcvnfplacement}. The objective is to minimize the average E2E latency between the users and the VNFs that compose the uRLLC service provided by the network, while considering the distribution of the available computational resources (CPU, memory, storage) at the network edge. Results highlight the advantages of the proposed solution over the baseline algorithm (that rejects any VNF placement request to an edge data center if it has reached 90\% utilization capacity) by achieving the least amount of SLA violations and the least number of VNF rejections at any traffic level. To enhance the proposed algorithm in future work, it is suggested to build the algorithm based on LSTM RNN to provide better insights into the usage trend of each unit in the network.

For AI/ML to work well in managing network services, it is important to have a good understanding of how resources are used by the network and its components. This will allow AI/ML to make better decisions and improve the user experience. Accordingly, Moazzeni \emph{et al.} introduced a Novel Autonomous Profiling method, known as NAP, that can be applied within the ambit of ZSM for the next generation of NFV orchestration \cite{novelnfvorch}. This NAP method encompasses three key acts:
\begin{enumerate}
    \item NAP utilizes a weighted resource configuration selection algorithm, which automatically generates a profiling dataset for VNFs by selecting the configuration of resources that have the greatest impact on the performance goals and Key Performance Indicator targets within a confined profiling time frame. 
    \item NAP creates a model to precisely predict the performance metrics for previously untested resource configurations, where the specified performance goals can be attained.
    \item NAP employs ML-based techniques to estimate the precise quantity of resources required to meet both the specified performance goals and the performance metrics in the target environments.
\end{enumerate}

The results obtained from real datasets pertaining to various profiled VNFs demonstrate that this NAP method can predict the untested configuration of resources as well as the performance metrics with remarkable accuracy. Therefore, the model generated by the predictor manager, in conjunction with the proposed NAP method, can be employed for the next generation of NFV orchestration throughout the entire life-cycle management of network slices. Future endeavors include expanding the current profiling work to encompass more resource types, thereby increasing the state space of profiling predictions exponentially. Additionally, it is intended to extend this autonomous profiling method to profile VNFs hosted at the edge, encompassing scenarios such as network slicing and mobility management.

In 5G networks, the ability to detect and diagnose faults is closely tied to the ability to allocate resources efficiently. For example, if a failure is detected in a specific part of the network, resources can be reallocated to other parts of the network to maintain service availability. Similarly, if the network is experiencing congestion, resources can be reallocated to alleviate the congestion and improve performance. Therefore, efficient and effective fault detection is necessary to ensure that resources can be reallocated quickly and efficiently in response to network failures or disruptions. This can be achieved through the use of advanced monitoring and troubleshooting tools, as well as automation and orchestration technologies, that can identify and diagnose faults in real-time and take appropriate action to mitigate them, signifying the self-healing aspect of ZSM. Sangaiah \emph{et al.} designed an automatic self-healing process that tackles both detection and diagnosis of faults in 5G+ networks using two sets of data collected by the network \cite{selfhealing}. The performance support system data is automatically collected by the network, while drive test data is manually collected in three call scenarios: short, long, and idle. The short call scenario is used to identify faults during call setup, the long call scenario is designed to identify handover failures and call interruption, and the idle mode is used to understand the characteristics of standard signals in the network. The complete and correct call rate quality criterion was utilized for identifying faults in the network.  Examples of detected and diagnosed faults include congestion and failures in traffic channel assignments. Recognized faults using this criterion were then processed further to determine the root cause of the fault. The data was separated into two categories, traffic and signaling data, and the issues associated with each section were identified individually. As the data available was unlabeled raw data, a clustering algorithm method (\emph{i.e.,} unsupervised learning approach) was employed. By applying different algorithms with varying numbers of clusters, five clusters with a Silhouette coefficient of 0.4509 were obtained for traffic data and 6 clusters with a Silhouette coefficient of 0.503 for signaling data. Each cluster represented a specific cause for a fault in the network. Finally, various classification algorithms were applied to the labeled data obtained from clustering to evaluate the results accurately. The best accuracy in test data was achieved by combining the results of different classifiers through opinion voting for both traffic and signaling data. One root cause of a fault is the lack of capacity in the traffic and signaling channel. In that case, the proposed solution is to increase the capacity (scaling) and allocate dynamic resources, specifically capacity, to the required channels according to the network traffic situation. Future work suggests examining subscriber complaint data in more detail, including the explanations that the subscribers provide to the complaints center, to identify the fault type and analyze its cause \cite{selfhealing}.

The realm of failure recovery in networks is comprised of two distinct approaches, known as Proactive Failure Recovery (PFR) and reactive failure recovery. This recovery process involves three key stages, including the deployment of backup VNFs and image migration, flow reconfiguration, and state synchronization. However, the execution of each stage incurs a significant delay, resulting in not only a decline in network performance but also a violation of SLAs due to prolonged interruption of service \cite{pfrDEF}. By utilizing failure prediction, the PFR approach can decrease recovery delay by initiating certain stages of the recovery procedure prior to the manifestation of the failure. For instance, PFR can save delays in flow rescheduling and backup launch by initiating these stages beforehand. In this manner, if we are able to recover failed VNFs using PFR, the performance of the network can be significantly improved by reducing interruption time during recovery. This motivates the proposal of a PFR framework for future 6G networks. Given the constraints of resources and the maximum allowable interruption time caused by failures, Shaghaghi \emph{et al.} established a network that is both highly reliable and resource-efficient by introducing Zero-Touch PFR (ZT-PFR) approach \cite{pfr}. This approach utilizes DRL to enhance the fault-tolerance of networks enabled by NFV. This is formulated as an optimization problem that aims to minimize a weighted cost, which takes into account factors such as resource costs and penalties for incorrect decisions. Shaghaghi \emph{et al.} adopted state-of-the-art DRL-based methods such as soft-actor-critic and proximal-policy-optimization as solutions to this problem. To train and test the proposed DRL-based framework, the authors construct an environment simulator using a simulated model of impending failure in NFV-based networks inspired by ETSI. Additionally, to capture time-dependent features, the agents are equipped with LSTM layers. Additionally, the concept of age of information is applied to balance between event-based and scheduled monitoring in order to ensure that network status information is up-to-date for decision-making. Given the ever-changing nature of NFV environments, it is important to develop learning methods that are online, fast, and efficient. Thus, further research in this direction could be of great interest \cite{pfr}.

To fulfill the 5G vision in terms of E2E automation and resource sharing/allocation, the 5GZORRO project \cite{5GZORRO} has been launched by the H2020 program. Its main objective is to utilize distributed AI and Distributed Ledger Technologies (DLTs) to design a secure and trusted E2E zero-touch service and network management and orchestration within the 5G network with a shared spectrum market for real-time trading on spectrum allocation. While AI is a pillar behind a zero-touch cognitive network orchestrator and manager, DLT (or blockchain technology) is a protocol that enables the secure and trusted functioning of a distributed 5G E2E service chain. Thus, the 5GZORRO framework creates a 5G service layer across different parties where SLAs are monitored, spectrum is shared, and orchestration is automated \cite{5GZORRO, zorroAIenabled}. Another project launched by the H2020 is the 6G BRAINS project which started on January 1\textsuperscript{st}, 2021, and is expected to run for 36 months \cite{6gbrains}. It focuses on developing an AI-driven multi-agent DRL algorithm for dynamic resource allocation exceeding massive machine-type communications with new spectrum links, including THz, Sub-6 GHz, and optical wireless communications. The aim is to improve the capacity, reliability, and latency for various vertical sectors, such as eHealth and intelligent transportation.

\begin{table}[ht]
\centering
\caption{Dynamic Resource Allocation Schemes}
\label{table:res_alloc_schemes}
\def\arraystretch{2}
\scriptsize
\begin{tabular}{|P{1.85cm} |P{4.15cm}| P{2.97cm}| P{2.97cm}|}
\hline
\textbf{Scheme} & \textbf{Description} & \textbf{Outcomes} & \textbf{Future Work} \\ \hline
\textbf{DRL in WLAN \cite{wlanDRL_RA}} &
\RaggedRight Implement a DRL-based dynamic radio resource allocation for WLAN management under varying traffic conditions. &
\RaggedRight Maximize E2E performance. &
\RaggedRight Establish sufficient trust for true fully-automated zero-touch operation. \\ \hline

\textbf{AREL3P \cite{rlvnfplacement}} & \RaggedRight Autonomously place VNFs in 5G networks using a Q-learning-based VNF performance prediction module. &
\begin{tabitemize}
    \item Adapt resiliently to network dynamics.
    \item Generalize better than SL algorithms. 
\end{tabitemize}&
\RaggedRight Address the slow convergence of this scheme in large-scale networks. \\ \hline

\textbf{DRL-based VNF Placement \cite{urllcvnfplacement}} & \RaggedRight Design an intelligent VNF placement solution in 5G networks using a deep deterministic policy gradient-based method. &
\begin{tabitemize}
    \item Reduced E2E latency
    \item Low SLA violations
    \item A low number of VNF rejections at any traffic level 
\end{tabitemize}
& \RaggedRight Explore the use of LSTM RNN to provide better insights. \\ \hline

\textbf{NAP	\cite{novelnfvorch}} & \RaggedRight Design an autonomous profiling method for the next generation of NFV orchestration in 5G networks. & \RaggedRight Accurately predict untested resource configurations. &	\RaggedRight Extend the work by covering additional types of resources. \\ \hline

\textbf{Self-healing Process \cite{selfhealing}} & \RaggedRight Autonomously carry out a self-healing process tackling both detection and diagnosis of faults in 5G+ networks. &	\RaggedRight This process improves network component understanding. & \RaggedRight Examine subscriber complaint data to identify the fault type and analyze its cause. \\ \hline

\textbf{ZT-PFR \cite{pfr}} & \RaggedRight Enhance the fault-tolerance of networks through a DRL-based zero-touch proactive failure recovery scheme & \RaggedRight Network status information is up-to-date for decision-making. & \RaggedRight Apply online learning ML models to address the ever-changing nature of NFV environments.
\\ \hline

\end{tabular}
\end{table}

\subsection{Network Slicing}
Network slicing is a powerful and innovative technology that allows for the creation of multiple virtual networks within a single physical infrastructure. This technology allows for a flexible and efficient allocation of resources, enabling service providers to offer customized services to different customers or types of traffic \cite{NSsurvey}. With network slicing, service providers create virtual networks, or “slices”, with unique characteristics and policies, such as different levels of security, reliability, and bandwidth. This allows for the creation of specialized networks for various use cases, such as IoT devices, industrial automation, and enhanced mobile broadband. In turn, this technology is deemed useful in the context of 5G networks, which are expected to support a wide range of use cases. SDN, NFV, and cloud computing are the key enablers needed to realize network slicing \cite{NSenablers}. 

Nonetheless, despite its many benefits, network slicing also presents a number of challenges \cite{NSchallenges}. One of the main challenges is the management and orchestration of the slices. As the number of slices increases, it becomes increasingly difficult to manage and monitor them all effectively. Additionally, managing the allocation of network resources across multiple slices can be complex and time-consuming, especially as the demand for each slice fluctuates over time. Accordingly, it is necessary to have a reliable management system to automate the process of creating, configuring, and deploying slices, monitor their performance, and troubleshoot any issues that may arise. Another challenge lies in the need for advanced security mechanisms to protect virtual networks from unauthorized access and malicious attacks. There exists another need to ensure the interoperability of virtual networks with existing networks and systems. This requires the development of new standards and protocols to ensure seamless communication between different virtual networks and existing systems. Business-wise, the deployment and maintenance of the network slicing technology can be costly, and service providers must find ways to effectively monetize the services they offer to recoup their investment.

Network slicing can unlock new opportunities for service providers, but it also poses a number of technical and operational challenges that are being addressed academically. One of the main areas of research has focused on developing automated methods for creating, configuring, and deploying network slices without any manual intervention. This includes the use of AI/ML algorithms to predict network resource demand and dynamically allocate resources to different slices. For instance, Casale \emph{et al.} proposed a ML-based approach to predict network resource demand and dynamically allocate resources to different slices \cite{casale2019autonomic}. The proposed approach is able to adapt to changes in network conditions and user demands, and make real-time decisions about the allocation of resources to different slices. In fact, the proposed algorithm is based on RL, which learns from past decisions to improve the performance of future decisions. The algorithm uses a combination of decision-making policies, including a greedy policy, a random policy, and a Q-learning policy. Results compare the performance of the proposed approach using simulations and compare it to traditional static allocation methods. The approach shows a better performance in terms of resource utilization, by allocating resources to the slices that need them the most. However, this approach has some limitations in terms of scalability, as it requires a large amount of data to train the ML models. Additionally, it assumes that the network conditions are static and do not change rapidly.

Another area of research has focused on developing zero-touch management and orchestration systems for network slicing in 5G+ networks. This includes the use of SDN and NFV technologies to enable the dynamic creation, configuration, and management of network slices. As such, Vittal \emph{et al.} presented HARNESS, a novel High Availability supportive self-Reliant NEtwork Slicing System for the 5G core, powered by the SON paradigm \cite{harness}. HARNESS intelligently handles control plane User Service Requests (USRs), ensuring uninterrupted high-availability service delivery for delay-tolerant and delay-sensitive slices. It addresses scaling, overload management, congestion control, and failure recovery of primary slice types, namely eMBB, uRLLC, and mMTC. The proposed HARNESS mechanism outperforms traditional scheduling methods, minimizing dropped USRs and improving response times. Experimentally, HARNESS achieved 3.2\% better slice service high-availability in a minimal active/active cluster configuration. Future work involves scaling the HARNESS framework and exploring the selective offloading of control plane USRs on smart network components for different slice types in a 5G system.

In the context of scaled systems, Chergui \emph{et al.} proposed a distributed and AI-driven management and orchestration system for large-scale deployment of network slices in 6G \cite{manoNS}. The proposed framework is compliant with both ETSI standards, ZSM and ENI, focusing on autonomous and intelligent network management and orchestration to enable autonomous and scalable management and orchestration of network slices and their dedicated resources. Future work suggests mapping the framework to different architectures to test its effectiveness. Another compliant framework was introduced by Baba \emph{et al.}, representing a resource orchestration and management architecture for 5G network slices. This framework comprises a per-MD resource allocation mechanism and an MD interworking function, aimed at facilitating the provision of E2E network services over network slices in the context of 5G evolution \cite{certifiedzsmns}. This proposed architecture is underpinned by a plethora of standard APIs and data models, and its efficacy is demonstrated through the successful orchestration across multiple domains, and the automation of closed-loop scenarios. The architecture has been verified and certified as a proof of concept by the ETSI ZSM.

Similarly, Afolabi \emph{et al.} proposed a novel and comprehensive global E2E mobile network slicing orchestration system (NSOS) that enables network slicing for next-generation mobile networks by considering all aspects of the mobile network spanning across access, core, and transport parts \cite{nsos}. The high-level architecture of the system comprises a hierarchical structure, including a global orchestrator and multiple domain-specific orchestrators and their respective system components. The focus of the system is on allowing customers to request and monitor network slices only, while the proposed Dynamic Auto-Scaling Algorithm (DASA) ensures that the system can react instantly to changes in workload. The DASA includes both proactive and reactive provisioning mechanisms, where the proactive mechanism relies on a workload predictor implemented using ML techniques, and the reactive provisioning module triggers asynchronous requests to scale in or out the different entities of the NSOS. The core of the solution is a resource dimensioning heuristic algorithm which determines the required amount of computational and virtual resources to be allocated to the NSOS for a given workload so that a maximum response time of the NSOS is guaranteed. Namely, the resource dimensioning algorithm is based on a queuing model and will be invoked when a provisioning decision is taken to decide how many resources have to be requested or released. The system's performance is evaluated through system-level simulations, showing that the algorithm is able to find the minimal required resources to keep the mean response time of the NSOS under a given threshold. The response time is defined as the sum of all processing and waiting times experienced by a slice orchestration request (\emph{e.g.}, slice creation or release) when passing through different NSOS's entities during its lifetime in the orchestrator. The simulation results also suggest that the request rejection rate during a given period is determined by the reaction time of the reactive provisioning mechanism, which is in turn affected by the slice's instantiation time \cite{nsos}. As CPU resources are the only resources taken into account, the inclusion of other resources, such as memory, is encouraged in the future.

As yet, the NSOS has been purely focused on the technical aspect. Breitgand \emph{et al.} delved into the issue of coordinating and orchestrating business processes across domains in order to facilitate efficient resource sharing among multiple Communication Service Providers (CSPs) \cite{csps}. The lack of a standard for this aspect of inter-CSP collaboration is identified as a major hurdle for achieving optimal resource utilization. To address this, Breitgand \emph{et al.} proposed a set of design principles that include autonomy for CSPs in their business and technical processes, non-intrusive extensions to existing NFV MANO frameworks, preservation of slice isolation, separation of concerns between technical and business aspects of orchestration, and a cloud-native declarative orchestration approach using Kubernetes as the cross-domain control plane. The proposed dynamic NS scaling occurs through collaboration between CSPs facilitated by DLT transactions. This approach utilizes ML techniques to automate the process of extending slices and ensuring QoS requirements, which is inspired by ETSI's ZSM closed-loop architecture. This orchestrator is demonstrated on the 5GZORRO virtual content delivery network use case scenario for highly populated areas. Content delivery networks are geographically distributed networks of computation and storage resources that offer high-availability and high-performance services such as web content, application data, and live/on-demand streaming media. The proposed approach has been validated in a development environment, and future work will involve evaluating it in a larger testbed and with additional use cases to quantify the benefits of inter-CSP slice scaling \cite{csps}.

\begin{table*}[htbp]
\centering
\caption{Network Slicing Schemes}
\label{table:ns_schemes}
\scriptsize
\def\arraystretch{1.6}
\begin{tabular}{|P{2cm} |P{3.78cm}| P{3.23cm}| P{3.125cm}|}
\hline
\textbf{Scheme} & \textbf{Description} & \textbf{Outcomes} & \textbf{Future Work} \\ \hline

\textbf{RL-based Resource Allocation \cite{casale2019autonomic}} &
\RaggedRight Predict network resource demand using RL and dynamically allocate resources to different slices. &
\RaggedRight Adapt to changes in network conditions and user demands. &
\RaggedRight Test the scalability of the approach for large-scale networks.
\\ \hline

\textbf{HARNESS \cite{harness}} &
\begin{tabitemize}
    \item Introduce a high-availability and self-reliant network slicing system for 5G core, built on the principles of SON.
    \item Intelligently schedule and serve control plane USRs, for seamless high-availability service delivery, in both delay-tolerant and delay-sensitive slices.
\end{tabitemize}
&
\begin{tabitemize}
    \item HARNESS outperforms least-loaded scheduling, minimizing dropped USRs that could impact slice high-availability.
    \item HARNESS optimizes the utilization of slice resources, ensuring uninterrupted slice services.
\end{tabitemize}
 &
\begin{tabitemize}
    \item Scale the proposed HARNESS framework.
    \item Evaluate the proposed algorithms by selectively offloading the scheduling of frequent and rare control plane USRs on smart network components for different slice types.
\end{tabitemize}
\\ \hline

\textbf{ML-based Distributed MANO System \cite{manoNS}} & \RaggedRight Design a distributed and AI-driven MANO system for large-scale deployment of network slices in 6G.  &
\begin{tabitemize}
    \item Scalable
    \item Compliant with both ETSI standards, ZSM and ENI
\end{tabitemize}
&
\RaggedRight Map the framework to different architectures to test its effectiveness. \\ \hline

\textbf{Resource MANO Architecture for 5G Network Slices \cite{certifiedzsmns}} &
\RaggedRight Propose a per-MD resource allocation mechanism and an MD interworking function, aimed at facilitating the provision of E2E network services over 5G network slices.  &
\begin{tabitemize}
    \item Architecture has been verified and certified as a proof of concept by the ETSI ZSM.
    \item It supports multiple use cases and services.
    \item It is a flexible and scalable architecture.
\end{tabitemize} &
\RaggedRight Address the challenges of security and privacy in network slicing. \\ \hline

\textbf{NSOS \cite{nsos}} & \RaggedRight Design a global E2E mobile network slicing orchestration system that enables network slicing for NGNs taking into account access, core, and transport parts of the network. &
\begin{tabitemize}
    \item Instantly react to workload changes with DASA.
    \item Find minimal resources to maintain the NSOS response time threshold.
\end{tabitemize}
&
\RaggedRight Account for additional resources other than CPU. \\ \hline

\textbf{Cross-Domain Orchestration and Dynamic Scaling among CSPs \cite{csps}} &
\RaggedRight Orchestrate cross-domain business processes for efficient resource sharing among CSPs using Kubernetes-based declarative orchestration and DLT-based dynamic slicing scaling. &
\begin{tabitemize}
    \item Enhance security and trust with smart multi-party contracts. 
    \item Enable efficient utilization of resources by scaling slices up or down based on traffic demands.
    \item Support multi-domain slicing, optimizing resource allocation across different network domains.
\end{tabitemize}
&
\RaggedRight Evaluate the model with a larger testbed and additional use cases to quantify the benefits of inter-CSP slice scaling. \\ \hline

\end{tabular}
\end{table*}

\subsection{Multi-access Edge Computing}
MEC represents a paradigm shift in the delivery of computing resources and services within networks. MEC leverages the inherent properties of 5G+ networks to provide low-latency, high-bandwidth data processing and storage resources that are situated in close proximity to mobile devices and users \cite{MEC}. This distributed deployment model enables the real-time processing of data, thereby eliminating the need for extensive data transfer over long distances, reducing the burden on the network, and leading to a significant improvement in the user experience and the overall network performance.

From a technical standpoint, MEC is based on the principles of cloud computing and NFV, and involves the placement of virtualized computational and storage resources at the network edge. This creates a highly distributed computing infrastructure, which can be dynamically reconfigured and optimized to meet the changing demands of mobile users \cite{MEC}. The use of NFV ensures that the MEC infrastructure is flexible, scalable, and easily manageable, and can support a wide range of services and applications. With the integration of MEC into 5G to support time-sensitive applications, it becomes essential to incorporate this decentralized computing infrastructure into the 5G network slicing framework \cite{mecNS}. To ensure low E2E latency, the management of slice resources must be comprehensive and extend throughout the entire application service. With this goal, Bolettieri \emph{et al.} investigated the integration of the 3GPP network slicing framework with the MEC infrastructure, in order to ensure efficient management and orchestration of latency-sensitive resources and time-critical applications \cite{mecNS}. To achieve this, the authors present a novel slicing architecture that manages and coordinates slice segments across all domains. This design approach also aligns with multi-tenant MEC infrastructure using nested virtualization, where each slice segment is assigned specific management and orchestration responsibilities. However, to improve the feasibility of MEC application relocation between tenants, a solution must be devised that enhances the interaction between the proposed architecture and the 5G CN functions, facilitating the synchronization of traffic forwarding rules across various administrative domains. Another framework leveraging MEC network technology was introduced by Wu \emph{et al.} to allow Autonomous Vehicles (AVs) to adapt to changing driving conditions by sharing their driving intelligence \cite{AVMEC}. In this framework, named Intelligence Networking (Intelligence-Net), driving intelligence refers to a trained neural network model for autonomous driving. Key features Intelligence-Net include:
\begin{itemize}
    \item Sharing of driving intelligence: A unique MEC network-assisted Intelligence-Net is proposed to facilitate real-time sharing of driving intelligence between AVs, allowing for adaptation to changing environmental conditions.
    \item Segmentation of roads: The road is divided into segments, each with its own dedicated driving model tailored to its specific environmental features, reducing the dimensionality of each road segment.
    \item Continuous model updates: Whenever a specific road segment experiences environmental changes, new data is collected and used to retrain the generic base driving model, improving its ability to adapt to the new conditions.
    \item Secure and efficient learning: To ensure security and efficiency, the framework implements blockchain-enabled federated learning, which combines the privacy benefits of federated learning with the reduced communication and computation costs of transfer learning. The blockchain technology authenticates learning participants and secures the entire learning process.
\end{itemize}
Simulation results indicate that this solution can produce updated driving models that better adapt to environmental changes compared to traditional methods. AVs can then adopt these changes by downloading the updated driving models. The proposed Intelligence-Net framework has yet to fully leverage the available resources. While it poses a challenge, utilizing heterogeneous edge computing resources to optimize the system remains a desirable objective in the future. 

The integration of MEC into 5G+ networks is a key enabler of ZSM, which represents a new approach to network management \cite{MECZSM}. With MEC, network functions can be deployed and managed dynamically, providing the necessary processing and storage resources to support the rapidly changing demands of mobile users. This enables ZSM to provide dynamic and efficient network management, improving the user experience and reducing operational costs. Following this, Sousa \emph{et al.} introduced a self-healing architecture based on the ETSI ZSM framework for multi-domain B5G networks \cite{MECZSM}. This architecture utilizes ML-assisted closed control loops across various ZSM reference points to monitor network data for estimating end-service QoE KPIs and to identify faulty network links in the underlying transport network. To demonstrate this architecture in action, the authors have instantiated it in the context of automated healing of Dynamic Adaptive Streaming over HTTP video services. Two ML techniques, online and offline, are presented for estimating an SLA violation through a QoE probe at the edge and identifying the root cause in the transport network.  Experimental evaluation indicates the potential benefits of using ML for QoS-to-QoE estimation and fault identification in MEC environments. Further work will consider improvements to the ML pipelines, such as model generalization.

\begin{table}[ht]
\centering
\caption{MEC-based Schemes}
\label{table:mec_schemes}
\def\arraystretch{2}
\scriptsize
\begin{tabular}{|P{2.5cm} |P{3.2cm}| P{3cm}| P{3.2cm}|}
\hline
\textbf{Scheme} & \textbf{Description} & \textbf{Outcomes} & \textbf{Future Work} \\ \hline
 
\textbf{MEC-enabled Network Slicing Framework \cite{mecNS}} &
\RaggedRight Manage and coordinate slice segments across all domains. &
\RaggedRight This framework aligns with the multi-tenant MEC infrastructure using nested virtualization. &
\RaggedRight Facilitate synchronization of traffic forwarding rules across domains to improve MEC application relocation feasibility between tenants. \\ \hline

\textbf{Intelligence-Net \cite{AVMEC}} &
\RaggedRight Share driving intelligence in AVs to adapt to changing driving conditions. &
\begin{tabitemize}
\item Continuous model updates
\item Secure and efficient learning    
\end{tabitemize}
&
\RaggedRight Utilize heterogeneous edge computing resources to optimize the system. \\ \hline

\textbf{Self-healing Architecture based on the ETSI ZSM Framework \cite{MECZSM}} &
\RaggedRight Use ML-assisted control loops across various ZSM reference points to monitor data and identify faulty links for B5G networks. &
\RaggedRight This approach is a non-invasive and encrypted network-level traffic monitoring approach. &
\RaggedRight Refine the ML pipeline in terms of model generalization, for example. \\ \hline

\end{tabular}
\end{table}

\section{Network Traffic Control} \label{sec:trafficcontrol}
Network traffic control is an essential aspect of modern networking that aims to effectively manage and regulate the flow of data packets in a network \cite{traf1}. Its primary goal is to ensure the efficient utilization of network resources, prevent congestion, and guarantee a high level of service quality for all network users.

One key component of network traffic control is traffic prediction, which uses advanced techniques such as ML to forecast future network traffic patterns. This information is used to proactively manage network congestion and optimize network resources, thereby ensuring that data is delivered in a timely and reliable manner \cite{trafpred1}. Another crucial aspect of network traffic control is intelligent routing. This process uses advanced algorithms and data analysis to determine the most efficient path for data to travel from its source to its destination. The routing algorithm takes into account various inputs, such as network conditions, available resources, and traffic patterns to make informed decisions on how to route data \cite{ears}.

When combined, traffic prediction and intelligent routing form a powerful system for network traffic control. By providing a comprehensive understanding of network traffic patterns and conditions, this system enables network administrators to make informed decisions on how to best manage and regulate the flow of data. This ultimately leads to a more efficient use of network resources, improved performance, and a higher level of service quality for all network users.

\subsection{Traffic Prediction \& Classification}
Network traffic prediction and classification involves forecasting the volume and nature of network traffic that will transpire in the future, as well as identifying and categorizing the various types of traffic that are currently traversing the network \cite{trafpred1}. This process is of paramount importance in the realm of network planning, optimization, and security, as it allows network administrators to anticipate and prepare for changes in network traffic, and to ensure that the network is operating at peak efficiency and security.

In the context of 5G networks, the importance of network traffic prediction and classification is accentuated even further. The advent of 5G networks foreshadows a new era of network connectivity, characterized by unprecedented speed and capacity, as well as the integration of a wide array of new technologies, such as virtual and augmented reality, autonomous vehicles, and IoT devices. With this increased complexity and diversity of network traffic, it becomes all the more crucial to be able to predict and classify network traffic with a high degree of accuracy. This is essential for ensuring that the network can handle the increased demand, and that the various types of traffic are properly managed and optimized. Furthermore, 5G networks are designed to be highly dynamic and adaptable, capable of adjusting to changes in traffic patterns in real-time. This makes it even more imperative to have accurate and up-to-date traffic predictions and classifications.

The integration of ZSM technology into 5G networks serves to elevate the already intelligent process of network traffic prediction and classification to new heights of precision and automation. ZSM, as a SON technology, enables 5G networks to automatically configure and optimize themselves in real-time, based on the predictions and classifications of network traffic \cite{zsmsurvey}. Through the utilization of advanced ML algorithms and analytics, ZSM conducts a thorough analysis of historical traffic data, which is then employed to predict future traffic patterns with a high degree of accuracy. These predictions, in turn, are utilized to optimize network resources and configure the network to handle the expected traffic with optimal efficiency. Furthermore, ZSM employs advanced analytics to classify the various types of traffic traversing the network, such as voice, video, and data, and to identify specific applications and services being utilized by the network's users. This information is then used to optimize network performance, ensuring that the different types of traffic are properly managed and delivered to their intended destinations with the utmost precision.

Several approaches have been proposed for predicting and classifying network traffic in 5G networks. Table \ref{table:traf_schemes} presents an overview of the traffic prediction schemes discussed in this survey. One of the most widely used methods is ML, which can be used to identify patterns and trends in network traffic data. Various ML algorithms, such as ANNs, decision trees, and SVMs, have been applied to network traffic prediction and classification in 5G networks. For example, in a study by Fan and Liu, the authors apply supervised SVM and unsupervised K-means clustering algorithms for network traffic classification \cite{SVMkmeans}. The dataset includes flow parameters directly obtained from packet headers, such as segment size and packet inter-arrival time. The dataset is manually labeled with ten traffic types, including multimedia, mail, database, and attacks. The comparison of these two algorithms highlights the enhanced performance of the SVM model compared to the K-means clustering algorithm. However, K-means is able to characterize new or unknown application types as training samples do not require manual labeling in advance. Future work aims to use the proposed model to predict the number of user plane 5G network functions needed to manage the data plane traffic in a virtualized environment. The next steps comprise applying the described classification models to real SDN traffic data.

DL is another approach that has been proposed for network traffic prediction and classification in 5G networks. DL algorithms, such as CNNs and RNNs, have been shown to be particularly effective for this task. The advantage of DL is that it can automatically learn features from raw data, which reduces the need for feature engineering. In a study by Jaffry \emph{et al.}, the authors proposed a RNN with an LSTM approach for cellular traffic prediction using real-world call data records \cite{LSTMTP}. The call data record utilized was published by Telecom Italia for the Big Data Challenge competition and collected for 62 days starting November 1\textsuperscript{st}, 2013 \cite{telecommitalia}. Sample data collected includes country code, inbound/outbound SMS activities, and inbound/outbound call activities. The proposed LSTM model has a hidden layer with 50 LSTM cells followed by a dense layer with one unit. Results highlight the enhanced performance of the proposed model over vanilla neural networks and statistical autoregressive integrated moving average model. This work can be extended by using this model to design an autonomous resource allocation scheme for 5G+ networks. Similarly, Alawe \emph{et al.} presented a combination of LSTM and deep neural networks to proactively predict the number of resources along with the network traffic to manage and scale the CN in terms of the resources used for the Access and Mobility Management Function (AMF) in 5G systems \cite{trafficforecastalawe}. Their experimental results reveal that the use of ML approaches improves the scalability and reacts to the change in traffic with lower latency. Gupta \emph{et al.} delved into the examination of various DL models, namely the Multi-Layer Perceptron (MLP), Attention-based Encoder Decoder, Gated Recurrent Unit (GRU), and LSTM, on the Dataset-Unicauca-V2 mobile-traffic dataset \cite{guptaDL}. This dataset comprises a compendium of six days of mobile traffic data, boasting a total of 87 features and 3,577,296 instances. The data presented was procured from the web division of Universidad Del Cauca, Colombia, Popayan, where it was meticulously recorded over a span of six days, specifically April 26, 27, 28 and May 9, 11, 15, 2017, at various hours, including both morning and evening.  The data has been meticulously classified into four distinct categories, namely streaming, messaging, searching, and cloud. Each sample contains comprehensive information about IP traffic generated by network equipment, including the IP address of origin and destination, port, arrival time, and layer 7 protocol (application). As for performance metrics, recall, precision, and f1-score were utilized to measure the performance of the models. Findings indicate that the MLP and Encoder-Decoder models yielded average results for mobile-traffic forecasting, while the GRU and LSTM models performed exceptionally well, with the latter yielding the optimal outcome. In the future, Gupta \emph{et al.} aim to investigate other time-prediction approaches for resources and to work in MEC in industrial IoT applications to support industry 4.0.

\begin{table}[htbp]
\centering
\caption{Traffic Prediction Schemes}
\label{table:traf_schemes}
\def\arraystretch{2}
\scriptsize
\begin{tabular}{|P{0.165\linewidth} |P{0.238\linewidth}| P{0.238\linewidth}| P{0.238\linewidth}|}
\hline
\textbf{Scheme} & \textbf{Description} & \textbf{Outcomes} & \textbf{Future Work} \\ \hline

\textbf{SVM and K-means Algorithms for Network Traffic Classification \cite{SVMkmeans}} &
\RaggedRight Predict and classify the network traffic in 5G networks. &
\begin{tabitemize}
    \item SVM shows enhanced performance compared to K-means. \item K-means is suitable for new/unknown application types.
\end{tabitemize}  &
\begin{tabitemize}
    \item Predict the number of user plane 5G network functions to manage the data plane traffic. \item Apply the proposed models to real SDN traffic data.
\end{tabitemize}
\\ \hline

\textbf{DL-based Cellular Traffic Prediction \cite{LSTMTP}} &
\RaggedRight Predict cellular traffic with RNN LSTM model using real world call data record. &
\RaggedRight The proposed model outperforms vanilla neural networks and autoregressive integrated moving average model. &
\RaggedRight Extend the work to design an autonomous resource allocation scheme for 5G+ networks. \\ \hline

\textbf{Scaling AMF in 5G Systems \cite{trafficforecastalawe}} &
\RaggedRight Proactively predict the number of resources in addition to the network traffic to manage resources and scale the AMF in 5G systems. &
\RaggedRight The proposed model reacts to the change in traffic with lower latency.
& \RaggedRight Utilize the proposed model to estimate the number of user plane 5G CN functions needed to handle the traffic. \\ \hline

\textbf{DL-based Mobile Traffic Prediction \cite{guptaDL}} &
\RaggedRight Examine various DL models on the Dataset-Unicauca-V2 mobile-traffic dataset. & 
\RaggedRight LSTM yields optimal outcomes followed by GRU, encoder-decoder, and MLP models, respectively. &
\begin{tabitemize}
    \item Investigate other time-prediction approaches.
    \item Work in MEC in industrial IoT applications to support industry 4.0.
\end{tabitemize}
\\ \hline
\end{tabular}
\end{table}

\subsection{Intelligent Routing}
With the advent of intelligent routing, networks are now equipped to adapt and optimize routes in real-time, taking the first step towards self-driving zero-touch networks. Accordingly, Hu \emph{et al.} unveiled EARS, an intelligence-driven experiential network architecture that seamlessly integrates the cutting-edge technologies of SDN and DRL to usher in the era of intelligent and autonomous routing \cite{ears}. Hu \emph{et al.} addressed the limitations of conventional routing strategies, which are heavily reliant on manual configuration, and introduced a DRL algorithm to optimize data flow routing. This DRL agent is highly adaptable, learning routing strategies through its interactions with the network environment. Furthermore, EARS incorporates advanced network monitoring technologies, such as network state collection and traffic identification, to provide closed-loop control mechanisms, enabling the DRL agent to optimize routing policies and enhance network performance. As such, EARS, an intelligence-driven experiential network architecture, harnesses the Deep Deterministic Policy Gradient (DDPG) algorithm to dynamically generate routing policies. DDPG utilizes a deep neural network to simulate the Q-table, and another neural network to produce the strategy function, allowing it to effectively tackle large-scale continuous control problems. Through relentless training, EARS can learn to make better control decisions by interacting with the network environment, and adjusting services and resources based on network requirements and environmental conditions. Simulations, which compare EARS with typical baseline schemes such as Open Shortest-Path First \cite{ospf} and Equal-Cost Multi-Path Routing \cite{multipath}, demonstrate that EARS surpasses these schemes by achieving superior network performance in terms of throughput, delay, and link utilization. As a future undertaking, the algorithm will be evaluated in a real-world SDN setup. Similarly, Tan \emph{et al.} introduced a load balancing algorithm, the Reliable Intelligent Routing Mechanism (RIRM), designed to optimize traffic data routing based on the traffic load in 5G CNs \cite{rirm}. This algorithm takes into account multiple factors, such as the shortest path, link latency, and node loading to prevent packet loss. It has been implemented on a 5G testbed, free5GC, by adding two elements, the RIRM traffic tracker and the RIRM traffic controller. The tracker continuously monitors user traffic data flow and reports to the controller, which then calculates the best route to avoid congestion. Experimental results demonstrate that the proposed algorithm surpasses traditional round-robin load balancing methods in terms of packet loss, latency, and average data throughput. This algorithm is specifically tailored for the uRLLC use case of 5G, and further research will explore its performance in other use cases, such as mMTC.

One application of 5G access networks is Flying Ad Hoc Networks (FANETs) that utilize unmanned aerial vehicles as nodes to provide wireless access. These networks are characterized by their limited resources and high mobility, which pose significant challenges for efficient routing \cite{fanets}. In this context, intelligent routing is a crucial component of FANETs, as it enables the network to adapt to the dynamic conditions of the environment and optimize the use of resources. The use of advanced techniques, such as DRL, is a promising approach to addressing these challenges. Deep Q-Network (DQN)-based vertical routing, proposed by Khan \emph{et al.}, is an example of such an intelligent routing approach, which aims to select routes with better energy efficiency and lower mobility across different network planes \cite{fanets}. DQN, the underlying mechanism of this proposed routing method, leverages the principles of RL and DL to empower an agent to make informed decisions based on the current state of the system. The integration of deep neural networks within DQN enables the agent to learn about various states, such as the residual energy and mobility rate, to predict optimal actions. The training process of DQN involves the use of mini-batches of experiences, and there are three vital features that aid in achieving an accelerated convergence toward the most optimal route:
\begin{enumerate}
    \item Delayed rewards from the replay memory allow for better prediction of state-action values in a dynamic environment.
    \item The decaying variable shifts the focus from exploration to exploitation as the number of episodes increases.
    \item Mini batches of run-time states from the replay memory are used to train and minimize a loss function.
\end{enumerate}
The main objective is to improve network performance by reducing frequent disconnections and partitions. The proposed method is a hybrid approach that utilizes both a central controller and distributed controllers to share information and handle global and local information, respectively. It is suitable for highly dynamic FANETs and can be applied in various scenarios, such as border monitoring and targeted operations. By clustering the network across different planes, this method offloads data traffic and improves network lifetime. Simulation results show that it can increase network lifetime by up to 60\%, reduce energy consumption by up to 20\%, and reduce the rate of link breakages by up to 50\% compared to traditional RL methods. Khan \emph{et al.} suggested that there are several open issues that can be explored in future research to improve the proposed routing method for FANETs in 5G access networks. These include:
\begin{itemize}
    \item Enhancing the vertical routing method to reduce the routing overhead incurred to establish and maintain inter- and intra-cluster, and inter- and intra-plane routes, by allowing clusters to adjust their sizes for optimal performance.
    \item Exploring other variants of DQN, such as DDPG, which is well suited for continuous action space, and double DQN, which addresses the issue of overestimation of Q-values.
    \item Examining the use of mini-batches from the replay memory to prevent overfitting and selecting distinctive experiences with equal chances.
    \item Investigating the method in different scenarios with varying amounts of types of terrain and types of obstacles.
\end{itemize}

As the demand for fast communication with minimal delay continues to rise in the field of intelligent transportation systems, Vehicle-to-Everything (V2X) communications have become increasingly crucial in enabling the seamless sharing of information among vehicles. The integration of 5G technology, particularly millimeter wave (mmWave), holds great promise in achieving these goals, but also presents unique challenges in terms of beam alignment and routing stability due to the dynamic nature of vehicle traffic \cite{v2x}. Addressing these challenges for V2X communications in 5G-based mmWave systems, Rasheed \emph{et al.} employed a 3D-based position detection scheme for beam alignment and a group-based routing algorithm for secure and stable data transmissions \cite{v2x}. Vehicles are grouped based on their distance, direction, and speed, and each group is headed by a leader who is responsible for authenticating the group members. Routing paths for packet transmission from the source vehicle are chosen based on the link weight, which takes into account the distance between neighbors and the direct/indirect trust degree. Data compression and encryption techniques are used to enhance the security of data transmissions. The effectiveness of the proposed approach is demonstrated through simulations, which show significant improvements over existing V2X communication schemes such as the Moving Zone-Based Routing Protocol. A summary of the schemes discussed in this section is presented in Table \ref{table:routing_schemes}.

\begin{table}[h!]
\centering
\caption{Intelligent Routing Schemes}
\label{table:routing_schemes}
\def\arraystretch{2}
\scriptsize
\begin{tabular}{|P{2cm} |P{3.2cm}| P{3.4cm}| P{3.4cm}|}
\hline
\textbf{Scheme} & \textbf{Description} & \textbf{Outcomes} & \textbf{Future Work} \\ \hline

\textbf{EARS \cite{ears}} & \RaggedRight Intelligence-driven experiential network architecture is designed to optimize data flow routing using DDPG. &
\RaggedRight EARS adjusts services and resources based on network requirements and environmental conditions. &
\RaggedRight Test the algorithm in a real-world SDN setup. \\ \hline

\textbf{RIRM \cite{rirm}} & \RaggedRight Design a load balancing algorithm to optimize traffic data routing based on the traffic load in 5G CNs for the uRLLC use case. &
\RaggedRight The proposed algorithm surpasses traditional round-robin load balancing methods in terms of packet loss, latency, and average data throughput.
& \RaggedRight Test the algorithm in other use cases, such as mMTC. \\ \hline

\textbf{DQN-based Vertical Routing \cite{fanets}} &
\RaggedRight Select routes with better energy efficiency and lower mobility across different network planes for FANETs. &
\begin{tabitemize}
    \item Improve the network performance by reducing frequent disconnections and partitions.
    \item Offload data traffic and improves network lifetime by clustering the network across different planes.
    \item Work with highly dynamic FANETs.
\end{tabitemize}
&
\begin{tabitemize}
    \item Reduce routing overhead via adjustable size clusters.
    \item Explore other variants of DQN, such as DDPG and double DQN.
    \item Examine the use of mini-batches from the replay memory to prevent overfitting.
    \item Evaluate the scheme with different types of terrains and obstacles.
\end{tabitemize}
 \\ \hline

\textbf{3D Position Detection \& Group-Based Routing for Secure Data Transmission \cite{v2x}} &
\RaggedRight Secure and stabilize data transmissions using a 3D-based position detection scheme for beam alignment and a group-based routing algorithm for V2X communications in 5G-based mmWave systems. &
\begin{tabitemize}
    \item Optimize packet transmission paths in view of the distance between neighbors and trust levels.
    \item Strengthen data transmissions via data compression and encryption techniques.
\end{tabitemize} &
\RaggedRight Utilize ECC-based session keys for secure communication between vehicles. \\ \hline
\end{tabular}
\end{table}

\section{Towards Energy Efficiency}\label{sec:energyefficiency}
Energy efficiency in networking is a crucial aspect that has gained significant attention in recent years. The rapid growth of data traffic and the increasing number of connected devices have led to a significant increase in energy consumption in networking systems. This has motivated researchers to focus on developing energy-efficient networking solutions to reduce the environmental impact and operating costs of networks.

One of the key approaches for achieving energy efficiency in networking is through the use of SDN and NFV technologies \cite{energyawaredef}. These technologies enable the dynamic provisioning and scaling of network resources, which can lead to significant energy savings. Additionally, the use of VNFs can reduce the energy consumption of network devices by consolidating multiple functions on a single physical platform. These enablers comprise 5G+ networks.

Energy efficiency in 5G+ networks is a critical aspect of the next generation of mobile networks as it directly impacts the environmental and economic sustainability of these networks. The deployment of 5G+ networks is expected to result in a significant increase in energy consumption due to the expansion of network infrastructure, the deployment of new technologies such as massive multiple-input/multiple-output and mmWave communications, and the increased demand for high-bandwidth applications \cite{energyawaredef}.

There are several key strategies that have been proposed to improve the energy efficiency of 5G+ networks. One of the key strategies is the use of energy-efficient network architecture and protocols. This includes the use of energy-efficient radio access technologies, such as energy-aware scheduling and power control, as well as the use of energy-efficient network functions and infrastructure. Another key strategy is the use of energy-efficient devices, such as energy-efficient smartphones and IoT devices, which can reduce the overall energy consumption of the network.

Zero-touch automation is another promising approach to improve the energy efficiency of 5G+ networks to automate the configuration, management, and optimization of network functions and infrastructure. This can help to reduce the energy consumption of the network by reducing the need for manual intervention and by enabling the network to adapt to changes in traffic demand and network conditions in a more efficient manner.

As such, Omar \emph{et al.} formulated an optimization problem to compute a green efficient solution to maximize energy efficiency under minimal area spectral efficiency and outage probability in a 5G heterogeneous network \cite{aseEE}. Using a convex method, this problem was solved. Results prove that network densification does not always result in the optimal efficient solution as the increase in mmWave base stations increases area spectral efficiency and decreases energy efficiency. As for introducing mmWave small cells, it has been established that they improve coverage, and consequently spectral efficiency. Future work suggests designing deployment strategies that have the environment in mind. Tackling a specific 5G use case, Dalgkitsis \emph{et al.} examined the impact of network automation on energy consumption and overall operating costs in the context of 5G networks, specifically for uRLLC services \cite{sche2ma}. A framework, known as Service CHain Energy-Efficient Management or SCHE2MA, is proposed, which utilizes distributed RL to intelligently deploy service function chains with shared VNFs across multiple domains. SCHE2MA framework is designed to be decentralized, eliminating the potential for central points of failure, allowing for scalability, and avoiding costly network-wide configurations. Parallelism is also achieved by introducing the auction mechanism, a system that enables inter-domain VNF migration in a distributed multi-domain network. Results show the reduction of average service latency with the enhancement of energy efficiency by 17.1\% compared to a centralized RL solution. This approach addresses the important challenge of balancing the performance constraints of uRLLC services with energy efficiency in the context of 5G+ networks, where reducing carbon emissions and energy consumption is of paramount importance. Future work will focus on implementing the auction mechanism in a fully decentralized manner by employing DLTs.

In the context of zero-touch, Rezazadeh \emph{et al.} proposed a framework for fully automated MANO of 5G+ communication systems which utilizes a knowledge plane and incorporates recent network slicing technologies \cite{ZTenergycontrolNS}. The knowledge plane plays the role of an all-encompassing system within the network by creating and retaining high-level models of what the network is supposed to do, in order to provide services to other blocks in the network. In other words, the knowledge plane joins the architectural aspects of network slicing to achieve synchronization in a continuous control setting by revisiting ZSM operational closed-loop building blocks. This framework, known as Knowledge-based Beyond 5G or KB5G, is based on the use of algorithmic innovation and AI, utilizing a DRL method to minimize energy consumption and the cost of VNF instantiation. A unique Actor-Critic based approach, known as the twin-delayed double-Q soft actor-critic method, allows for continuous learning and accumulation of past knowledge to minimize future costs. This stochastic method supports continuous state and action spaces while stabilizing the learning procedure and improving time efficiency in 5G+. It also promotes a model-free approach reinforcing the dynamism and heterogeneous nature of network slices while reducing the need for hyperparameter tuning. This framework was tested on a 5G RAN NS environment called smartech-v2, which incorporates both CPU and energy consumption simulators with an OpenAI Gym-based standardized interface to guarantee the consistent comparison of different DRL algorithms. Numerical results demonstrate the advantages of this approach and its effectiveness in terms of energy consumption, CPU utilization, and time efficiency. Future directions suggest the inclusion of different resources, such as memory. Table \ref{table:energy_schemes} summarizes the mentioned schemes, including future work to enhance their performance and utility.

\begin{table}[ht]
\centering
\caption{Energy Efficiency Schemes}
\label{table:energy_schemes}
\def\arraystretch{2}
\scriptsize
\begin{tabular}{|P{1.85cm} |P{0.28\linewidth}| P{0.23\linewidth}| P{0.22\linewidth}|}
\hline
\textbf{Scheme} & \textbf{Description} & \textbf{Outcomes} & \textbf{Future Work} \\ \hline

\textbf{Green Efficiency Optimization \cite{aseEE}} &
\RaggedRight A green efficient solution is designed to maximize energy efficiency under minimal area spectral efficiency and outage probability in a 5G heterogeneous network. &
\begin{tabitemize}
    \item Network densification does not always result in the best solution.
    \item Incorporating mmWave small cells improve coverage and spectral efficiency.
\end{tabitemize}
&
\RaggedRight Design deployment strategies that have the environment in mind. \\ \hline

\textbf{SCHE2MA \cite{sche2ma}} &
\RaggedRight Utilize distributed RL to intelligently deploy service function chains with shared VNFs across multiple domains for uRLLC services. &
\begin{tabitemize}
    \item Eliminate the potential for central points of failure.
    \item Achieve parallelism through the auction mechanism.
\end{tabitemize}
&
\RaggedRight Employ DLTs to implement the auction mechanism in a fully decentralized manner. \\ \hline

\textbf{KB5G \cite{ZTenergycontrolNS}} &
\begin{tabitemize}
    \item A framework for fully automated MANO of 5G+ communication systems is proposed.
    \item A twin-delayed double-Q soft actor-critic algorithm is designed to minimize energy consumption and the cost of VNF instantiation.
\end{tabitemize}
&
\begin{tabitemize}
    \item Support continuous state and action spaces.
    \item Stabilize the learning procedure.
    \item Improve time efficiency in 5G+.
\end{tabitemize}
&
\RaggedRight Take into account resources other than CPU, such as memory. \\ \hline
\end{tabular}
\end{table}

\section{Network Security \& Privacy}\label{sec:networksecurity}
Network security and privacy are fundamental components of modern communication networks, ensuring the confidentiality, integrity, and availability of data and information transmitted over these networks. They are essential for protecting against unauthorized access, malicious attacks, and data breaches, and for preserving the privacy of network users.

\subsection{Safeguarding 5G+ Networks: Security Measures \& Weaknesses}
5G+ networks are characterized by a highly distributed architecture, which consists of multiple network elements such as the RAN, the CN, and the transport network \cite{zsm5g6g}. This architecture poses new security challenges, as it increases the attack surface and creates new attack vectors for malicious actors. 5G+ networks also have a greater number of connected devices compared to previous generations of communication networks, which further exacerbates the security and privacy risks. Additionally, the transmission of data in 5G+ networks is characterized by elevated levels of speed, bandwidth, and connectivity. This has dramatically increased the volume of data transmitted over these networks. As such, it is imperative to implement robust security and privacy measures to protect against the threat of cyberattacks and promote trust in 5G+ networks.

5G+ networks integrate a plethora of security mechanisms to secure the transmission of data, protect network infrastructure, and prevent unauthorized access to sensitive information \cite{zsmsecurity}. These mechanisms include, but are not limited to, cryptographic algorithms, firewalls, Virtual Private Networks (VPNs), Intrusion Detection Systems (IDSs), NFV, and SDN. Cryptographic algorithms, such as Advanced Encryption Standard, ensure the confidentiality and integrity of the data transmitted over the network. Firewalls act as a barrier between the internal and external networks, providing a line of defense against unauthorized access and malicious attacks. VPNs allow secure communication over an insecure network, and IDSs detect and respond to security threats in real-time. NFV and SDN technologies abstract the underlying hardware from the network services, enabling the automation of network management and reducing the attack surface.

Despite these robust security measures, 5G+ networks remain vulnerable to a wide range of threats. Threats to 5G networks can be distinguished based on the technological domains that they impact \cite{insliceddos}.
\begin{enumerate}
    \item User Equipment (UE) Threats: These attacks are aimed at targeting mobile devices of end-users such as smartphones and laptops. This can encompass the utilization of mobile botnets to launch Distributed Denial of Service (DDoS) attacks on various network layers, with the objective of disrupting and shutting down services.
    \item RAN Threats: These attacks focus on the RAN, which is responsible for the wireless connection between the UE and the network. The presence of rogue base stations that launch Man in the Middle (MitM) attacks can compromise user information, break privacy, track users, and cause Denial of Service (DoS).
    \item CN Threats: The CN is responsible for the management and direction of data traffic within the network, and is therefore a target for security threats. These can encompass attacks on elements such as SDN and VNF components, as well as Network Slicing, leading to DoS, eavesdropping, interception, or hijacking.
    \item Network Slicing Threats: Network slicing involves creating virtual networks within the network to meet specific needs. These attacks target this concept and can compromise the isolation between slices, thereby compromising security and privacy.
    \item SDN Threats: SDN separates control and user (data) planes, making it a potential target for malicious actors. These attacks target the link between the control and user planes and can take the form of distributed DoS attacks or gaining control over network devices through Topology Poisoning attacks.
\end{enumerate}
As such, it is imperative for network operators to remain vigilant and proactive in their approach to network security, and to continuously monitor and update their security solutions to stay ahead of evolving security threats.

\subsection{ZSM Security Threats}
ZSM is a key aspect of network security and privacy for 5G+ networks. As previously iterated, it enables the automated deployment and management of network services, reducing the risk of human error and enhancing network security. The implementation of ZSM allows network operators to manage network configurations and deploy network services with unparalleled speed, scalability, and accuracy, minimizing the potential for security breaches and ensuring the confidentiality and integrity of the data transmitted over the network \cite{zsmsecurity}. Nevertheless, the very factors that enable a ZSM system may also breed various security threats that could obstruct its functionality. These threats are summarized in Table \ref{table:zsmSecurity} \cite{zsmsecurity, survey5GP}.

\begin{table}[h!]
\caption{Security Threats within the ZSM Context \cite{zsmsecurity, survey5GP}}
\label{table:zsmSecurity}
\centering
\scriptsize
\bgroup
\def\arraystretch{1.5}
\begin{tabular}{|P{0.12\linewidth}|  P{0.19\linewidth}  P{0.6\linewidth}|}
\hline
\cellcolor[HTML]{FFFFFF}\textcolor{black}{\textbf{Enabler}}&
  \cellcolor[HTML]{FFFFFF}\textcolor{black}{\textbf{Attack}} &
  \multicolumn{1}{c|}{\cellcolor[HTML]{FFFFFF}\textcolor{black}{\textbf{Description}}} \\ \hline
\multirow{13}{*}{\textcolor{black}{\textbf{Open API}}} & 
  \cellcolor[HTML]{E7E7E7} Script Insertion &
  \cellcolor[HTML]{E7E7E7} The attack exploits vulnerabilities in systems that treat the inputted parameter as a script. \\  
 &
  SQL Injection &
  Malicious code is inserted into a database via a vulnerable input, compromising the database's security \\ 
 &
  \cellcolor[HTML]{E7E7E7} Buffer Overflow Attack &
  \cellcolor[HTML]{E7E7E7} The attack is activated by data that falls outside the expected types or ranges, causing the system to malfunction and granting access to its memory areas. \\
 &
  Identity Attack & Attack attempts to access a targeted API by using a list of previously compromised passwords, stolen credentials, or tokens. \\
 &
  \cellcolor[HTML]{E7E7E7} DoS Attack &
  \cellcolor[HTML]{E7E7E7} An attacker floods the API with a high volume of requests, rendering it unavailable. \\ 
 &
  Application and Data Attack &
  They involve unauthorized access to data, alteration/deletion of data, insertion of malicious code, and disruption of scripts. \\
 &
  \cellcolor[HTML]{E7E7E7} MiTM Attack &
  \cellcolor[HTML]{E7E7E7} An attacker intercepts the communication between the API client and server to steal confidential information \\ \hline
\multirow{2}{*}{\cellcolor[HTML]{FFFFFF} \begin{tabular}[c]{P{0.89\linewidth}@{}c@{}}\end{tabular}}

\multirow{2}{*}{\textcolor{black}{\textbf{Intents}}} &
  Data Exposure &
  An unauthorized individual intercepts information related to the application's purpose (\emph{e.g.,} advertising content), exposing the system's goals to risks and triggering additional attacks. \\ \cellcolor[HTML]{FFFFFF} 
 &
  \cellcolor[HTML]{E7E7E7} Tampering &
  \cellcolor[HTML]{E7E7E7} The attacker makes physical modifications to a connection point or interface.\\ \hline
\multirow{3}{*}{\cellcolor[HTML]{FFFFFF} \begin{tabular}[]{p{0.88\linewidth}@{}c@{}}\end{tabular}} &
  MITM Attack &
  An attacker intercepts messages between two entities in order to remotely eavesdrop on or alter the traffic. \\ \cellcolor[HTML]{FFFFFF} \textcolor{black}{\textbf{Automated Closed-Loop}} 
 &
  \cellcolor[HTML]{E7E7E7} Deception Attack &
  \cellcolor[HTML]{E7E7E7} The deceiver convinces the target to believe a false version of the truth and manipulates the target's actions to benefit the deceiver. \\ \cellcolor[HTML]{FFFFFF}
 &
  DoS Attack & A DoS attack overloads the network with a high volume of traffic, making the network unavailable to its users.\\ \hline
\multirow{2}{*}{ \cellcolor[HTML]{FFFFFF}} 

\multirow{2}{*}{\textcolor{black}{\textbf{SDN/NFV}}} &
  \cellcolor[HTML]{E7E7E7} Privilege Escalation &
  \cellcolor[HTML]{E7E7E7} An intruder gains access to a target account, bypassing authorization and gaining unauthorized access to data. \\   
 &
  Spoofing &
  An attacker sends false address resolution protocol messages over a local area network.  \\ \hline
\multirow{5}{*}{\textcolor{black}{\textbf{AI/ML}}} &
  \cellcolor[HTML]{E7E7E7} Adversarial Attack &
  \cellcolor[HTML]{E7E7E7} A malicious actor tampers the training data and/or inserts small perturbations into the test instances. \\  
 &
  Model Extraction Attack &
  The attack attempts to steal the model's parameters to recreate a similar ML model. \\ \cellcolor[HTML]{FFFFFF}
 &
  \cellcolor[HTML]{E7E7E7} Model Inversion Attack &
  \cellcolor[HTML]{E7E7E7} The attack aims to recover the training data or the underlying information from the model's outputs. \\ \hline
\end{tabular}
\egroup
\end{table}

\subsection{Advances in 5G+ Network Trust Management}
To enhance trust in 5G+ networks, Benzaïd \emph{et al.} introduced a blockchain-based data integrity framework \cite{trust5Gblockchain} that ensures the security and privacy of the data processed by ML pipelines. This framework records and stores inputs and outputs of ML pipelines on a tamper-evident log, and uses smart contracts to enforce data quality requirements and validate the data processed. By providing transparency and verifiability in the data used by ML pipelines, the blockchain technology helps detect and correct any data tampering or manipulation, thus improving trust and reliability in ML results and contributing to the overall security and privacy of 5G+ networks. Additionally, Palma \emph{et al.} enhanced the security and trustworthiness of 5G+ networks by integrating Manufacturer Usage Description (MUD) and Trust and Reputation Manager (TRM) into the INSPIRE-5GPlus framework. \cite{MUDtrm}. MUD is a standard that provides access control by specifying the type of access and network functionalities available for different devices in the infrastructure. It helps to configure monitoring tools and learn about the normal behavior of devices, which enables the identification of abnormal events on the 5G infrastructure. TRM assesses trust in the infrastructure using multiple values and enables MUD security requirements to be enforced in a trustworthy manner. The integration of MUD and TRM into INSPIRE-5GPlus enhances security and trust by enforcing security properties and continuously auditing the infrastructure and security metrics to compute trust and reputation values. These values are used to enhance the trustworthiness of zero-touch decision-making, such as the ones orchestrating E2E security in a closed-loop. Future research aims to develop trust tools at each domain level to create a comprehensive trust framework. Another study by Niboucha \emph{et al.} incorporated a zero-touch security management solution tackling the problem of in-slice DDoS attacks in mMTC network slices of 5G \cite{insliceddos}. The proposed solution employs a closed-control loop that monitors and detects any abnormal behavior of MTC devices, and in the event of an attack, it automatically disconnects and blocks the compromised devices. This was achieved by following 3GPP traffic models and training a ML model using gradient boosting to identify normal and abnormal traffic patterns. The detection algorithm then calculates the detection rate for each device, and the decision engine takes the necessary steps to mitigate the attack by severing the connection between the devices and the network, thereby safeguarding against any potential reoccurrence of similar attacks. Results show its effectiveness in detecting and mitigating DDoS attacks efficiently. Future lines of research will focus on utilizing online learning techniques in the event of encountering new forms of attacks. 

In regards to the ZSM paradigm, the use of ML models in the architecture raises privacy and resource limitations concerns. In turn, Jayasinghe \emph{et al.} presented a multi-stage federated learning-based model that incorporates ZSM architecture for network automation \cite{FLzsm}. The proposed model is a hierarchical anomaly detection mechanism consisting of two stages of network traffic analysis, each with an federated learning-based detector to remove identified anomalies. The complexity and size of the detector's database vary depending on the stage. The authors simulate the proposed system using the UNSW-NB 15 network dataset and demonstrate its accuracy by varying the anomaly percentage in both stages. The results show that the model reaches a minimum accuracy of 93.6\%. Future work aims to increase the accuracy of the model and apply it in a security analytics framework in the ZSM security architecture. To summarize, the key findings and future directions from the schemes analyzed in this section are presented in Table \ref{table:sec_schemes}.

\begin{table}[ht]
\centering
\caption{Network Trust Management Schemes}
\label{table:sec_schemes}
\def\arraystretch{2}
\scriptsize
\begin{tabular}{|P{2.4cm} |P{3.8cm}| P{2.78cm}| P{2.8cm}|}
\hline
\textbf{Scheme} & \textbf{Description} & \textbf{Outcomes} & \textbf{Future Work} \\ \hline

\textbf{Blockchain-based Data Integrity Framework \cite{trust5Gblockchain}} &
\RaggedRight A framework is designed to ensure the security and privacy of data processed by ML pipelines in 5G+ networks. &
\RaggedRight Enhance trust and reliability in ML results.
&
\RaggedRight Design liability-aware trust schemes to enable liable E2E service delivery in NGNs. \\ \hline

\textbf{MUD and TRM Integration \cite{MUDtrm}} &
\RaggedRight Integrate MUD and TRM into INSPIRE-5GPlus. &
\RaggedRight The proposed scheme enforces security properties. &
\RaggedRight Incorporate trust tools at each domain level to create a comprehensive trust framework. \\ \hline

\textbf{Zero-touch Security Management Solution for In-slice DDoS Attacks \cite{insliceddos}} &
\begin{tabitemize}
    \item Monitor and detect abnormal mMTC device behavior.
    \item Automatically disconnect and block compromised devices during an attack.
\end{tabitemize}
&
\RaggedRight Detect and mitigate DDoS attacks efficiently. &
\RaggedRight Apply online learning techniques to tackle new forms of attacks. \\ \hline

\textbf{Multi-stage Federated Learning-based Model \cite{FLzsm}} &
\RaggedRight A two-stage anomaly detection mechanism is proposed to analyze network traffic, where each stage is equipped with a federated learning-based detector to remove identified anomalies. &
\begin{tabitemize}
    \item Decentralized processing
    \item Higher privacy
    \item Communication efficiency
\end{tabitemize}  &
\RaggedRight Apply it in a security analytics framework in the ZSM security architecture. \\ \hline
\end{tabular}
\end{table}

\section{Network Automation Solutions}\label{sec:automationsolutions}
The evolution of networking technology has triggered a fundamental shift in network management, owing to the growing scale and complexity of networks, making traditional manual configuration and management techniques inefficient and error-prone. In this regard, network automation solutions, such as the ZSM framework, offer a revolutionary approach to network management, leveraging cutting-edge technologies such as ML, AI, and SDN to streamline operations and enhance network performance \cite{zsmsurvey}. These methodologies play a pivotal role in enabling automated self-management functionalities of ZSM, thus leading to an enhancement in service delivery and a reduction in operating expenses.

At the heart of network automation lies the use of software to automate tedious and time-consuming tasks like device configuration, policy enforcement, and network monitoring. By exploiting programmable interfaces and open standards, network automation solutions facilitate seamless interoperability and integration across multi-vendor environments, providing a unified network view, and enabling fast issue resolution and troubleshooting.

Furthermore, network automation solutions leverage the power of AI and ML to facilitate self-healing and self-optimizing networks \cite{MLCH1}. By monitoring network performance in real-time and detecting anomalies, automation solutions can trigger automated workflows that mitigate potential issues and restore normal operation. This capability substantially reduces downtime and boosts network resiliency, while the identification of under/over-utilized areas provides efficient resource optimization. Additionally, ML algorithms enhance security by flagging and mitigating potential threats through traffic patterns and user behavior analysis, offering a proactive response to security threats such as malware or cyberattacks.

Nevertheless, the application of AI techniques in automation solutions presents challenges and risks. Although AI and ML techniques enable cognitive processing in the ZSM system, resulting in complete automation, the performance of such an implementation is not entirely satisfactory \cite{survey5GP}. Network operators expect superior service availability and reliability to minimize network outages and SLA violations, which could result in significant financial losses.

To address these challenges, one can leverage the power of DTs and AutoML. DTs facilitate the creation of virtual replicas of the physical network infrastructure, enabling operators to test and validate network configuration changes in a simulated environment before deploying them to the live network \cite{DT2}. This reduces the risk of misconfiguration, human errors, and downtime. Moreover, AutoML algorithms enable the automatic processing of data and the discovery of optimal configurations and models, reducing the need for manual intervention \cite{autoML}. This section highlights significant ML challenges and proposes solutions to address them.

\subsection{AI/ML Challenges}
The path toward integrating ML into network automation solutions is not without its obstacles. To unlock the full potential of ML for network management, it is crucial to identify and tackle the unique challenges that arise with this innovative technology. By analyzing and understanding the potential complexities and roadblocks associated with ML integration, organizations can overcome them and leverage the power of ML using DTs and AutoML to achieve optimal network performance.

\subsubsection{Need for Skilled Personnel}
The application of ML in network automation requires a substantial level of data science expertise, coupled with specialized knowledge of network infrastructure. However, finding and retaining such skilled personnel is a formidable task, with the demand for data scientists being high \cite{MLCH2}. This challenge calls for innovative solutions that reduce the level of expertise required while maintaining optimal performance.

DTs and AutoML represent two such solutions. DTs simulate network behavior and generate large amounts of training data, automating the data preparation process. AutoML can automatically select the best-performing algorithm and parameters without the need for human intervention, eliminating the requirement for a highly skilled workforce. In this way, DTs and AutoML address the need for data science experts, making network automation solutions more accessible and cost-effective. 

\subsubsection{Data Processing Challenges}
Data processing is a critical step in ML, as high-quality data is necessary for training effective models. However, in this era, network data is complex, heterogeneous, and voluminous, making it challenging to process and analyze manually \cite{MLCH1}. Data processing involves data collection, cleaning, and feature extraction. Data collection involves acquiring the data from the network, and cleaning involves removing unwanted or corrupted data points. Feature extraction involves selecting and transforming the relevant features of the data. These tasks require a significant amount of time and expertise, making them a bottleneck for efficient ML.

DTs can generate realistic, simulated network data, enabling the automated processing and cleaning of data through AutoML. By automating these tasks, organizations can reduce the time and resources required for data preparation, freeing them up for more complex tasks.

\subsubsection{Model Selection Challenges}
The selection of an appropriate ML model is also critical for the success of network automation. With numerous ML models available, selecting the best one for a specific task is not a trivial task. Model selection entails evaluating, comparing, and choosing the best model based on performance, complexity, and other factors, which can be time-consuming and require specialized expertise \cite{autoML}.

DTs can help address model selection challenges by providing an accurate representation of the network, enabling the comparison and evaluation of different models based on their performance. Furthermore, AutoML can automatically identify the best model for a specific network automation task, based on the characteristics of the dataset and the desired performance metrics.

\subsubsection{Model Training Challenges}
Training an ML model on a large amount of data is a time-consuming task that requires significant resources, including processing power and memory \cite{MLCH1,MLCH2}. Model training involves data augmentation, hyperparameter tuning, and regularization. Data augmentation generates additional training data to supplement the existing data, while hyperparameter tuning optimizes the parameters of the model to achieve the best performance. Regularization adds constraints to the model to prevent overfitting.

DTs can provide a simulated network environment that can generate additional data to supplement real-world data, reducing the burden on real-world resources. AutoML can also help organizations optimize the training process by automatically selecting the most efficient and effective models and hyperparameters for a given task.

\subsubsection{Dynamic Nature of Wireless Networks}
The uncertainty and dynamic nature of wireless networks represent significant challenges in the context of network automation. Wireless networks are subject to a range of factors that can impact network performance, including interference, noise, and mobility of devices. The ever-changing nature of these networks makes it difficult to build accurate and reliable models that can effectively control and predict network behavior \cite{survey5GP,zsmsurvey}. Specifically, unpredictable changes in data streams can impair the performance of ML models. This challenge can be particularly problematic in today's fast-paced environment, where rapid response to network issues is critical.

To overcome these challenges, DTs can be employed to simulate changes in the wireless network and generate new training data to retrain ML models. With AutoML, a new model can be trained and selected automatically to replace the outdated one, without the need for manual intervention. This approach ensures that the ML model remains effective and up-to-date, despite the non-stationary environment.

\subsection{Digital Twins}

As industries continue their digital transformation, the deployment of cutting-edge technologies is paramount. A high-performance network connection that harnesses state-of-the-art networking technologies is critical to facilitating the transfer of data from physical systems to cloud-hosted databases for data analytics and the deployment of AI algorithms. This connection also links physical systems to web or mobile interfaces, allowing users to monitor and control physical systems in real-time remotely \cite{DT1}. This ground-breaking deployment is known as the Digital Twin, a technology trend that has been identified as a top strategic initiative by Gartner in 2017 \cite{DTgartner,DT7}.

A DT entails creating a real-time digital replica of a physical system that synchronizes with its physical counterpart through bidirectional data and control information flows \cite{DT2}. It is more sophisticated and capable than a surveillance system or a simple model, and unlike simulation, it represents an actual asset with as few assumptions or simplifications as possible \cite{DT3}. DTs focus on maintaining the full history and up-to-date information of the assets or systems to facilitate intelligent and data-driven decision-making.

The roots of the DT concept can be traced back to the 1960s when NASA first pioneered "twining" for their Apollo program \cite{DT1960, DT1}. By creating physical duplicates on earth that mirrored their systems in space, they were able to simulate a variety of scenarios and test different conditions to analyze performance and behavior. The idea gained even more traction when it proved vital in resolving technical problems during the infamous Apollo 13 mission. Fast-forward to the early 2000s, Michael Grieves introduced the concept of DTs for the manufacturing industry, creating virtual replicas of factories \cite{DT5M1,DT6M2}. These DTs serve as an impeccable tool for monitoring processes, predicting failures, and increasing productivity, forever changing the landscape of industrial innovation.

The adoption of DTs represents a remarkable leap forward for industries seeking to thrive in the digital era. Real-time monitoring, control, and data acquisition are just some of the advantages that DTs bring to the table \cite{DT1, DT9}. These sophisticated tools enable remote access, ensuring business continuity and increasing overall efficiency. With DTs, decision-making is based on highly-informed predictions that consider both the present and the future, and the risks associated with each course of action can be assessed and mitigated in real-time. By testing and optimizing solutions in a virtual environment, DTs increase overall system efficiency and minimize potential disruptions. A DT allows virtual testing of various solutions to perform what-if analysis to evaluate these solutions without affecting the physical system \cite{DT4}. In addition, all data is easily accessible in one platform, allowing for faster and more efficient business decisions by data analytics tools.

\begin{figure}[H]
\centering
\includegraphics[width=8cm]{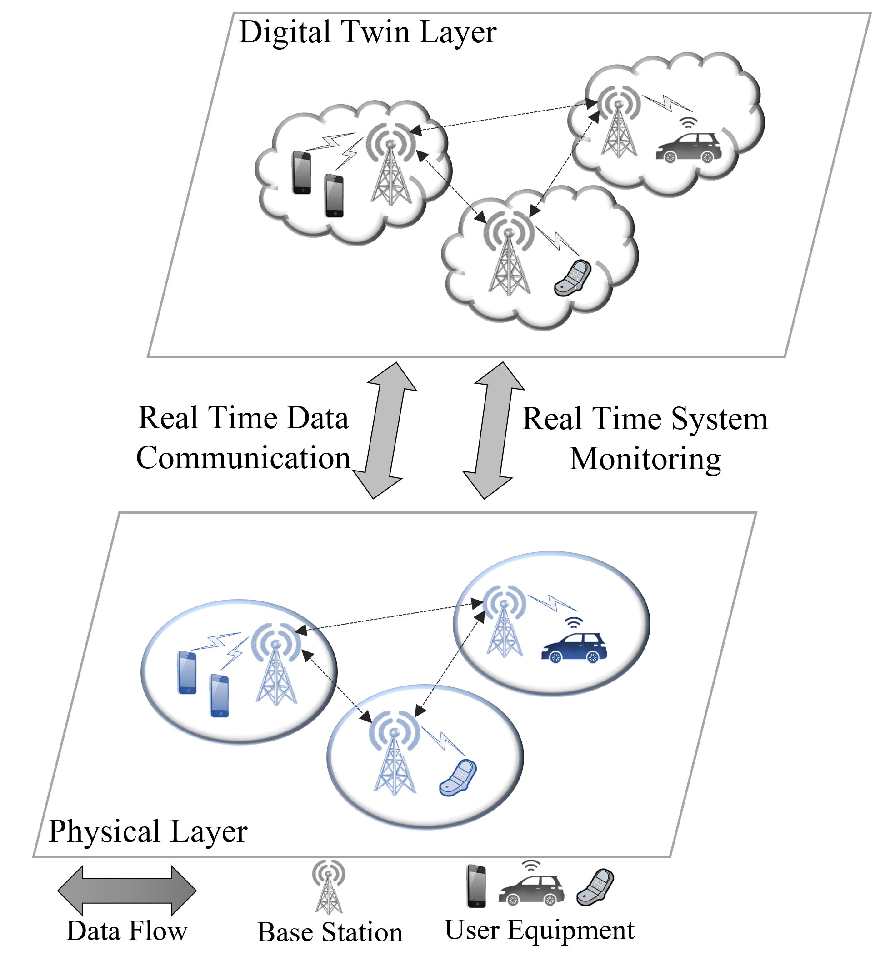}
\caption[]{Sample Architecture of a DT network for a MEC Network}
\label{fig:digitaltwin}
\end{figure}

A DT network is composed of three pillars, namely physical, digital/virtual, and connection pillars, as illustrated in Figure \ref{fig:digitaltwin} \cite{DT3}. The physical pillar represents the physical asset, the virtual/digital pillar represents the DT, and the connection pillar allows for the exchange of data and control commands among them. The system's modularity enables the system to evolve as the technology on each component evolves. A DT can be highly modular, which allows for the rapid reproduction of processes and knowledge transfer \cite{DT3,DT4}. In addition, the modularity of a DT allows for creating hybrid simulation and prototyping systems, which can accelerate the design process.

Key enablers for DT networks include reliable and high-performance network connections, cloud computing, big data analytics, AI/ML, and IoT \cite{DT8enabler1}. The integration of these technologies allows for the seamless exchange of data and control information between physical and virtual systems, providing a platform for experimentation, monitoring, and optimization of physical systems.

One of the main enablers of DT networks is the availability of high-performance, reliable network connections \cite{DT2, DT1}. DT networks require a lot of data to be transmitted in real-time between physical and digital systems. This data needs to be accurate and synchronized in both directions so that the DT can provide an accurate representation of the physical system at all times. Thus, it is critical to have a network infrastructure that is capable of handling large amounts of data and providing fast and reliable connections.

Another important enabler is the use of cloud-based technologies for data storage, processing, and analysis \cite{DT2}. With the vast amounts of data generated by DTs, it is often not feasible to store or process it all locally. Cloud-based technologies allow for scalable and flexible storage and processing, enabling the analysis of large datasets using advanced analytics techniques such as ML.

Additionally, the development and deployment of DTs require the use of a range of other technologies, including sensors, edge computing, and AI \cite{DT2}. Sensors are used to capture real-time data from physical systems. These sensors collect a wide range of data, such as temperature, pressure, vibration, and more, enabling accurate and comprehensive modeling of the physical system \cite{DT1}. Edge computing allows for data processing and analysis to take place closer to the source of the data. Additionally, ML algorithms enable the DT network to learn and adapt to new operating conditions and provide automated and proactive responses to system events.

DTs can be used in various applications, including manufacturing, transportation, and healthcare \cite{DT1}. For instance, in manufacturing, DT networks can be used to simulate the behavior of a production line, predict machine failure, and optimize production efficiency. In transportation, DT networks can be used to simulate traffic patterns, optimize traffic flow, and predict road accidents. In healthcare, DT networks can be used to simulate the behavior of the human body, predict disease progression, and optimize treatment plans.

Nevertheless, the high demand for throughput, reliability, resilience, and low latency required by DT technology goes beyond what is currently offered by 5G \cite{DT3}. Although DT technology already exists in some industrial applications supported by 5G or even 4G, it has not been widely adopted in other sectors, and has not reached its full potential. Therefore, 6G can be considered an enabler for the massive adoption of DTs, particularly in high-connectivity-demanding and rapidly emerging applications of aerospace, Industry 4.0, and healthcare.

\subsection{Automated Machine Learning} \label{subsec:automl}
AutoML, the cutting-edge technology that automates the laborious and intricate process of building and deploying ML models, is a true marvel of modern AI. AutoML has become increasingly popular in recent years due to the growing demand for ML applications and the scarcity of data science and ML expertise.

One of the key benefits of AutoML is that it can greatly reduce the time and resources required to develop and deploy ML models, as it automates time-consuming and resource-intensive tasks \cite{autoMLtoDate}. These tasks include everything from data preparation and feature engineering to hyperparameter tuning and model selection \cite{autoMLsurvey}. With AutoML, these processes can be performed in a fraction of the time it would take to do them manually, freeing up valuable time and resources for more complex and strategic work.

AutoML is not just a productivity tool, but a solution that democratizes access to the power of ML. It makes the process of building and deploying ML models accessible to a broader audience, without requiring extensive expertise in the field. This capability has the potential to revolutionize the way businesses operate, giving them the ability to leverage data-driven insights for decision-making and product development.

\begin{figure*}[ht]
\centering
\includegraphics[width=\textwidth,height=\textheight,keepaspectratio]{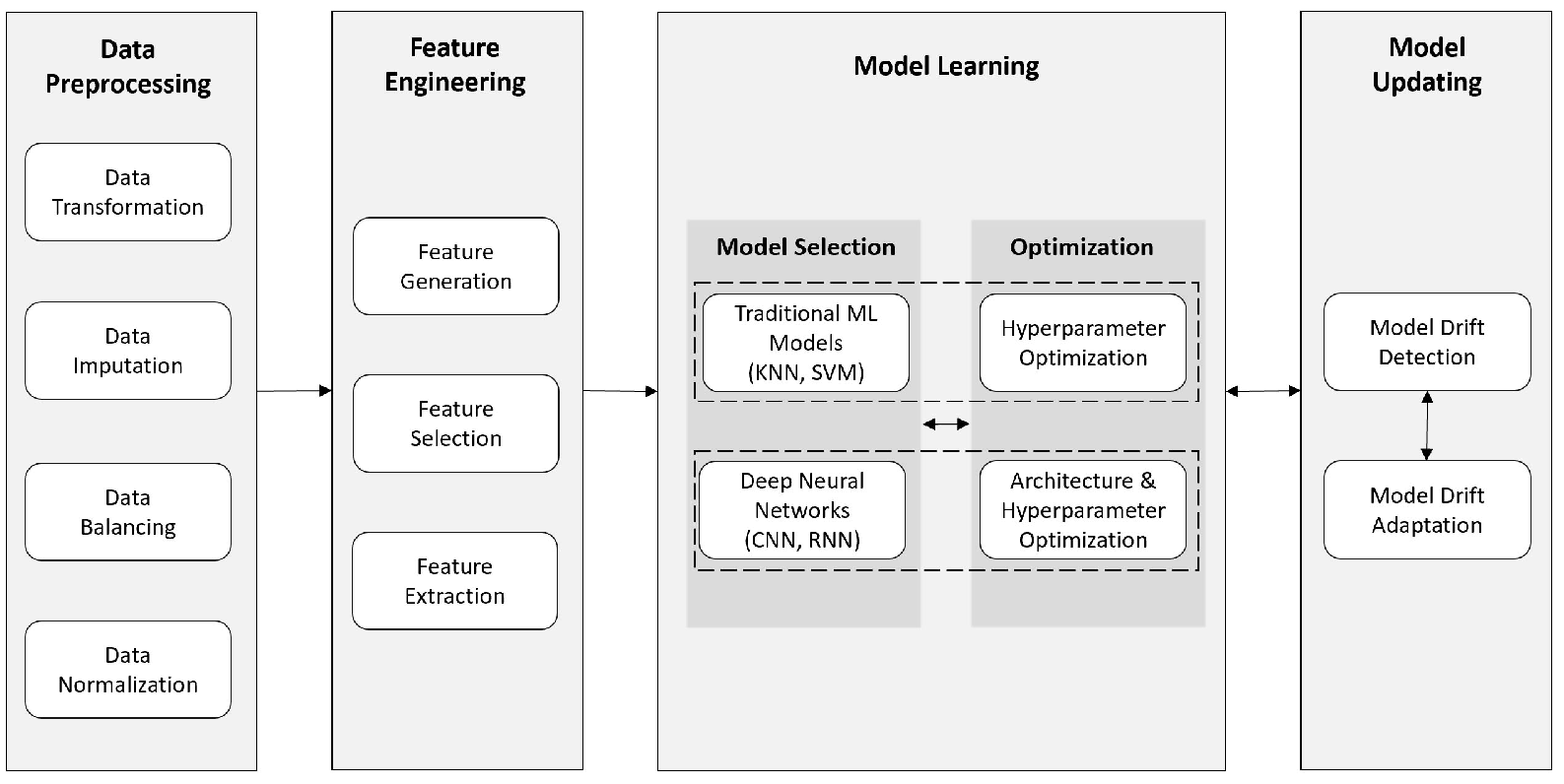}
\caption[]{An Overview of the AutoML Pipeline}
\label{fig:automlpipeline}
\end{figure*}

At its core, AutoML is a comprehensive and integrated solution that covers the entire ML pipeline, from data preprocessing to model updating, as illustrated in Figure \ref{fig:automlpipeline} \cite{autoML}. This pipeline operates on search and optimization algorithms to identify the optimal model and corresponding hyperparameters for a given problem.

\subsubsection{Automated Data Preprocessing}
Data preprocessing is a critical step in AutoML as it directly affects the performance of the model. The main goal of the data preprocessing phase is to transform raw data into a format that is suitable for training a ML model. This includes data transformation, imputation, balancing, and normalization \cite{autoML}.
\begin{enumerate}[label=(\roman*)]
\item Data Transformation: Data transformation involves converting between numerical and categorical features. In real-world applications, data is often represented as strings, requiring encoding to make it machine-readable. Techniques such as label encoding and one-hot encoding are used to assign values or columns to categorical features, but lack meaningful information. Target encoding, on the other hand, replaces categorical values with meaningful values, such as the median or mean of that variable, creating better features for ML models \cite{autoMLwave}.

\item Data Imputation: Missing data is a ubiquitous problem in ML, and is often encountered in real-world datasets due to factors such as data collection issues, measurement errors, or intentional missingness. Missing data can pose significant challenges to ML tasks, as many learning algorithms require complete and consistent data to achieve optimal performance \cite{dataImputation1}. 

While it may be tempting to simply drop any observations or features with missing data, this approach can result in significant data loss \cite{autoML}. Instead, imputation techniques, whether model-free or model-based, can be used to replace missing values with reasonable estimates.

Model-free imputation techniques are relatively straightforward and do not require extensive computation, making them a popular choice. This category includes both basic (\textit{e.g.,} zero, mean, and median imputation) and advanced (\textit{e.g.,} backward/forward filling) methods \cite{autoML}. On the other hand, model-based imputation methods utilize ML models to estimate missing values by learning from existing feature values. Examples of such techniques include K-Nearest Neighbors, XGBoost, linear regression, and Datawig (\textit{i.e.,} a DL-based approach) \cite{dataImputation2,datawig}. These methods typically yield more accurate results than model-free techniques, but they require more computational resources and time to train the ML models.

\item Data Balancing: Maintaining balanced data is a crucial aspect of ML, especially with the increase in size and complexity of datasets. The occurrence of class imbalance in datasets is marked by highly skewed class distributions, leading to the degradation of ML models. Learning on imbalanced datasets can lead to unwarranted bias in the majority classes, negatively impacting the prediction accuracy of minority classes. Hence, resampling techniques that involve either reducing the number of samples in majority classes (under-sampling) or increasing the number of samples in minority classes (over-sampling) are essential to address this class imbalance issue \cite{autoMLwave, idsiov}.

\item Data Normalization: Data normalization is a process that scales features to make them comparable and ensure that no feature is given more weight than another. When features have vastly different scales, some may overshadow others, and the model may be skewed towards the more significant ones. Two of the most commonly used normalization techniques in ML are z-score and min-max normalization \cite{dataNormalization, autoMLwave}. Z-score normalization transforms the data to have a mean of zero and a standard deviation of one. Min-Max normalization scales the data by mapping the range of values in the feature to the range between 0 and 1 \cite{unsuperAnomalyDetect}.
\end{enumerate}

\subsubsection{Automated Feature Engineering}
It is widely recognized that the quality of data and the selection of relevant features play a fundamental role in achieving optimal performance in ML applications. Hence, feature engineering has emerged as a critical component of the ML pipeline, aimed at maximizing the extraction of informative features from raw data for use by ML algorithms and models. This crucial process can be decomposed into three core sub-topics: feature generation, feature selection, feature extraction, and feature construction \cite{autoML}.

Feature generation is the art of creating new features from existing ones to enhance the accuracy and robustness of a ML model. Some feature generation methods include unary (\textit{e.g.,} an exponential transformation of a feature), binary (\textit{e.g.,} multiplying two features to create a new feature), and high-order operations (\textit{e.g.,} an average of a group of records) \cite{autoML}.

Feature generation can spawn numerous features, yet some may be irrelevant or redundant, bearing minimal or negative effects on prediction. Feature selection helps identify the best-suited features to improve model performance and training speed. Selection methods can be classified into filter, wrapper, and embedded categories \cite{autoMLsurvey}. Filter methods assign a score to each feature and select a subset based on a threshold, wrapper methods make predictions based on a selected subset of features and evaluate feature sets, and embedded methods include the selection process in the learning process of ML models.

 Feature extraction (\textit{e.g.,} PCA) employs mapping functions to reduce dimensionality \cite{autoMLsurvey}. It alters the original features to extract more informative features that can replace the original features. Although not mandatory in the feature engineering process, feature extraction can be useful when the produced feature set is high dimensional or underperforming. 
 
As such, automated feature engineering can be seen as a dynamic and synergistic combination of these three processes to optimize the feature set for ML models.

\subsubsection{Automated Model Learning}
In this AutoML process, various search and optimization algorithms are used to identify the best ML for a given dataset. These algorithms implement a range of search and optimization techniques to scour through different model architectures and parameters in search of the best fit for the data. A summary of some of the most commonly used optimization techniques, along with their respective advantages and disadvantages, are provided in Table \ref{table:optMethods} \cite{autoMLtimeseries,autoMLwave,hyperopt_li,bayOpt,gradientOpt,hyperband}.

\begin{table*}
\centering
\caption{An Overview of Common Optimization Techniques \cite{autoMLtimeseries,autoMLwave,bayOpt,gradientOpt,autoMLtimeseries,hyperband,hyperopt_li}}
\def\arraystretch{2}
\scriptsize
\begin{tabular}{|P{1.85cm} P{3.3cm} P{3.1cm} P{3.4cm}|}\hline
\textbf{Optimization Algorithm} & \textbf{Description} & \textbf{Advantages} & \textbf{Limitations} \\ \hline

\textbf{Grid Search} & \RaggedRight Exhaustive search over a defined hyperparameter space & \begin{tabitemize}
    \item Simple and easy to implement
\end{tabitemize} & \begin{tabitemize}
    \item Computationally expensive
    \item Not ideal for high-dimensional search spaces
\end{tabitemize} \\ \hline

\textbf{Random Search} & \RaggedRight Random sampling of hyperparameters from a defined search space & \begin{tabitemize} \item More efficient than grid search for high-dimensional search spaces \end{tabitemize} & \begin{tabitemize}
    \item Can still be computationally expensive for large search spaces
    \item Can not make use of previous observations
\end{tabitemize} \\ \hline

\textbf{Bayesian Optimization} & \RaggedRight Sequential model-based optimization using Bayesian inference & \begin{tabitemize} \item Fast convergence to the optimal solution \item Make use of prior observations \item \end{tabitemize} & \begin{tabitemize} 
    \item Limited capacity for parallelization
\end{tabitemize} \\ \hline

\textbf{Gradient-Based Optimization} & \RaggedRight Gradient descent-based optimization using first-order derivative information &  \begin{tabitemize}
    \item Efficient for optimizing differentiable objective functions
\end{tabitemize} & \begin{tabitemize}  
    \item Can get stuck in local optima
    \item Not ideal for non-differentiable objective functions
\end{tabitemize} \\ \hline

\textbf{Evolutionary Algorithms} & \RaggedRight Stochastic optimization inspired by biological evolution &  \begin{tabitemize}
    \item Scale well to higher dimensional problems
    \item Can handle both continuous and discrete variables
\end{tabitemize} & \begin{tabitemize}  
    \item Depend heavily on the selection of algorithmic parameters
    \item Susceptible to premature convergence
\end{tabitemize} \\ \hline

\textbf{Hyperband} & \RaggedRight Sequential halving algorithm that adapts to performance of trials and allocates resources accordingly & \begin{tabitemize}
    \item Efficient use of resources
\end{tabitemize} & \begin{tabitemize}
    \item Need subsets with small budgets to be representative
\end{tabitemize} \\ \hline

\end{tabular}
\label{table:optMethods}
\end{table*}

Grid search, for example, involves exhaustively searching through a pre-defined set of hyperparameters, while random search randomly selects a set of hyperparameters for evaluation \cite{autoMLtimeseries,autoMLwave,hyperopt_li}. Bayesian optimization uses probabilistic models to predict the performance of different hyperparameter configurations and selects the next configuration to evaluate based on this prediction \cite{bayOpt}. Gradient-based optimization methods, such as gradient descent and its variants, use the gradient of the loss function with respect to the hyperparameters to iteratively update the values of the hyperparameters until convergence \cite{gradientOpt}. Evolutionary algorithms, such as genetic algorithms, use evolutionary principles, such as mutation and crossover, to search for optimal hyperparameters \cite{autoMLtimeseries}. Hyperband is a type of early stopping algorithm that uses a successive halving approach to quickly identify the most promising hyperparameter configurations \cite{hyperband}. 

AutoML tools also evaluate multiple models across a range of algorithm families, including linear models, decision trees, SVMs, and neural networks. These models differ in their underlying assumptions, which can make some models better suited to certain types of data or problems.

Once the ideal model has been identified, AutoML tools employ optimization algorithms to tune its hyperparameters. Hyperparameters, such as the learning rate, regularization strength, number of hidden layers, and dropout rate, are essential configuration settings that influence the performance of the model but are not learned from the data \cite{autoML, autoMLtimeseries}. AutoML tools use a range of search and optimization techniques to search for the most effective combination of hyperparameters for the model. \cite{autoMLToApplication}

With the model architecture and hyperparameters optimized, AutoML tools initiate the training process, whereby the model is trained on the provided dataset. This involves utilizing the optimization algorithm to minimize the objective function, which typically measures the difference between the predicted output and the true output. Advanced optimization techniques, such as stochastic gradient descent and Adam, are utilized to speed up the training process and improve the accuracy of the model \cite{autoML,autoMLsurvey}. Additionally, regularization techniques such as L1 and L2 regularization are used to prevent overfitting, and ensemble methods such as bagging and boosting can be employed to improve model robustness and performance \cite{autoMLsurvey}.

The resulting trained model is then evaluated on a validation dataset to estimate its generalization performance. Evaluation metrics, such as accuracy, precision, recall, and F1 score, are computed to compare the performance of different models and hyperparameters and identify the best-performing model \cite{autoMLsurvey}. The best model is then selected based on its ability to generalize well to unseen data and meet the desired performance criteria.

\subsubsection{Automated Model Updating}
After the AutoML pipeline completes the training process, the final step is to deploy the trained model into production. This entails packaging the model as a software component, like a web service or an API, and integrating it into a larger system \cite{autoMLsurvey}. AutoML tools facilitate this deployment process by providing automated model versioning, monitoring, and maintenance, ensuring that the model remains accurate and current over time \cite{autoML}. However, one of the challenges that may arise is model drift, which can be further categorized into two types: data drift and concept drift. Data drift occurs when the underlying data distribution shifts over time, rendering the existing model obsolete. Concept drift occurs when the task that the model was designed to perform changes over time \cite{drift2}. To overcome this challenge, two approaches are commonly utilized: model drift detection and adaptation.

Model drift detection involves analyzing the statistical properties of the data and identifying any changes over time. This is done through two primary methods: distribution-based and performance-based methods \cite{conceptDrift2}.

Distribution-based methods assume that changes in the data distribution indicate model drift \cite{conceptDrift2}. For instance, the Kolmogorov-Smirnov compares the empirical distribution of a feature or target variable over time with a reference distribution, such as the distribution at the start of the data stream \cite{KStest}. A significant change in the distribution suggests the presence of model drift.

Performance-based methods, on the other hand, assume that model drift leads to a decline in model performance. Statistical tests, window-based approaches, and ensemble-based approaches are common techniques for detecting changes in model performance \cite{performanceBasedConceptDrift, autoML}. Statistical tests (\textit{e.g.,} Friedman test) compare the current model's performance with a reference dataset. Window-based approaches track model performance over a rolling window of data and compare it to the previous window. Ensemble-based approaches compare the current model's performance with an ensemble of previous models trained on different time intervals.

Once model drift is detected, the AutoML pipeline can adapt the model to the new data distribution. This is done through a process called model adaptation, which involves updating the model to reflect the changes in the data \cite{conceptDrift1,drift3}. There are several techniques for model adaptation, including:
\begin{itemize}
    \item Updating the model incrementally using small data batches, instead of retraining on the entire dataset at once \cite{conceptDrift1, autoML, drift1}
    \item Retraining the model on the new data to ensure that it remains accurate and effective \cite{autoML}
    \item Transferring knowledge from the old model to a new model trained on the new data, thereby minimizing the amount of training required \cite{TLconceptdrift}
    \item Creating an ensemble of models trained on different time intervals and combining their predictions to achieve better performance on the new data, such as the weighted probability averaging ensemble framework proposed by Yang \emph{et al.} \cite{drift4}
\end{itemize}

\section{Case Study: Online AutoML for Application Throughput Prediction}\label{sec:casestudy}
In this case study, we explore the application of online AutoML for predicting application throughput in dynamic network environments. Through a comprehensive analysis, we highlight the advantages of using AutoML in this context. The subsections that follow provide an overview of the use case, dataset, AutoML framework, 5G system architecture, and present insightful results and analysis. These include a comparison with traditional ML approaches, a complexity-accuracy trade-off analysis, and periodic AutoML model drift monitoring. This study demonstrates the potential of AutoML in optimizing application throughput prediction for dynamic networks.

\subsection{Use Case Overview}
The emergence of 5G+ networks has ushered in a new era of opportunities and challenges for the telecommunications industry. The explosive growth of mobile data traffic and the increasing demand for high-speed and low-latency services have made optimizing network resources and handling traffic loads more critical than ever. Furthermore, the emergence of new use cases, such as IoT applications, has resulted in a massive increase in connected devices, making reliable and efficient network performance essential to support the data traffic generated by these devices \cite{iottrend}.

5G+ networks are designed to cater to a wide range of applications, including high-definition video streaming, virtual and augmented reality, and real-time gaming, each with varying requirements. Network operators must meet a diverse set of KPIs for different use cases, including eMBB, mMTC, and uRLLC, necessitating a thorough understanding of network performance and user behavior \cite{kpisperf}. Therefore, predicting application throughput is crucial to ensure that these applications deliver high-quality service to end-users.

Application throughput is a crucial metric used to evaluate the performance of applications running over a network. It reflects the amount of data that can be downloaded per unit of time at the application layer, which directly impacts the user experience.

Throughput is influenced not only by network infrastructure but also by user traffic. Network congestion can occur, leading to reduced application throughput, as more users connect to the network and start using data-intensive applications. Predicting application throughput can help network operators optimize network resources, identify potential network bottlenecks, and troubleshoot network issues before they affect end-users. Accurately predicting application throughput can also improve user experience by ensuring that applications operate efficiently, delivering fast and reliable service to end-users. Therefore, predicting application throughput is crucial for ensuring network performance, enhancing user experience, and meeting business objectives.

With the deployment of new technologies such as network slicing and edge computing in 5G+ networks, predicting application throughput becomes even more critical. Network slicing allows operators to create customized virtual networks to support specific services and applications, and predicting application throughput is essential to ensure that the resources allocated to each slice are sufficient to meet the performance requirements of the applications running on that slice. Edge computing reduces latency and improves application performance, and predicting application throughput in such an environment can help operators decide where to deploy their computing resources to achieve optimal performance.

Real-time monitoring and adjustments to network resources are necessary to meet a diverse set of KPIs. ZSM is one promising solution for automating the process of predicting application throughput and scaling network resources accordingly to ensure a seamless user experience. ZSM proactively detects and resolves potential network issues before they impact service quality by predicting and optimizing network performance metrics such as download rate at the application layer. By leveraging AutoML algorithms to generate up-to-date predictive models, ZSM can autonomously adapt to changes in traffic patterns and automate the optimization and management of network services, resulting in improved service quality and increased operational efficiency. This is crucial in a dynamic and fast-changing environment like 5G+ networks.

In this study, we utilize an open-source production dataset to predict application throughput autonomously using AutoML. To demonstrate the effectiveness of AutoML in generating up-to-date predictive models, we will simulate a real-world scenario where the application throughput experiences a sudden and significant change due to traffic congestion. This scenario is inspired by the increasing usage of clouds to upload and access one's files. Model drift is not restricted to changes in user behavior. It can also occur due to changes in the application infrastructure. For instance, deploying additional servers to handle increased traffic will result in an increase in application throughput, which in turn can lead to a decrease in the model's predictive accuracy. In order to mitigate the effects of model drift, the model is monitored and adjusted based on the new data that reflects the current state of the application infrastructure.

Overall, our use of AutoML and the open-source production dataset will allow us to accurately predict application throughput and respond to changes in network conditions in real-time. By continuously monitoring the network and training the model on the latest data, we can ensure that our predictive model stays up-to-date to accurately reflect the current state of the network, even in the face of rapidly changing conditions.

\subsection{Dataset Overview}
In this study, we employ a publicly available production dataset collected from a major mobile operator in Ireland \cite{dataset}. The dataset comprises client-side cellular KPIs obtained from G-NetTrack Pro, a widely used Android network monitoring tool. The data were generated from two mobility patterns, stationary and in-motion, across two distinct application patterns, on-demand video streaming and file downloading (of file size $>$ 200 MBs), with a total duration of 3142 minutes. The video streaming dataset provides a direct measurement of popular over-the-top services, namely Netflix and Amazon Prime, which are representative of the typical user behavior when watching on a mobile device. 

The dataset captures traces pertaining to both 4G and 5G networks, encompassing a range of channel-, context-, and cell-related metrics, in addition to throughput information. Table \ref{table:datasetFeatures} provides a depiction of the manifold metrics contained within the dataset, thereby enriching our understanding of the network's performance \cite{dataset, Gyokov2021}.

\newcommand{\mysubscript}[1]{\raisebox{-0.34ex}{\scriptsize#1}}
\begin{table*}
\centering
\caption{Production Dataset Metrics \cite{dataset, Gyokov2021}}
\scriptsize
\def\arraystretch{2}
\begin{tabular}{|P{4cm}  P{8.34cm}|}\hline
\textbf{Feature} & \textbf{Description} \\ \hline

\textit{Timestamp} & Timestamp of sample  \\ \hline

\multirowcell{2}{\textit{Longitude} \\[0.7em] \textit{Latitude}} & \multirow{2}{*}{GPS coordinates of the device} \\ & \\ \hline

\textit{Speed} & Speed of mobile device $(km/h)$  \\ \hline

\textit{Operator Name} & Anonymized cellular operator name  \\ \hline

\textit{Network Mode} & Mobile communication standard  \\ \hline

\textit{Node\mysubscript{hex}} & Radio network controller ID in hexadecimal  \\ \hline

\textit{LAC\mysubscript{hex}} & Location area code in hexadecimal \\ \hline

\textit{State} & State of the download process ('I' for idle \& 'D' for downloading) \\ \hline

\multirowcell{2}{\textit{DL\textunderscore bitrate} \\[0.7em] \textit{UL\textunderscore bitrate}} & \multirowcell{2}{Download/Upload rate measured at the device \\ (application layer) $(kbps)$} \\ & \\ \hline

\textit{CellID, CellID\mysubscript{hex}, CellID\mysubscript{raw}} & ID of serving cell for mobile along with its hexadecimal and raw formats  \\ \hline

\multirowcell{2}{\textit{Ping\mysubscript{avg}, Ping\mysubscript{min}, Ping\mysubscript{max}} \\[0.7em] \textit{Ping\mysubscript{std}, Ping\mysubscript{loss}}} & \multirowcell{2}{Ping statistics \\(average, minimum, maximum, standard deviation, and loss,\\ respectively)} \\ & \\ \hline

\textit{Channel Quality Indicator} & Feedback provided by the UE to the base station \\ \hline

\textit{Signal-to-Noise Ratio} & Difference between the received wireless signal and the noise floor $(dB)$ \\ \hline

\textit{Received Signal Strength Indicator (RSSI)} & Measurement of the power present in a received radio signal $(dBm)$ \\ \hline

\multirowcell{2}{\textit{Reference Signal Received} \\[0.7em] \textit{Power (RSRP)}}
&  \multirowcell{2}{Linear average of power for resource elements carrying cell-\\specific reference signals $(dBm)$} \\ &  \\ \hline

\textit{Reference Signal Received Quality (RSRQ)} & Ratio between RSRP and RSSI $(dB)$ \\ \hline

\multirowcell{2}{\textit{NRxRSRP} \\[0.7em] \textit{NRxRSRQ}} & \multirow{2}{*}{RSRP and RSRQ values for the neighboring cell} \\ & \\ \hline

\end{tabular}
\label{table:datasetFeatures}
\end{table*}

We leverage this comprehensive dataset to investigate download rate prediction at the application layer, commonly referred to as application throughput, in the context of 4G and 5G networks. To this end, the dataset is partitioned into discrete 4G and 5G sets for both streaming and downloading applications, by merging both traffic conditions, stationary and in-motion, for each permutation of network mode and application. Such an approach enables us to explore the complex interplay between network mode, application type, and traffic conditions in determining the application throughput.

The goal is to provide network operators with valuable insights into application throughput prediction, leading to improved network management, resource optimization, and enhanced user experience. The decision to choose this dataset was driven by its following attributes.

\begin{itemize}
    \item Real-world Relevance: The dataset is derived from a major Irish mobile provider, containing real-world 5G network data. This real-world data ensures the relevance and practicality of the predictions made by the AutoML pipeline.
    \item Comprehensive Metrics: The dataset includes diverse cellular KPIs obtained from G-NetTrack Pro. These metrics cover various network aspects, enabling a comprehensive analysis of network performance.
    \item Application-specific Insights: The dataset focuses on two distinct application types, video streaming and file downloading. Accordingly, the AutoML pipeline can provide tailored insights and predictions relevant to these applications.
    \item Mobility Patterns: The dataset further includes two mobility patterns, static and driving, providing insights into bandwidth changes in different scenarios. This provides a more accurate model of real-world network conditions.
    \item Network Mode and Application Segmentation: The dataset is divided into different segments based on network mode (4G and 5G) and application (streaming and downloading). This enables the AutoML pipeline to capture and analyze variations in performance across different network modes and applications.
\end{itemize}

\subsection{Dataset Distribution}
\begin{figure*}[htbp]
\centering
\subfigure[Distribution of Application Throughput on Saturday, 2019-12-14]{\label{fig:142019}\includegraphics[width = 14cm, keepaspectratio]{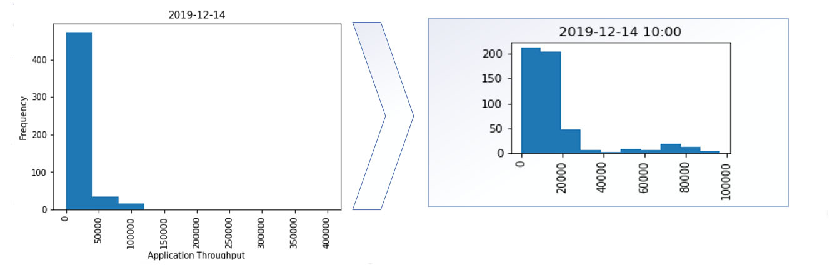}}
\hfill
\subfigure[Distribution of Application Throughput on Monday, 2019-12-16]{\label{fig:162019}\includegraphics[width = 13cm, keepaspectratio]{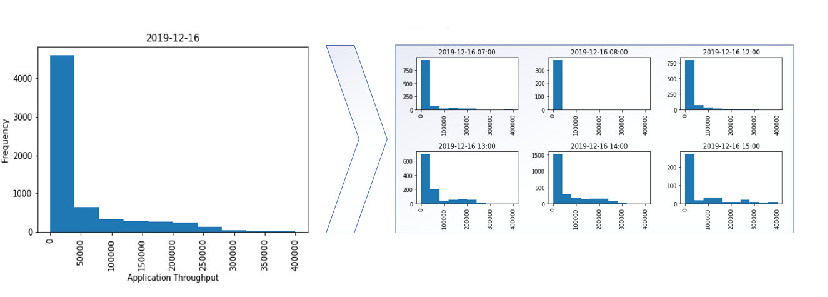}}
\hfill
\subfigure[Distribution of Application Throughput on Tuesday, 2019-12-17]{\label{fig:172019}\includegraphics[width = 13cm, keepaspectratio]{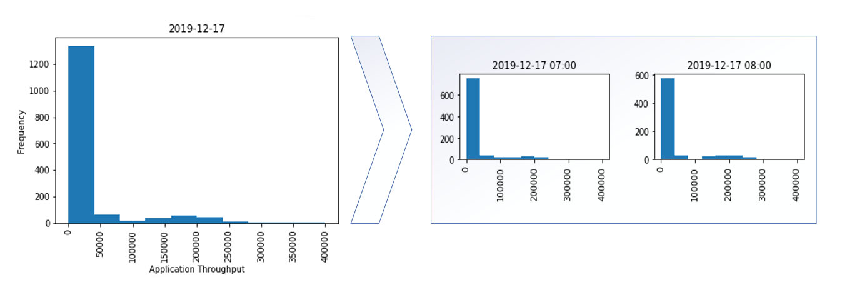}}
\hfill
\caption[]{Distribution of Application Throughput ($kbps$) for 5G Data from Downloading a File by Day and Hour in 2019}
\label{fig:5GDistDownload2019}
\end{figure*}

\begin{figure}[htbp]
\centering
\subfigure[Distribution of Application Throughput on Thursday, 2020-01-16]{\label{fig:162020}\includegraphics[width = 13cm, keepaspectratio]{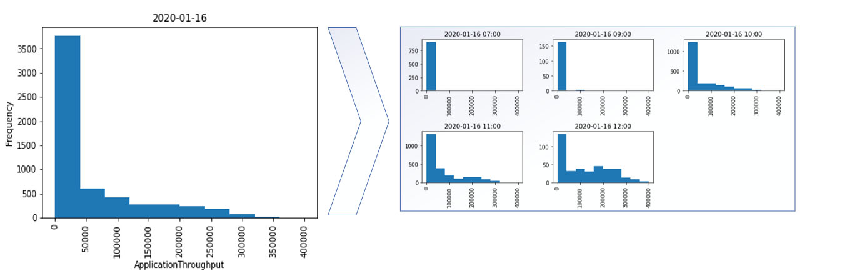}}
\hfill
\subfigure[Distribution of Application Throughput on Thursday, 2020-02-13]{\label{fig:132020}\includegraphics[width = 13cm, keepaspectratio]{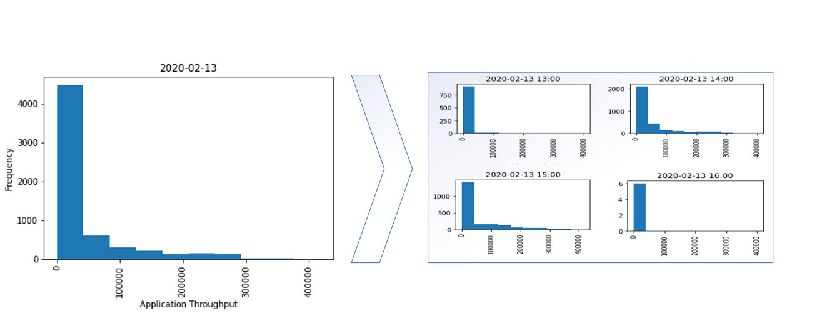}}
\caption[]{Distribution of Application Throughput ($kbps$) for 5G Data from Downloading a File by Day and Hour in 2020}
\label{fig:5GDistDownload2020}
\end{figure}

To start with, we utilize these datasets to explore the dynamic changes in the application throughput. To showcase this variability, we focus on a specific example from one of the four merged datasets, namely the 5G data pertaining to file downloads. Through this example, we examine the distribution of the data by grouping it based on the day and hour of collection. Figures \ref{fig:5GDistDownload2019} and \ref{fig:5GDistDownload2020} provide a visual representation of the distribution of the application throughput across different days and hours in 2019 and 2020, respectively. Specifically, the 5G traces were captured on three days in December 2019, including Saturday the 14\textsuperscript{th} (Figure \ref{fig:142019}), Monday the 16\textsuperscript{th} (Figure \ref{fig:162019}), and Tuesday the 17\textsuperscript{th} (Figure \ref{fig:172019}), as well as on two Thursdays in January 2020 (Figure \ref{fig:162020}) and February 2020 (Figure \ref{fig:132020}).

Upon inspecting the distribution of application throughput across different days, we observe that despite differences in sample sizes, each day exhibits a similar shape, suggesting the existence of a shared underlying structure. Taking this analysis a step further, we proceed to examine the distribution of application throughput across different hours of the day, focusing our attention on data collected on weekdays during 2019 (as illustrated in Figures \ref{fig:162019} and \ref{fig:172019}). Our examination reveals that the majority of throughput values for downloading a file in the early morning hours (specifically, at 7:00 a.m. and 8:00 a.m.) are concentrated within the first bin of the histogram. Such an observation suggests that the distribution of throughput is heavily skewed towards low values, characterized by a download rate of 50 Mbps or less. Such a phenomenon could be a result of lower network traffic or reduced user activity during this particular time window. A similar pattern is observed in Figure \ref{fig:162020} for the early morning hours, 7:00 a.m. and 9:00 a.m., of January 16, 2020.

At present, we have yet to detect any data drift, which indicates a consistent pattern of application usage given the same network operator. Nonetheless, we establish a baseline by building a model and assessing its performance on a testing set. Even in the absence of detected model drift, continuous monitoring of the model is vital to detect any potential drops in performance due to changes in user behavior and adapt accordingly.

Consider, for instance, a renowned music streaming platform that allows users to download songs and albums for offline listening. As the platform expands to incorporate fresh artists and genres, users may alter their download behavior, transitioning to larger file downloads, such as complete albums or extended playlists. This trend becomes more pronounced during the release of a highly anticipated album. Such a shift in user behavior may cause a corresponding impact on the application's throughput, resulting in longer wait times and slower download speeds for users.

This change may impair the model's ability to accurately predict the application throughput and adjust the download speed to user behavior (\textit{e.g.,} by allocating network resources accordingly). If the model is not updated to reflect the new user behavior, it may overestimate or underestimate the required throughput, leading to sub-optimal download speeds and user experience.

To evaluate the model's ability to adapt to changes in the underlying data distribution, we will simulate the aforementioned scenario by introducing model drift, specifically data drift. Through periodic evaluation on the incoming data, we will track the model's performance and determine whether it can maintain high performance levels in the presence of data drift.

\subsection{AutoML Framework}
\begin{figure*}[ht]
\centering
\includegraphics[width=\textwidth,height=\textheight,keepaspectratio]{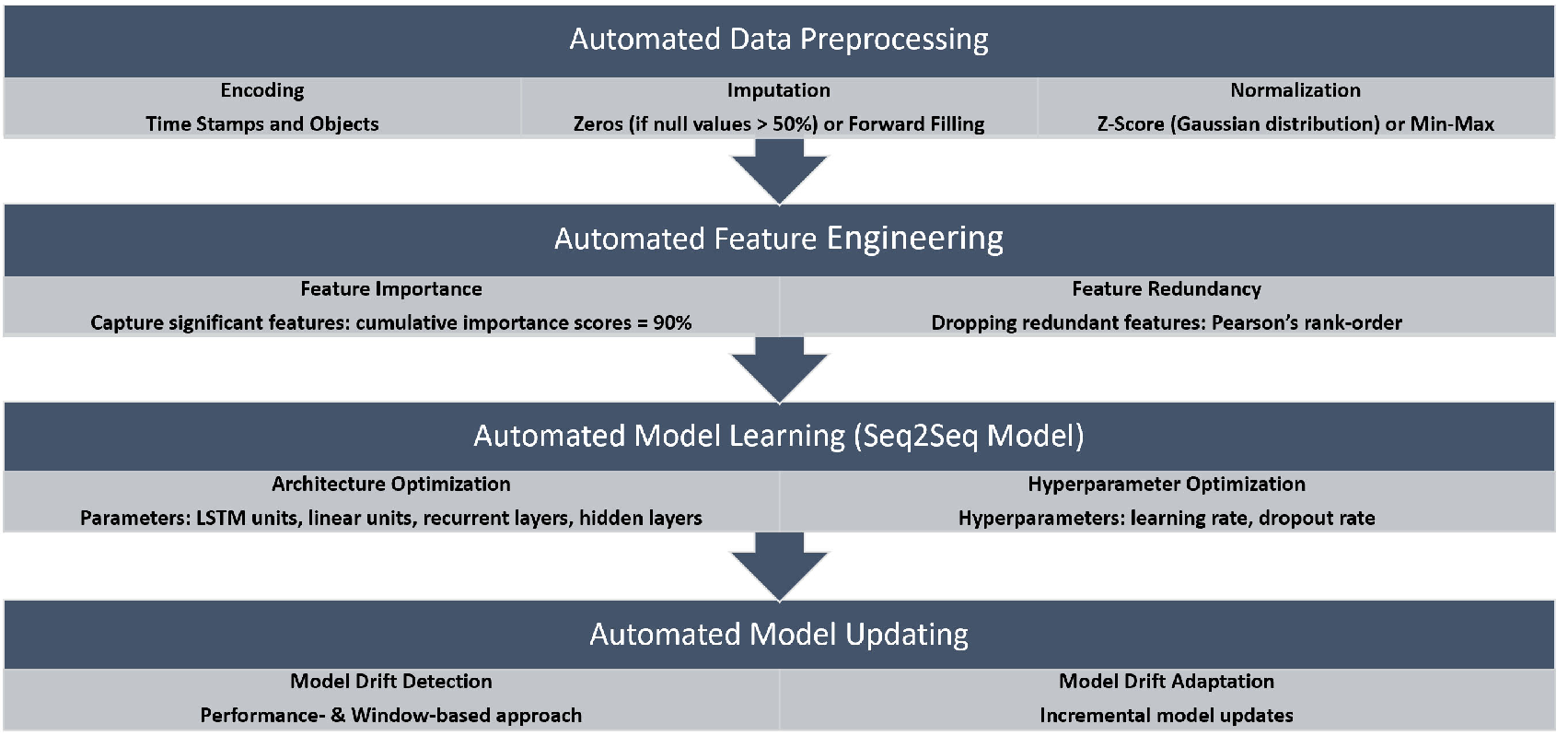}
\caption[]{Respective AutoML Framework for Case Study}
\label{fig:usecaseAutoMLFramework}
\end{figure*}
In this work, we tailored the generic AutoML framework demonstrated in Figure \ref{fig:automlpipeline} and elaborated in Section \ref{subsec:automl} to fit the nature of the case study. Figure \ref{fig:usecaseAutoMLFramework} provides a glimpse into the tailored AutoML pipeline, which commences with the data preprocessing phase. This involves encoding non-numerical values and timestamps, substituting null values with either zeros or forward-filled values, and applying standard or min-max normalization to enhance data quality. Moving onto the next step, feature engineering, insignificant features are dropped based on the cumulative importance scores obtained through the light gradient boosting machine algorithm. Subsequently, we remove redundant features based on Pearson's rank-order. These two steps comprehensively manipulate the data to extract valuable insights and prepare it for the model learning phase.

\begin{figure}[htbp]
\centering
\includegraphics[width =\linewidth, keepaspectratio]{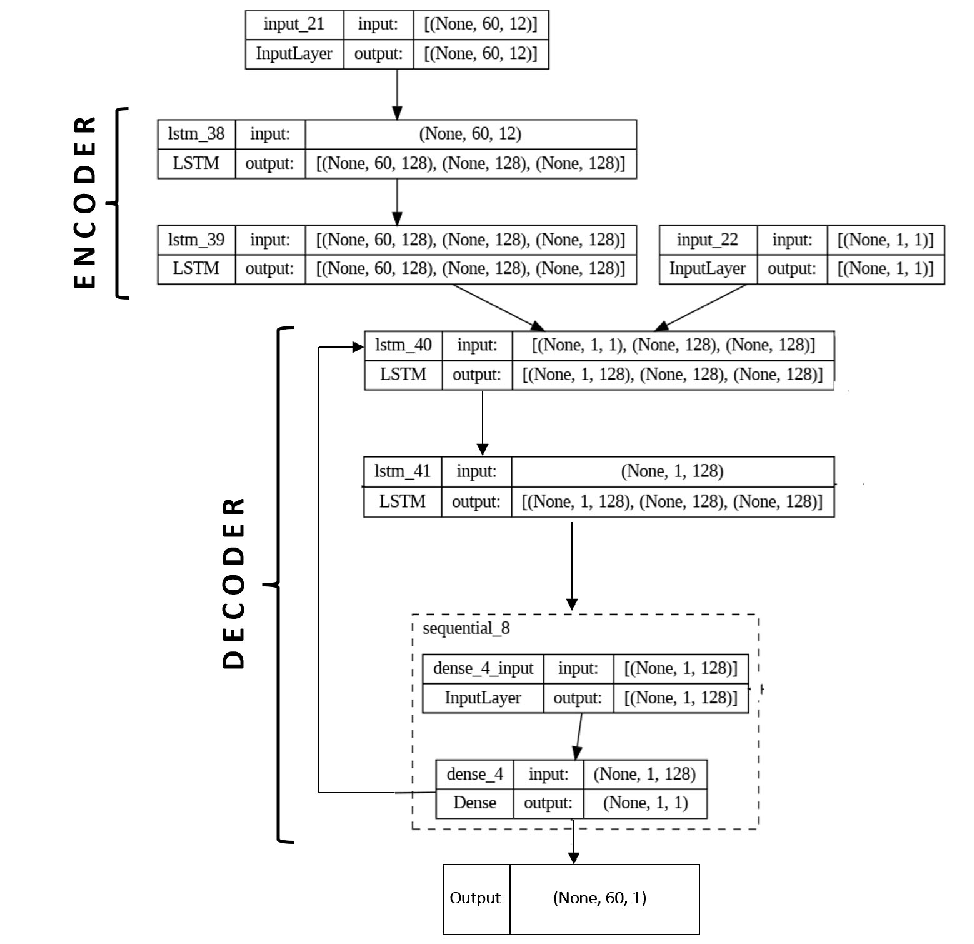}
\caption[]{Sample Encoder-Decoder Architecture}
\label{fig:seq2seq}
\end{figure}

The third step, model selection and hyperparameter tuning, involves optimizing the neural network architecture of a Sequence-to-Sequence (Seq2Seq) model. Seq2Seq models have gained widespread attention in recent years due to their ability to learn complex patterns in sequential data \cite{seq2seq}. In particular, the encoder-decoder architecture of Seq2Seq models has been shown to be effective for multi-step prediction in various applications, such as machine translation, speech recognition, and network traffic prediction. This architecture is especially suited for application throughput prediction, as it can identify the temporal dependencies and patterns in the input sequence. The encoder component of the model maps the input sequence, usually historical throughput data, into a fixed-length vector representation that captures the relevant information, also known as context vector. This vector is then passed to the decoder, which generates the output sequence, one element at a time. The decoder leverages the encoder's context vector and the previously generated output elements as context to predict the subsequent output element in the sequence. The Seq2Seq models are equipped with LSTM units in both the encoder and decoder components to further enhance the model's performance. A sample encoder-decoder architecture is illustrated in Figure \ref{fig:seq2seq}. The encoder component incorporates two LSTM layers, with the second output serving as the context vector. This context vector, along with the current time-step, is then passed to the decoder. The decoder, likewise, comprises two LSTM layers. The final output from the decoder represents the subsequent time-step, which is then fed back into the decoder to predict the succeeding time-step. This iterative process continues until the entire sequence is predicted. Both the architecture selection and hyperparameter tuning phases are conducted concurrently via Bayesian optimization. The model architecture is optimized in terms of the number of hidden dense and recurrent layers in the encoder and decoder, as well as the number of dense and LSTM units. The hyperparameters that are tuned are the learning rate of the Adam optimizer and the dropout rate.

The final step in the pipeline involves monitoring the model's performance over time to detect any potential data drift. We utilize a performance-based approach, specifically a window-based method, where we check the model's performance at regular intervals (\textit{e.g.,} every ten minutes with a window size of 600 seconds). This allows us to detect any significant changes in the data distribution, which can cause the model's performance to deteriorate. Data drift is detected if the performance falls below a dynamic threshold, which is a certain percentage of the current score. The choice of window size and threshold percentage involves a trade-off between accuracy and computational complexity. The more strict the values, the more accurate the model is, but the more computationally complex it becomes. In case of data drift, we update the model's weights incrementally to adapt to the new data distribution.

\subsection{5G System Architecture}

\begin{figure*}[ht]
\centering
\includegraphics[width=\textwidth,height=\textheight,keepaspectratio]{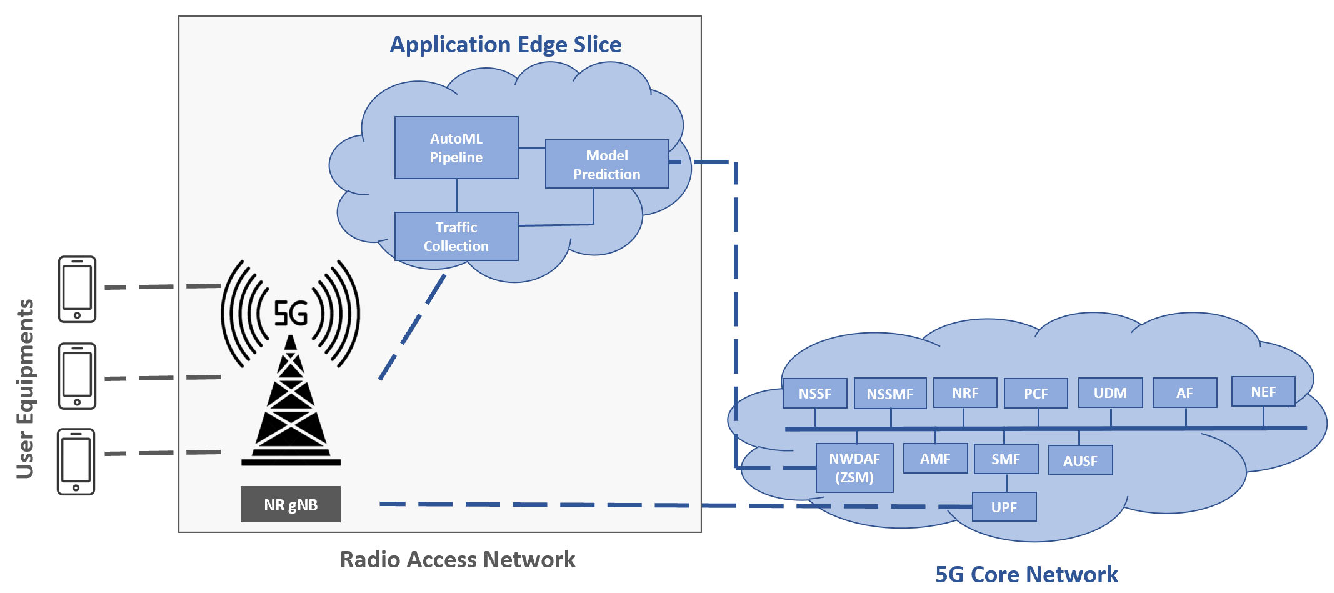}
\caption[]{5G System Architecture}
\label{fig:5gSystemArchitecture}
\end{figure*}

The E2E 5G system architecture, as depicted in Figure \ref{fig:5gSystemArchitecture}, is composed of various components that work together to provide a seamless user experience. At the heart of this architecture are the UEs, which are connected to the gNodeB (gNB) in the RAN. The gNB is also connected to the User Plane Function (UPF) in the CN. However, what sets this architecture apart is the presence of the Application Edge Slice (AES), located in the RAN, which plays an integral role in predicting application throughput for specific applications, such as Netflix.

The AES is designed to be in close proximity to the UE and has enough resources to collect and analyze data, making it ideal for analyzing traffic patterns. Within the AES, three primary entities work together to collect data and generate predictions. These entities are the traffic collection entity, the AutoML pipeline, and the model prediction entity.

The traffic collection entity is responsible for collecting data on UE traffic,  including information about the type of traffic, expected traffic volume, and QoS requirements. To achieve this, the traffic collection entity must be configured to identify relevant metrics, such as DL rates, UL rates, and coordinates. 

Once the traffic collection entity has collected the necessary data, it is sent to the AutoML pipeline. This pipeline preprocesses the data and trains a suitable ML model with periodic monitoring to ensure its accuracy.

The model prediction entity is the final piece of the puzzle in the AES, responsible for generating predictions based on the model created through the AutoML pipeline. These predictions are then sent to the Network Data Analytics Function (NWDAF), which houses the ZSM framework \cite{nwdaf}.

The ZSM in the NWDAF helps optimize resource allocation for the network slices based on the predictions made by the model in AES. Once the decision is made, it is sent to the Network Slice Selection Function (NSSF) and Network Slice Subnet Management Function (NSSMF), which select the appropriate network slice for the UE and manage the resources within the network slice, including resource allocation and orchestration.

Finally, the AMF and UPF coordinate resource allocation to guarantee that each UE receives the requested QoS for the selected network slice. In this way, the AES entities and the ZSM framework work together to collect data, generate predictions, and optimize resource allocation to ensure an optimal user experience for this specific application on the network. 

\subsection{Results and Analysis }

We implemented the proposed AutoML pipeline in Python 3.10.11 to predict the application throughput based on 4 sub-datasets categorized by network mode (4G and 5G) and application (streaming and file downloading). Specifically, the experiments were conducted on a machine equipped with a 14 Core i9-12900HK processor and 32 GB of memory. The experiments harnessed the computational power of the NVIDIA GeForce RTX 3060 GPU, featuring 6 GB of GDDR6 memory.

To evaluate the performance of our pipeline, we used Mean Absolute Error (MAE) and Mean Absolute Percentage Error (MAPE), which are commonly used metrics in network traffic prediction, to measure the accuracy of the predicted values. Accurate prediction of network traffic is crucial for optimizing network resources and improving overall network performance. MAE measures the average magnitude of the errors in the predictions, while MAPE measures the size of the errors relative to the actual values. These metrics are preferred over other metrics like root mean square error as they are less sensitive to outliers and do not penalize large errors as heavily, which can significantly impact the network's performance. Moreover, they are easy to interpret by network operators and non-experts, making them useful for validating the models.

\subsubsection{AutoML vs. Traditional ML}
To validate our proposed AutoML pipeline's efficiency, we compare its performance with a basic LSTM model and a basic Seq2Seq model without hyperparameter tuning. The LSTM model has one layer with 128 LSTM cells followed by a dense output layer whose number of neurons depends on the prediction horizon. The model is compiled with the MAE loss function and the Adam optimizer. Similarly, the Seq2Seq model has one LSTM encoder and decoder, each with 128 cells, and one dense layer. The decoder would then generate the output sequence one step at a time, taking the previous prediction as input for the next step.

Default parameters are used for both traditional ML models, and we set the forecast horizon to 60 seconds, a parameter that can be adjusted based on application requirements. Our results, presented in Table \ref{table:comparison}, demonstrate that AutoML with Seq2Seq outperforms both basic LSTM and encoder-decoder models in terms of MAE and MAPE for all four network traffic datasets. To illustrate, in the case of file downloading in a 5G network, Upgrading from the LSTM model to the encoder-decoder model reduces the loss from 6.52 \% to 4.76 \% for MAPE and from around 0.17 to 0.11 for MAE. This trend was observed consistently across all the other datasets. This is because AutoML models, such as those based on NAS, can lead to better accuracy and lower error rates compared to manually designed models, due to their ability to automatically search and optimize the model architecture and hyperparameters for a given task.

Furthermore, our analysis revealed that the basic encoder-decoder model performed better than the LSTM model. This is due to the fact that encoder-decoder models are specifically designed for sequence-to-sequence learning tasks, making them more suitable for the application throughput prediction task in our study. The encoder-decoder model's superiority is because it is a more advanced form of LSTM that can handle both input and output sequences, allowing it to capture more information and produce better predictions.

Overall, our findings emphasize the importance of selecting the appropriate model architecture for application throughput prediction and demonstrate the superiority of AutoML models in selecting the optimal model. Using interpretable performance metrics such as MAE and MAPE enables network operators to easily validate the models and make informed decisions, enhancing overall network performance.

\begin{table*}[htbp]
\centering
\caption{MAPE \& MAE: Comparison of Three Models on 4G and 5G Datasets}
\label{table:comparison}
\def\arraystretch{2}
\scriptsize
\begin{tabular}{|P{0.22\linewidth} |P{0.1\linewidth}| P{0.12\linewidth} P{0.12\linewidth} |P{0.12\linewidth} P{0.12\linewidth}|} \hline

\multirow{2}{*}{\textbf{Model}} & \multirow{2}{*}{\textbf{Metric}} & \multicolumn{2}{c|}{\textbf{4G}} & \multicolumn{2}{c|}{\textbf{5G}}\\

& & \textbf{Video Stream} &\textbf{ File Download} & \textbf{Video Stream} & \textbf{File Download} \\ \hline

\multirow{2}{*}{\textbf{Traditional LSTM}}
& \textbf{MAE} & 0.2377	& 0.1827 & 0.1949 & 0.1663 \\
& \textbf{MAPE} & 9.76 \% & 7.28 \%	& 10.87 \%	& 6.52 \% \\\hline

\multirow{2}{*}{\shortstack{\textbf{Traditional Seq2Seq} \\\textbf{Encoder-Decoder}}}
& \textbf{MAE} & 0.1892 & 0.1302 & 0.1478 & 0.1066\\
& \textbf{MAPE} & 7.71 \% & 6.26 \% & 8.91 \% & 4.76 \% \\\hline

\multirow{2}{*}{\textbf{AutoML}}
& \textbf{MAE} & 0.0473 & 0.0209 & 0.0309 & 0.0166 \\
& \textbf{MAPE} & 5.33 \% & 4.59 \% & 5.56 \%	& 3.58 \% \\\hline

\end{tabular}
\end{table*}

\subsubsection{Complexity-Accuracy Trade-off Analysis}
In this section, we employ the AutoML pipeline to predict the application throughput and present a detailed analysis of its performance. We investigate the impact of utilizing different past and future sequences of varying lengths on the prediction accuracy. Our analysis primarily focuses on the 5G dataset for file downloading.

\begin{figure}[htbp]
	\centering
	\includegraphics[width = 0.8\linewidth]{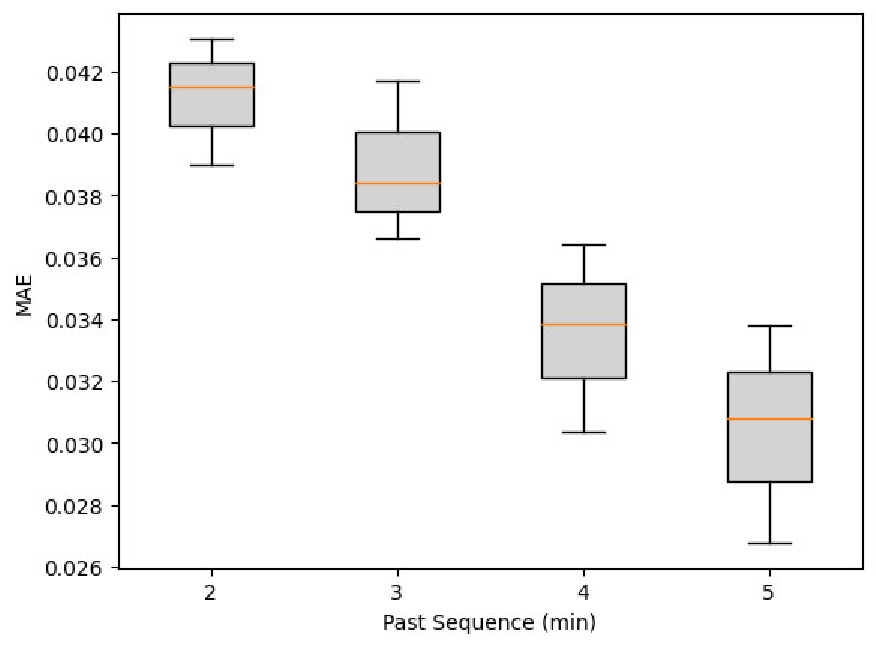}
	\caption[]{MAE among Varying Past Sequences}
	\label{fig:result_boxplot}
\end{figure}

To begin with, we employ a fixed look-ahead of five minutes to predict the application throughput using the previous \textit{n} minutes. We vary the length of the past sequence from 2 to 5 minutes and observe the corresponding changes in the prediction accuracy. Our results, as illustrated in Figure \ref{fig:result_boxplot}, show that utilizing fewer past timesteps leads to a higher MAE, indicating worse predictive performance. Conversely, employing more past timesteps results in a lower MAE but is computationally complex in terms of time and resources. As a longer sequence is utilized, the computational time required to train the model also increases, resulting in higher resource utilization. Thus, there exists a trade-off between the prediction accuracy and complexity, and a balance must be struck to optimize the model's performance.

\begin{figure}[htbp]
	\centering
	\includegraphics[width = 0.8\linewidth]{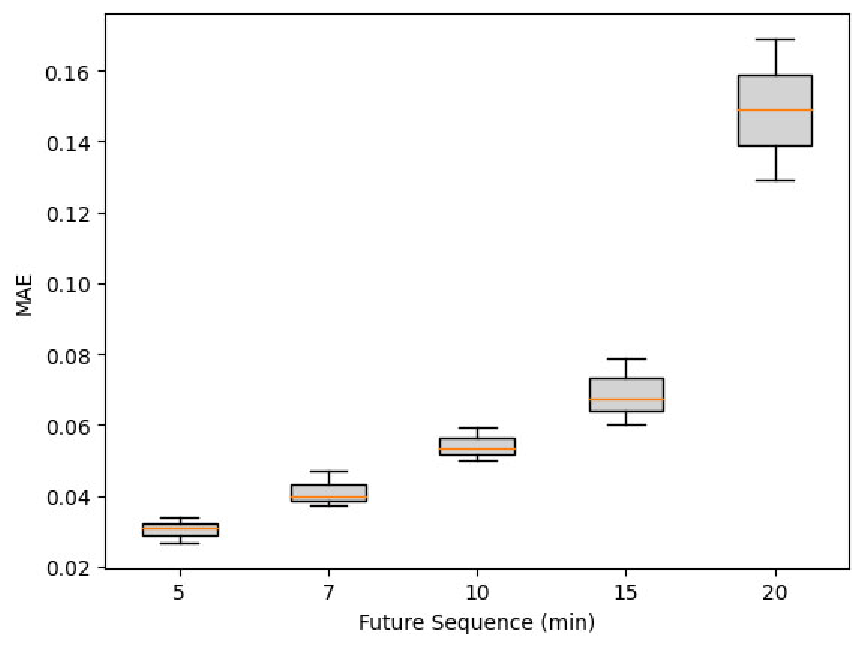}
	\caption[]{MAE vs. Look-ahead Sequence Length}
	\label{fig:result_boxplot2}
\end{figure}

Furthermore, we fix the look-back to 5 minutes and explore the impact of utilizing different future sequences of varying lengths, namely 5, 7, 10, 15, and 20 minutes, on the prediction accuracy. Our results, as illustrated in Figure \ref{fig:result_boxplot2}, show that predicting longer sequences leads to a higher MAE, indicating that more past information is needed to provide better future knowledge. However, increasing the past sequences results in a higher computational complexity. This finding confirms the trade-off between the prediction accuracy and complexity that we previously observed.

Overall, the AutoML pipeline can be effectively utilized to predict the application throughput in the 5G network. However, the prediction accuracy is highly dependent on the number of past and future timesteps utilized and the associated computational complexity. Careful consideration must be given while selecting the appropriate sequence lengths to achieve the desired prediction accuracy while optimizing the computational resources utilized.

\subsubsection{Periodic AutoML Model Drift Monitoring}
This section covers the final step of the AutoML pipeline, which involves model updating. We will focus again on the 5G dataset for file downloading. In this step, we set the forecast horizon to 5 minutes, with resulted in an MAE of $0.0213$. To detect model drift, we monitor the ML model periodically every 10 minutes. If the MAE exceeds $0.02556$, which corresponds to 20\% of the baseline MAE, model drift is detected, and the model weights are adjusted.

\begin{figure}[htbp]
	\centering
	\includegraphics[width = 0.8\linewidth]{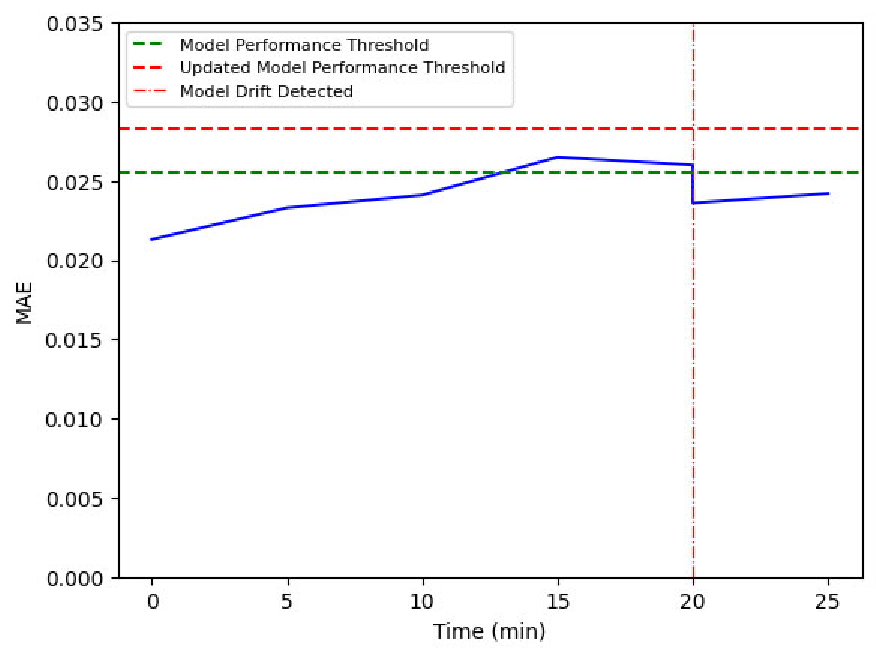}
	\caption[]{Periodic AutoML Monitoring for Drift Detection and Adaptation}
	\label{fig:model_drift}
\end{figure}

It is important to note that the threshold percentage can be adjusted accordingly and is just a parameter. Figure \ref{fig:model_drift} illustrates the timeline of the model monitoring process. For the first 10 minutes ($0 \leq t < 10 mins$), MAE falls below the threshold. At 10 minutes, the model is checked, and no model drift is detected. Between 10 and 14 minutes, MAE still falls below the threshold. At this point in the process, a segment of the data is randomly sampled and intentionally manipulated to mimic the occurrence of data drift.

At 14 minutes, the MAE surpasses the threshold, but the model isn't checked yet. At $t = 20 mins$, the model is checked, and model drift is detected. The weights are updated accordingly, and the MAE falls back to $0.0236$. The threshold is also updated to $0.02832$, which is 20\% of the new MAE, to account for the new data distribution. After minute 20, the MAE doesn't exceed the threshold.

It is essential to note that model drift did occur before it is detected, at $t = 14 mins$, due to the periodic nature of monitoring. Decreasing the monitoring period would have led to earlier detection. However, decreasing the period means checking more often, which may be computationally exhaustive. Therefore, there is a trade-off between the periodic interval and model accuracy.

Ultimately, the model updating step of the AutoML pipeline plays a crucial role in ensuring the model's accuracy over time. By monitoring the model for drift and updating its weights accordingly, we can ensure that the model remains relevant in the face of changing data distributions. However, determining the appropriate monitoring interval is essential to balance the trade-off between model accuracy and computational resources.

\section{Open Challenges \& Future Directions}\label{sec:challenges}
The emergence of cutting-edge innovations such as ZSM and AutoML has brought exciting new opportunities to the networking world. However, despite the headway made in these areas, several challenges remain to be addressed to fully unleash their potential. A summary of these challenges can be found in Table \ref{table:future_work}.

\begin{table*}[htbp]
\centering
\caption{Challenges \& Future Directions}
\label{table:future_work}
\bgroup
\def\arraystretch{1.7}
\scriptsize
\begin{tabular}{|P{1.5cm}|  P{1.9cm} P{5.2cm}  P{3.4cm}|}
\hline
\textbf{Category} & \textbf{Challenge} & \textbf{Description} & \textbf{Future Work} \\
\hline

\multirow{4}{*}{\textbf{ZSM}}
& Explainable Zero-Touch Management &
\begin{tabitemize}
    \item Data-driven decisions must be human-understandable.
    \item XAI enhances trustworthiness.
    \item XAI enables human oversight, allowing network operators to review and approve the AI-driven actions.
\end{tabitemize}
& 
\begin{tabitemize}
    \item Generate an intelligible ML model.
    \item Utilize metrics to quantify the degree of AI explainability. 
\end{tabitemize}
\\ \cline{2-4}

& Trustworthy ZSM &
\begin{tabitemize}
    \item Shared data, among different stakeholders, may contain sensitive information like network topology, user data, and operational data.
    \item Confidentiality, integrity, and availability of the data are crucial.
\end{tabitemize}
& 
\begin{tabitemize}
    \item Utilize blockchain technology.
    \item Encrypt data in transit and at rest.
    \item Incorporate TEEs.
\end{tabitemize}\\ \cline{2-4}

& Computational Complexity in ZSM &
\begin{tabitemize}
    \item ZSM networks struggle with computational complexity in handling large data volumes.
    \item ML algorithms require significant computational resources, which can conflict with the efficiency needs of ZSM networks.
\end{tabitemize}
& 
\begin{tabitemize}
    \item Develop optimization techniques to minimize complexity.
    \item Implement hardware-based strategies (\textit{e.g,} FPGA-based acceleration and GPU
processing)
\end{tabitemize}\\\hline

\multirow{4}{*}{\textbf{AutoML}}
& Interpretability &
\begin{tabitemize}
    \item Current AutoML approaches prioritize accuracy over interpretability, resulting in complex models.
    \item Interpretability is essential for ethics and regulatory compliance in networking.
\end{tabitemize}
&
\begin{tabitemize}
    \item Incorporate XAI paradigms.
    \item Add constraints (\textit{e.g.,} sparsity, monotonicity, and causality).
\end{tabitemize}\\ \cline{2-4}

& Scalability &
\begin{tabitemize}
    \item Large datasets and the need for extensive model training pose scalability challenges for AutoML.
    \item Data processing and model generation times impact network performance, causing latency issues for users.
\end{tabitemize}
& 
\begin{tabitemize}
    \item Employ parallel and distributed algorithms.
    \item Efficiently sample and partition data.
    \item Dynamically adapt hyperparameters.
\end{tabitemize}  \\ \cline{2-4} 

& Robustness & 
\begin{tabitemize}
    \item AutoML struggles with adversarial attacks. 
    \item Non-robust models in ZSM systems lead to unreliable performance and potential security breaches.
\end{tabitemize} & 
\begin{tabitemize}
    \item Apply adversarial training.
    \item Incorporate defensive distillation.
\end{tabitemize}\\ \cline{2-4}

& Cold-Start &
\begin{tabitemize}
    \item Search process begins with sub-optimal models or bad configurations, resulting in inefficient resource usage and prolonged search times.
    \item Resulting delays in ZSM services can have an impact on the user experience.
\end{tabitemize} & 
\begin{tabitemize}
    \item Utilize meta-learning.
    \item Exploit domain-specific knowledge.
    \item Apply transfer learning.
\end{tabitemize}\\
\hline

\end{tabular}
\egroup
\end{table*}

\subsection{ZSM Challenges}
ZSM systems have emerged as a promising solution to automate network operations and improve service delivery and management. However, their adoption comes with significant hurdles, including explainability, trustworthiness, and computational complexity. Addressing these challenges is critical for the future success of ZSM systems, and it requires further research and development efforts to ensure the effective integration of ZSM in NGNs.

\subsubsection{Explainable Zero-Touch Management}
To successfully employ ZSM, its data-driven decisions must be human-understandable. In this light, the European Commission approach to AI centers on excellence, trust, and transparency, which will play a crucial role in NGNs and the establishment of quality of trust \cite{EUwhitepaper, qot}. Explainable Artificial Intelligence (XAI) is essential for ZSM as it provides a rationale for the actions taken by AI systems, making them more trustworthy. This is essentially important in services like remote surgery, where network management decisions can have a significant impact on human lives \cite{zsm5g6g}. XAI allows networking experts to understand the input that drove the decisions made by ML models and approve their actions following a human-in-the-loop model.

However, making XAI a reality in the ZSM paradigm requires an intelligible ML model in addition to specific metrics to measure the level of AI explainability \cite{zsm5g6g}. Techniques to generate an intelligible ML model include using inherently interpretable ML models (\textit{e.g.}, random forests) or statistical procedures to describe the features on which a prediction was based. As for the metrics, there are human-grounded evaluations and functionality-grounded evaluations \cite{xaimetrics}. The former assesses the qualitative aspects of the resulting explanations, such as their ability to assist humans in completing tasks and the impact of such decisions on the system. The latter relies on formal definitions and quantitative methods to verify data-driven decisions, such as service migration. 

\subsubsection{Trustworthy ZSM}
One of the major challenges in ZSM is the secure data sharing among different stakeholders, including network operators, service providers, and third-party vendors. The data shared between these stakeholders can include sensitive information, such as network topology, user data, and operational data, making it critical to ensure the confidentiality, integrity, and availability of the data.

To establish trust, different approaches can be utilized, such as blockchain technologies, encryption techniques, and Trusted Execution Environments (TEEs). Blockchain technology, for instance, not only ensures data governance but also promotes multi-party trust and data usage accountability \cite{distributedledger}. Additionally, encrypting data in transit and at rest provides an extra layer of protection against unauthorized access \cite{encyrptchallenge}. Incorporating TEEs, which are secure areas of a processor that enable the execution of trusted code and data, is another promising solution \cite{tee}. By integrating hardware-based TEEs into network infrastructures, network operators can ensure that critical operations, such as data sharing and network management, are performed with a high degree of trust. 

\subsubsection{Computational Complexity in ZSM}
ZSM networks can be severely challenged by the computational complexity of managing the massive amount of generated data. This includes real-time data analysis, network resource optimization, and coordination of various network functions. ML algorithms, in particular, demand a high level of computational resources to operate efficiently, which can conflict with the needs of ZSM networks, where computation efficiency is as essential as communication performance. In NGNs, the high latency associated with complex operations is incompatible with time-sensitive services, making ML algorithm optimization a key factor. Accordingly, it is essential to develop optimization techniques to minimize the complexity of these models without jeopardizing accuracy. To this end, implementing hardware-based strategies, such as FPGA-based acceleration and GPU processing, can reduce the computational complexity of ZSM \cite{aizsm5gp}.

\subsection{AutoML Challenges}
Despite the many advantages of AutoML, it is not a silver bullet, and there are several challenges that need to be addressed. These include devising an efficient search process, building a scalable system to handle big data, addressing security concerns posed by adversarial attacks, and ensuring that the models are interpretable and transparent. Nevertheless, the potential benefits of AutoML are significant, and it is an area that is likely to see continued growth and development in the years to come.

\subsubsection{Interpretability}
AutoML solutions are often seen as black boxes, which makes it difficult for users and experts to fully understand how they work and the rationale behind their solutions. However, interpretability is essential for building trust and ensuring ethical considerations, especially in highly regulated domains, such as the networking domain.

The lack of interpretability in AutoML models can lead to difficulties in deploying and using these models in ZSM services. For example, it may be difficult to diagnose and correct biases in the model, or to identify the root cause of unexpected behaviors. Additionally, the lack of interpretability can make it difficult to validate the model's accuracy, which is critical for ensuring the reliability and performance of the network. Therefore, the development of transparent AutoML systems with mechanisms for explaining and understanding their decisions is necessary. Unfortunately, many current AutoML approaches prioritize accuracy over interpretability, resulting in complex models that are difficult to comprehend.

To address this challenge, interpretable models leveraging XAI paradigms, such as Shapley additive explanations, local interpretable model-agnostic explanations, RuleFit, and partial dependence plots, can be used to increase transparency and credibility \cite{XAI}. Additionally, constraints such as sparsity (i.e., low number of features), monotonicity, and causality can improve interpretability \cite{autoMLchallengesInterp}. Monotonicity guarantees that the relationship between an input feature and the target outcome always goes in the same direction, aiding in the understanding of feature-target relationships. Causality constraints ensure that only causal relationships are identified, promoting effective interactions between humans and ML systems. Incorporating these approaches can greatly improve AutoML's accessibility and the ability of network operators and other stakeholders to understand and trust the decisions made by AutoML.

\subsubsection{Scalability}
The growing size of datasets, coupled with the need for an overwhelming number of model trainings to determine the optimal final learner, present significant scalability challenges for AutoML. This can be particularly challenging in ZSM services, where the large volume of network-generated data requires fast processing to meet network demands. As model complexity increases, so do computational requirements, making it challenging to deploy models in resource-constrained environments. Additionally, network performance can be impacted due to the time required to process data and generate models, leading to latency and responsiveness issues that affect the user experience.

To address this issue, future research can focus on developing parallel and distributed AutoML algorithms that can harness modern hardware such as graphic processing units. Additionally, techniques that can more efficiently sample the data, leverage data partitioning, or dynamically adapt the algorithm's hyperparameters can significantly reduce the computational overhead. Such strategies will ensure that AutoML remains a powerful tool for ML, irrespective of the dataset's size.

\subsubsection{Robustness}
AutoML, particularly NAS, has shown remarkable performance on well-labeled datasets such as ImageNet \cite{imagenet}. However, real-world datasets inevitably contain noise and adversarial examples, which can significantly undermine the performance of AutoML models \cite{autoMLsurvey}. Adversarial attacks can be specifically designed to fool the model, compromising its performance.

The deployment of non-robust models in ZSM systems can lead to unreliable network performance, which can have a significant impact on the overall user experience. Additionally, the lack of robustness can result in potential security breaches. Adversarial attacks can be used to exploit vulnerabilities in non-robust models, allowing attackers to gain unauthorized access to the network or manipulate network behavior. This can have serious consequences, such as compromising the confidentiality and integrity of user data, disrupting network operations, and causing financial losses. Therefore, ensuring the robustness of AutoML models is critical to their successful application and safe deployment in ZSM systems.

AutoML systems can improve their robustness to adversarial attacks by incorporating techniques such as adversarial training and defensive distillation in their pipelines. Adversarial training can enhance robustness by training the model with a combination of clean and adversarial data \cite{advtrain}. This exposes the model to a range of adversarial attacks during training, making it more robust to such attacks at inference time. On the other hand, defensive distillation is a technique that distills knowledge from a large robust model (teacher model) into a smaller target model (student model) \cite{defdist}. This enables the robustness of the teacher model to be transferred to the student model through knowledge distillation, yielding a more robust student model.

\subsubsection{Cold-Start}
AutoML systems may undergo a cold-start, where the search process starts with a sub-optimal model or a bad configuration, resulting in inefficient resource usage and prolonged search times \cite{metalearning1}. This can be particularly problematic in ZSM services, where processing time is crucial, and delays can have a significant impact on the user experience. One possible technique to warm-start the search process is meta-learning. By leveraging prior knowledge from similar datasets, meta-learning can initialize the search process with a promising configuration obtained from that previous knowledge. Accordingly, the search space is reduced, and the search process is accelerated \cite{metalearning2}.

Furthermore, incorporating domain-specific knowledge can design a more efficient search space, improving the initialization of the search process. For example, in image classification tasks, domain-specific knowledge can be utilized to define a search space that contains CNNs with specific architectures.

Transfer learning can also aid in warm-starting the search process by providing a good initialization of the model's weights, reducing the time required for model training. Through transfer learning, AutoML systems can leverage knowledge learned from related domains (pre-trained model) to improve the performance of the target model.

\section{Conclusion} \label{sec:conc}
Next-Generation Networks (NGNs) have unleashed a remarkable shift in the telecommunications industry, opening up an array of possibilities for applications and service areas with diverse needs. While NGNs hold tremendous promise to fulfill the demanding requirements of future use cases, they must be devised as highly-adaptable infrastructures using cutting-edge technologies, such as software defined networking, network function virtualization, and network slicing.

Nevertheless, as networks grow more complex, traditional manual approaches for network management become less efficient. Consequently, Zero-touch network and Service Management (ZSM) has emerged as a fully automated management solution designed to introduce intelligence into mobile networks for the purpose of automation and optimization. As explored in this survey, ZSM has the potential to optimize network resources, boost energy efficiency, enhance security, and manage traffic in NGNs. However, it also confronts significant ML challenges, such as the need for effective model selection and hyperparameter tuning. The paper explores viable network automation solutions, specifically Automated Machine Learning (AutoML) and digital twins.

AutoML is one solution to these issues by automating the ML pipeline within ZSM itself and thus increasing its efficiency. This paper thoroughly analyzes the AutoML pipeline, providing insights into the techniques utilized at each step. The practical application of AutoML is demonstrated through a case study that predicts application throughput for 4G and 5G networks using an online AutoML pipeline. Simulation results prove the superiority of AutoML over ML approaches. By leveraging AutoML algorithms to generate  up-to-date predictive models, ZSM can adapt to changing traffic patterns. This facilitates the automation of network service management, leading to improved service quality and enhanced operational efficiency.

While ZSM has shown promise across diverse domains, much work remains to be done to refine and incorporate this framework. Nonetheless, the potential for NGNs to revolutionize the way we live, work, and communicate remains as high as ever. ZSM and AutoML will play a pivotal role in realizing this potential.

\bibliographystyle{elsarticle-num}
\bibliography{ms}
\end{document}